\documentclass{emulateapj}
 
\shorttitle{Stellar obliquities and magnetic activities from Transit Chord Correlation}

\shortauthors{Dai et al.}

\usepackage{amsmath}
\usepackage{epstopdf}
\usepackage{apjfonts}
\usepackage{xspace}
\usepackage{hyperref} 
\usepackage{bm}
\usepackage{graphicx}
\usepackage{rotating}
\begin{document}


\title{Stellar obliquities and magnetic activities of Planet-Hosting Stars and Eclipsing Binaries
  based on Transit Chord Correlation}


\author{
  Fei\ Dai\altaffilmark{1,2},
  Joshua N.\ Winn\altaffilmark{2},
  Zachory Berta-Thompson\altaffilmark{3},
  Roberto Sanchis-Ojeda\altaffilmark{4},
  Simon Albrecht\altaffilmark{5}
  }


\altaffiltext{1}{Department of Physics and Kavli Institute for
  Astrophysics and Space Research, Massachusetts Institute of
  Technology, Cambridge, MA 02139, USA {\tt FD:284@mit.edu}}

\altaffiltext{2}{Department of Astrophysical Sciences, Peyton Hall, 4
  Ivy Lane, Princeton, NJ 08540 USA}

\altaffiltext{3}{Department of Astrophysical and Planetary Sciences,
  University of Colorado at Boulder, Duane Physical Laboratories,
  Colorado Ave, Boulder, CO 80305 USA}

\altaffiltext{4}{Netflix, 100 Winchester Circle, Los Gatos, CA 95032 USA}

\altaffiltext{5}{Stellar Astrophysics Centre, Department of Physics
  and Astronomy, Aarhus University, Ny Munkegade 120, DK-8000 Aarhus
  C, Denmark}


\begin{abstract}
  
  \noindent The light curve of an eclipsing system shows anomalies
  whenever the eclipsing body passes in front of active regions on the
  eclipsed star.  In some cases, the pattern of anomalies can be used
  to determine the obliquity $\Psi$ of the eclipsed star.  Here we
  present a method for detecting and analyzing these patterns, based
  on a statistical test for correlations between the anomalies
  observed in a sequence of eclipses.  Compared to previous methods,
  ours makes fewer assumptions and is easier to automate.  We apply it
  to a sample of 64 stars with transiting planets and 24 eclipsing
  binaries for which precise space-based data are available, and for
  which there was either some indication of flux anomalies or a
  previously reported obliquity measurement.  We were able to
  determine obliquities for ten stars with hot Jupiters. In particular
  we found $\Psi \lesssim$10$^\circ$ for Kepler-45, which is only the
  second M dwarf with a measured obliquity.   The other
  8 cases are G and K stars with low obliquities.  Among the eclipsing
  binaries, we were able to determine obliquities in 8 cases, all of
  which are consistent with zero.  Our results also reveal some common
  patterns of stellar activity for magnetically active G and K stars,
  including persistently active longitudes.

\end{abstract}


\section{Introduction}

The Solar System is nearly coplanar, with the Sun's equator and the
planetary orbits aligned to within a few degrees.  This makes sense
because the Sun and the planets likely formed from a flattened disk of
material with a well-defined sense of rotation.  However, the
discovery of planets with orbits nearly perpendicular to the star's
equator \citep[e.g., HAT-P-11b,][]{Winn2010}, or nearly retrograde
\citep[e.g., WASP-17b,][]{Triaud2010} show that planet formation and
evolution can be more complicated than in this simple picture.  There is
hope that the distribution of obliquities, and its dependence on
stellar and planetary properties, will help to elucidate the formation
history and orbital evolution of planets \citep[as recently reviewed
  by][]{WinnFabrycky2015,Triaud2017}.  Many methods are available
for measuring stellar obliquities: the Rossiter-McLaughlin effect
\citep{Queloz2000}; the $v \rm{sin} i_\star$ method
\citep{Schlaufman2010, Winn2017vsini}; the photometric variability
method \citep{Mazeh+2015}; the gravity-darkening method
\citep{Barnes2009, Masuda2015}; the asteroseismic method
\citep{Chaplin2013, Huber2013,VanEylen2014}; and the method based on
spot-crossing anomalies \citep{Sanchis-Ojeda2011WASP, Desert2011,
  Mazeh2015, Holczer2015}, the subject of this paper.

When a transiting planet is blocking a dark starspot, the loss of
light is smaller than when it is blocking an unspotted portion of the
photosphere.  This produces a positive glitch in the light curve.
Similarly, transits over bright regions (plages and faculae) produce
negative glitches.  In some cases the stellar obliquity can be deduced
from the pattern of recurrence of spot-crossing anomalies in a
sequence of transits.  As a simple example, consider a low-obliquity
star with long-lived starspots and a rotation period $P_{\rm rot}$
much longer than the planet's orbital period $P_{\rm orb}$.  In such a
case, the anomalies that are seen during one transit will recur in the
next transit, but shifted forward in time because the active regions
have advanced across the the visible hemisphere of the star due to
stellar rotation.  Conversely, if the stellar obliquity is high, the
rotation of the star carries the active regions outside of the
``transit chord'', the narrow strip on the stellar disk that is
traversed by the transiting planet.  In this case, the anomalies
produced by a given spot will not recur in subsequent transits.

Unlike some of the other methods for measuring obliquities, the
spot-crossing method does not require high-resolution spectroscopy,
and can therefore be applied to relatively faint stars.  However, the
spot-crossing method does require a large number of consecutive light
curves with a high signal-to-noise ratio (SNR) to allow spot-crossing
events and their recurrence to be detected.  As a result, it has only
been applied to about a dozen systems.  Most of these were discovered
by the space missions {\it CoRoT} and {\it Kepler}.  The method of
analysis has generally relied on the identification of discrete
anomalies, often through visual inspection, a laborious and somewhat
subjective procedure.  Most of the modeling has been performed with
idealized assumptions such as circular spots of uniform intensity.

The motivation for our work was to develop a more objective method
that does not rely on explicit spot modeling, and can be more easily
applied to a large sample of systems.  Instead of assuming that the
active regions are discrete dark and bright spots, we treat the
transit light curve as a measure of the intensity distribution of the
stellar photosphere along the transit chord. We do not model the
intensity distribution with discrete spots, although we still must
hope that the active regions persist and are nearly stationary in
the rotating frame of the star for at least a few orbital periods.
Given the values of $P_{\rm rot}$ and $P_{\rm orb}$, we can calculate
the angle by which the active regions should advance in between
transits if the obliquity is low.  Then we can seek evidence for
correlations --- with the appropriate lag --- between the anomalies
observed in a series of transits.  A significant correlation implies a
low obliquity.  We can search for evidence of retrograde rotation in a
similar way.

For convenience we call this the Transit Chord Correlation (TCC)
method, although we do not claim it is a completely new concept.  It
is closely related to eclipse mapping \citep{Horne1985}, which has
long been used to probe the brightness distribution of stars and
accretion disks.  Eclipse mapping has also been applied to a couple of
stars with transiting planets \citep{Huber2010corot, Scandariato2017}.
The main difference is that the previous investigators assumed zero
obliquity and sought to determine the brightness distribution across
the transit chord, while we use the method to try and determine the
obliquity.

Similar in spirit to our method is the one described by
\citet{Mazeh2015} and \citet{Holczer2015}.  They were able to
distinguish prograde and retrograde motion in a few {\it Kepler}
systems by searching for a significant correlation between the
observed transit timing variations (TTV) and the time derivative of
the stellar flux immediately before and after transits.  Stars with
prograde rotation should show a correlation if a single active region
is responsible for both the apparent TTV and the out-of-transit
variation, while retrograde stars should show an anti-correlation.
Two advantages of this method are its simplicity and the freedom from
the requirement that the active regions have a lifetime of a few
orbital periods.  One problem is that the out-of-transit variations
are the net effect of all the active regions on the star, and are not
necessarily dominated by the region that is responsible for the
transit anomalies.  Our TCC method does not rely on the out-of-transit
variations, and is also able to provide higher precision in the
obliquity determination.

In this paper, we explain the TCC method, validate it through
application to systems for which the stellar obliquity has been
measured by independent methods, and apply it to all the transiting
planets for which the method is currently feasible.  We also apply the
TCC method to a sample of eclipsing binaries drawn from the {\it
  Kepler} survey.  Section 2 describes the TCC method in greater
detail.  Section 3 describes the target selection and light curve
preparation and analysis.  Section 4 presents the results for the
obliquities, as well as some interesting features we noticed in the
pattern and time evolution of active regions.  Section 5 discusses
these results in the context of what was previously known about both
stellar obliquities and stellar activity.

\section{Method}
\label{Method}

\begin{figure*}
\begin{center}

\includegraphics[width = 1.\columnwidth]{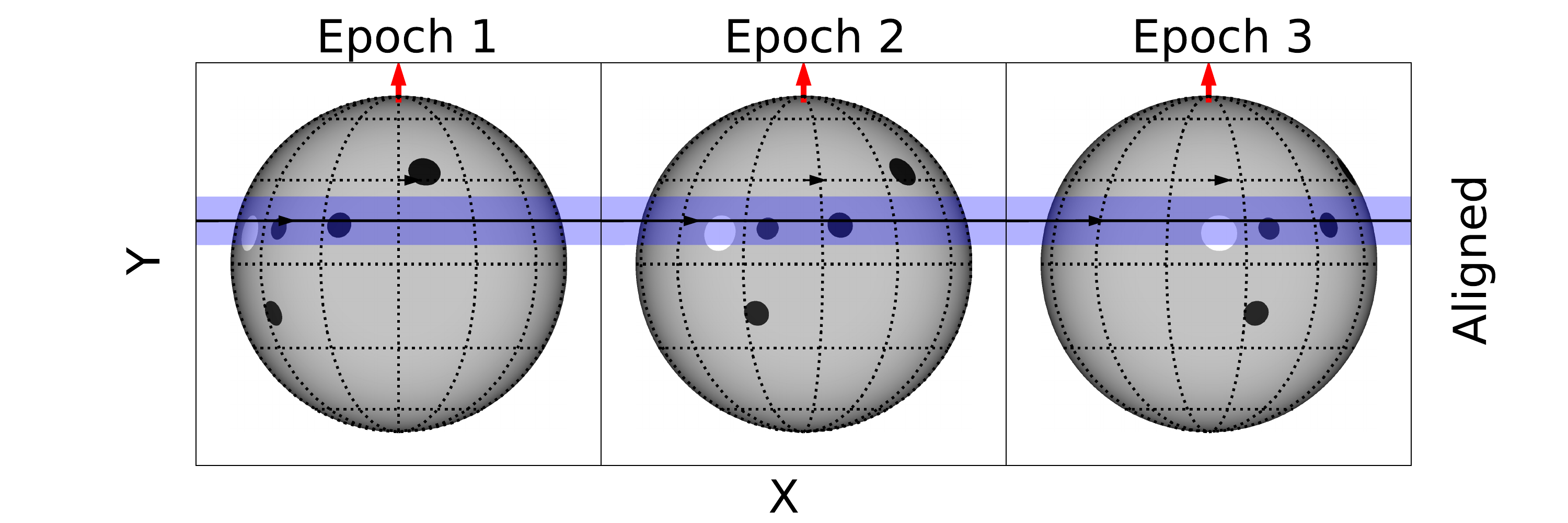}
\includegraphics[width = 1.\columnwidth]{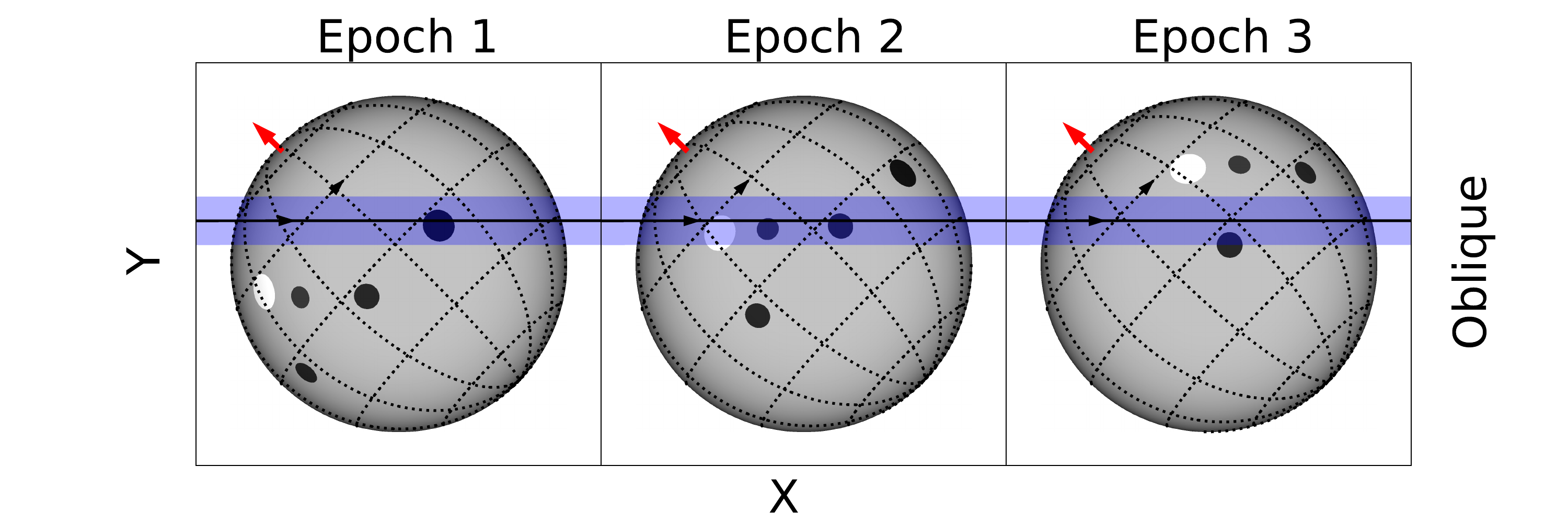}
\includegraphics[width = 1.\columnwidth]{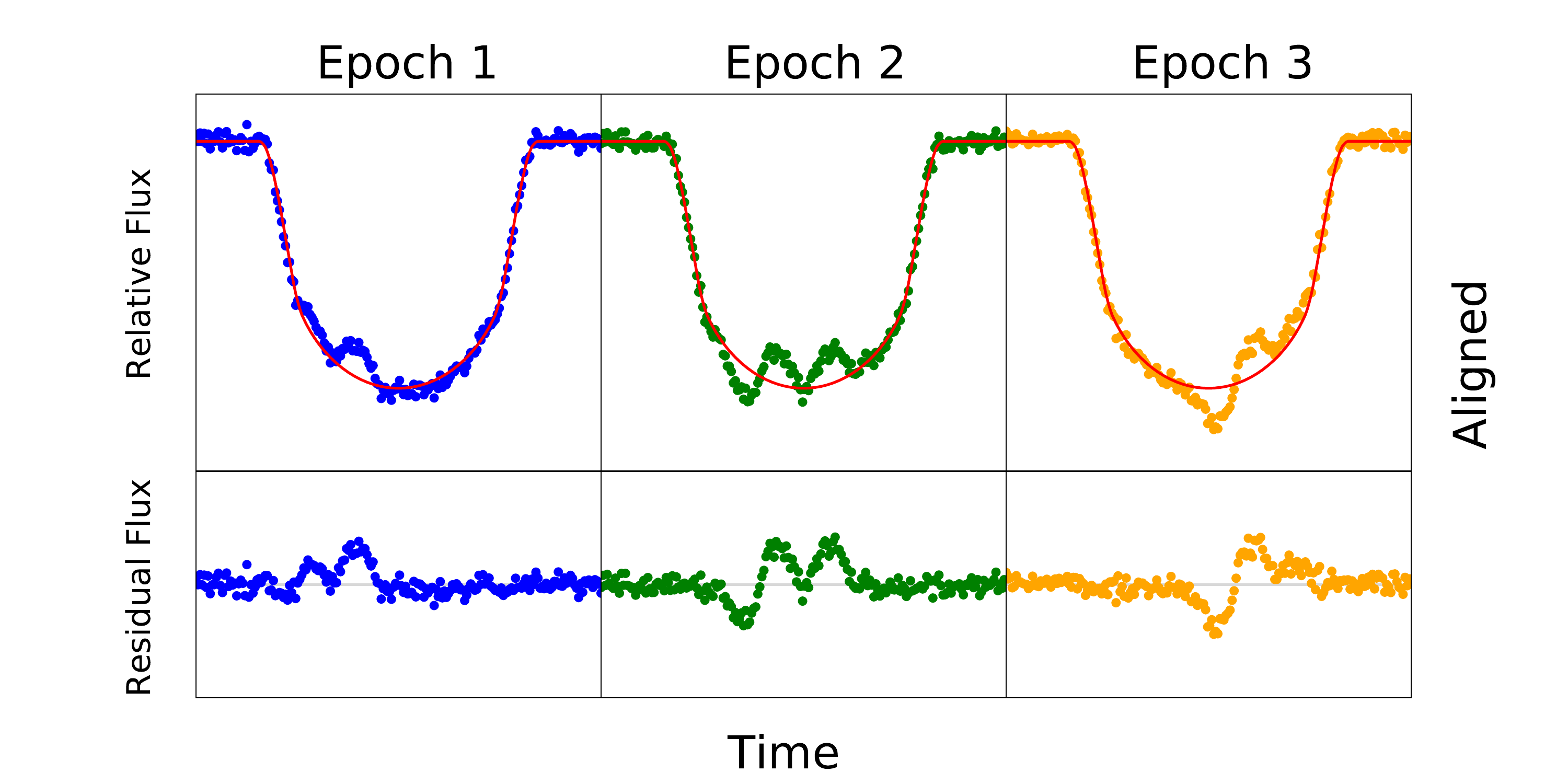}
\includegraphics[width = 1.\columnwidth]{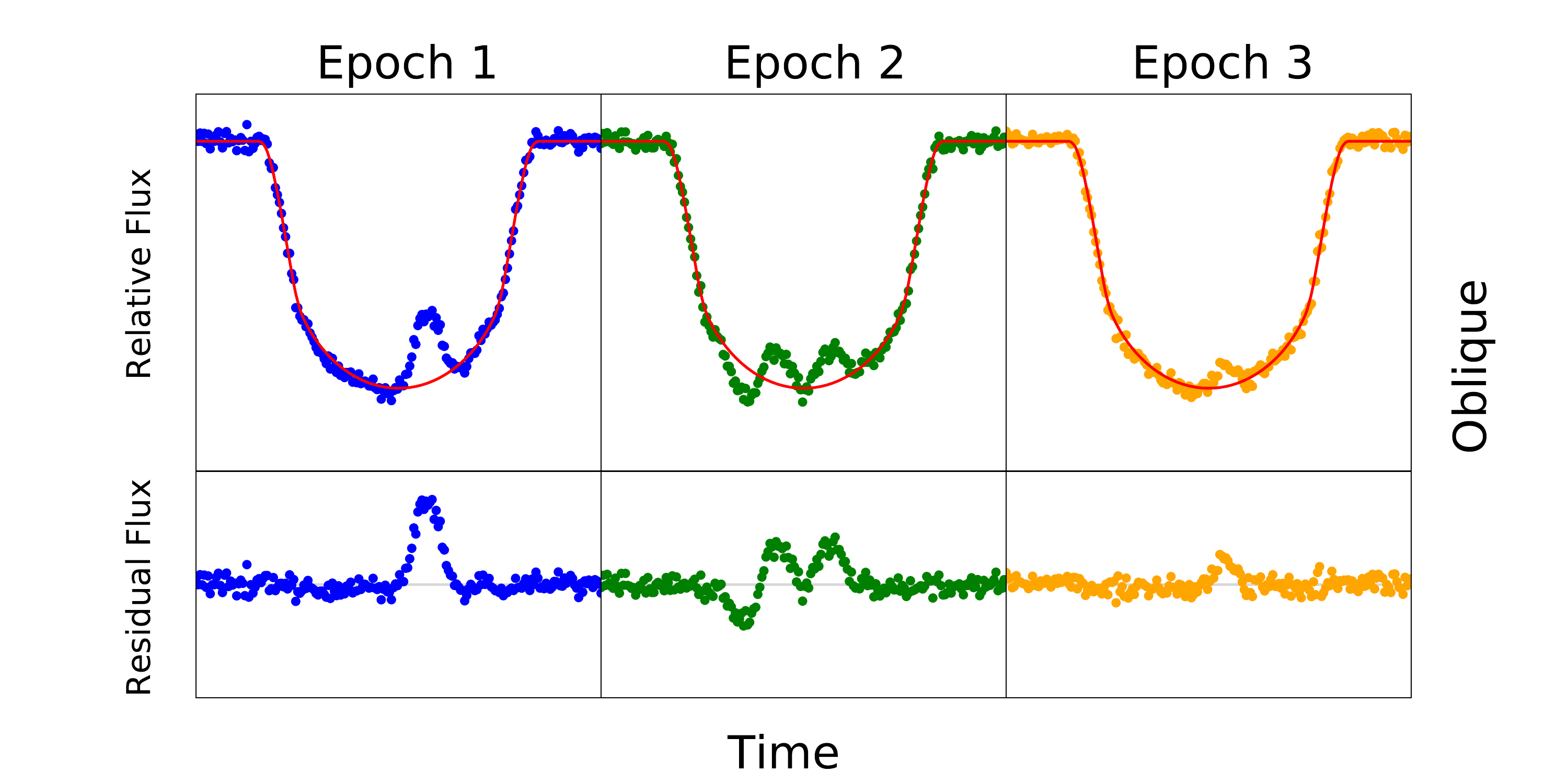}
\includegraphics[width = 1\columnwidth]{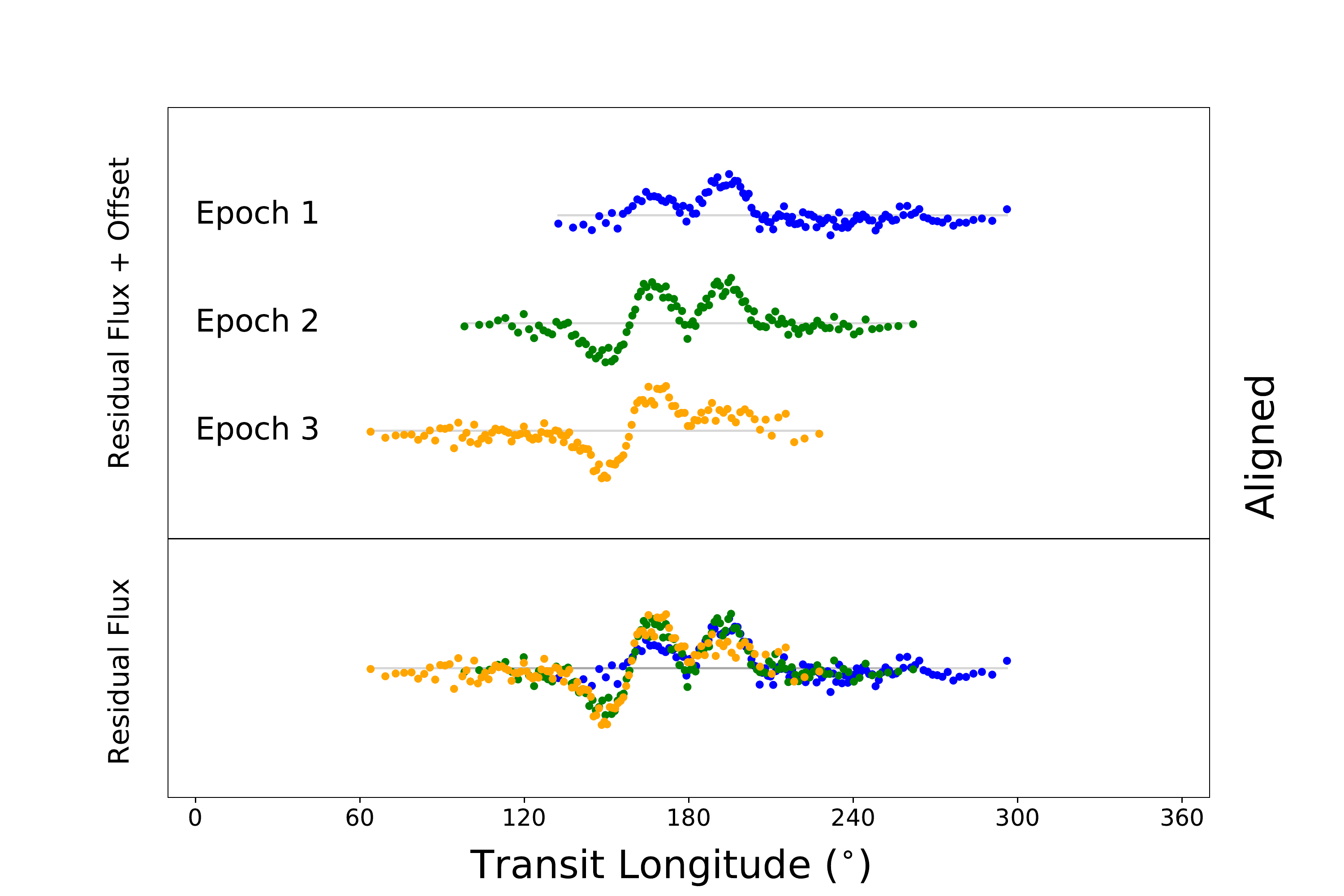}
\includegraphics[width = 1\columnwidth]{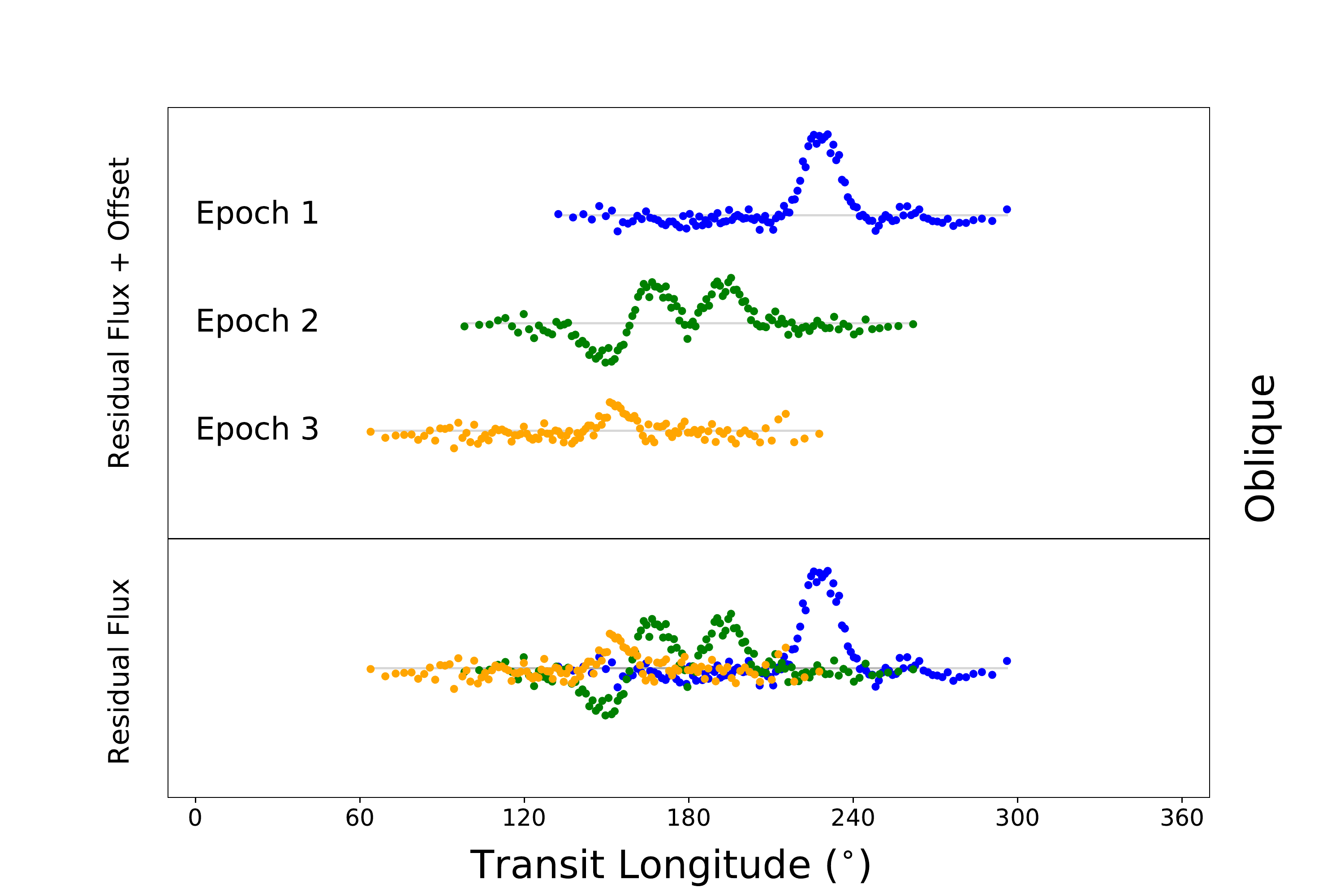}

\caption{{\bf Top: Transits of a planet across a star with active
    regions.}  The red arrows and dotted lines convey the orientation
  of the rotating host star. The solid line is the trajectory of the
  transiting planet and the blue bar is the
  transit chord. The black and white circles are active regions
  (spots and faculae) on the host star. Three consecutive transits are
  shown, for each of two different geometries: ``Aligned'' with
  stellar obliquity $\Psi \approx 0^{\circ}$, and ``Oblique'' with $\Psi \approx 45^{\circ}$.
  The two cases are initialized such that the light
  curves are identical in Epoch 2.\\
  {\bf Middle: Corresponding light curves.}
  A glitch occurs when the planet transits across
  an active region.  The red solid curves are the best-fitting transit
  models, which do not account for the active regions.  When
  $\Psi \approx 0^\circ$, the anomalies advance in time from one transit
  to the next, due to stellar rotation.\\
  {\bf Bottom: Residual flux as a function of transit longitude.}
  Knowing the stellar rotation
  period, we transform the time stamps of the data into the
  transit longitude (see Equation~\ref{eqn:longitude}).
  In the Aligned case, the planet repeatedly eclipses
  the same set of active regions,
  leading to recurring glitches in the
  residual flux at the same transit longitudes.
  In the Oblique case, the active
  regions rotate outside of the transit chord.  No recurring
  patterns are observed.  The Transit
  Chord Correlation method looks for statistically significant correlations
  in the residuals as a function of transit longitude.}
\label{sky_model}
\end{center}
\end{figure*}

\subsection{Conceptual illustration}

Figure~\ref{sky_model} illustrates the concept of the TCC method.
This figure shows the face of a star and the corresponding light
curves of three consecutive transits.  On the left are the light
curves for a well-aligned star ($\Psi \approx 0^{\circ}$) with a random
pattern of active regions that slowly rotates across the visible
hemisphere.  The anomalies are seen to move progressively to the right
along the time axis.  We isolate the anomalies by subtracting the
best-fitting transit model.  Then we transform the time coordinate
into the longitude $\Phi$ of the star that is being crossed by the
planet, using the equation
\begin{equation}
  \label{eqn:longitude}
  \Phi(t) = \sin^{-1} \left[ \frac{x(t)}{\sqrt{1-y(t)^2}} \right] -
  \frac{2\pi(t-t_{\rm ref})}{P_{\text{rot}}}
\end{equation}
where $x(t)$ and $y(t)$ are the sky-plane coordinates of the planet in
the system defined by \citet{Winn2010b}, in which the origin is at the
center of the stellar disk and the $x$ direction is aligned with the
planet's trajectory.  The reference time $t_{\rm ref}$ is arbitrary;
we adopt the usual {\it Kepler} convention of BJD\,2454833.  Note that
$\Phi$ is the longitude of the star in its own rotating frame.  Thus,
active regions maintain a constant value of $\Phi$ even as they rotate
across the star's visible hemisphere, as long as they do not evolve or
migrate significantly on the timescale of the orbital period of the
planet (typically 1--10 days, for the planets considered in this
work).  We also note that this transformation assumes that $P_{\rm
  rot}$ is a constant and thereby neglects the effects of latitudinal
differential rotation.  When the star has a low obliquity, $P_{\rm
  rot}$ should be understood as the rotation period at the latitude of
the transit chord.

Now that we have calculated the stellar longitude that is being
blocked by the planet as a function of time, we can plot the
light-curve residuals as a function of stellar longitude. This is
shown in the bottom panels of Figure \ref{sky_model}.  The three
patterns of residuals are seen to be very similar, as expected, since
we have assumed that the active regions are stationary in the star's
reference frame. The slight change of the observed pattern between
transits are caused by geometric foreshortening and limb darkening of
the active regions.  In this well-aligned case, we would observe
strong correlations between the residuals of successive transits.

The right side of Figure~\ref{sky_model} shows similar illustrations
for a star that has an obliquity $\Psi$ of 45$^\circ$.  In this case,
the active regions rotate across the transit chord, rather than along
the chord.  When a naive observer transforms time into stellar
longitude under the assumption $\Psi \approx 0^\circ$, the residuals show no
correlation.  This is because the anomalies that are seen in
successive transits are produced by different active regions.  In this
case $\Phi$, as calculated by Equation~\ref{eqn:longitude}, is not
really the stellar longitude.  To be more general we will refer to
$\Phi$ as the {\it transit longitude} rather than the stellar
longitude; it corresponds to the stellar longitude only when the star
has a low obliquity.

The TCC (defined below) is a statistic that quantifies the degree of
correlation between the residuals of successive transits as a function
of transit longitude.  If the obliquity is low and the SNR is high
enough, the TCC should be high when the correct stellar rotation
period $P_{\rm rot}$ is used to calculated $\Phi$ in Equation
\ref{eqn:longitude}. The high TCC indicates there is a pattern of
active regions beneath the transit chord that is stationary in stellar
longitude, which can only happen for a low obliquity.  A similar test
for perfectly retrograde obliquities can be performed by switching the
sign of the second term in Equation~\ref{eqn:longitude}.

\begin{figure*}
\begin{center}
\includegraphics[width = 1.5\columnwidth]{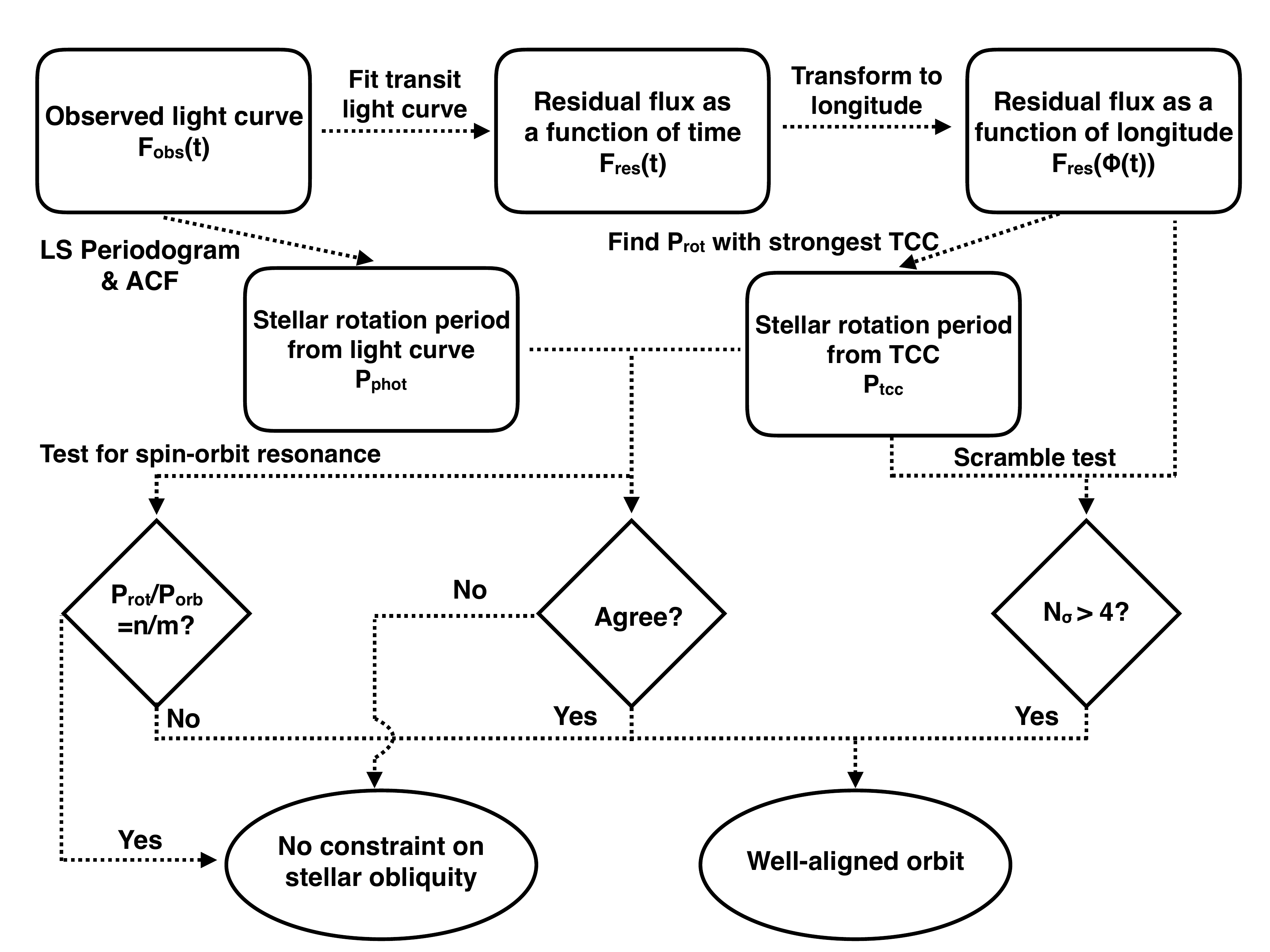}
\caption{{A flow chart illustrating the major steps in the TCC method.}}
\label{flow_chart}
\end{center}
\end{figure*}

\subsection{Light curve fitting}

Figure~\ref{flow_chart} summarizes the TCC method with a flow chart.
We begin with the observed light curve.  We isolate the individual
transits, retaining only the segments of data spanning twice the
transit duration and centered on each transit midpoint.  The transit
data are used for the TCC computation.  The rest of the data are used
only to estimate the stellar rotation period.  For this purpose we
look for the strongest peak in the Lomb-Scargle periodogram of the
light curve, and designate this as $P_{\rm phot}$, the photometric
rotation period, with an uncertainty equal to the full width at half
maximum of the peak.  We also estimate the stellar rotation period
based on the auto-correlation function \citep{McQuillan2013,
  McQuillan2014}, and consider $P_{\rm phot}$ to be securely measured
when the results from the periodogram and the auto-correlation
function are in agreement.

We then fit a simple transit model assuming no active regions, and
compute the residuals between the data and the best-fitting model.
For the model we use the {\tt Batman} code \citep{Kreidberg2015}, with
the following parameters: the orbital period $P_{\rm orb}$; the
planet-to-star radius ratio ($R_p/R_\star$); the scaled orbital
distance ($a/R_\star$); the impact parameter ($b \equiv a\cos
i/R_\star$); the quadratic limb-darkening coefficients ($u_1$ and
$u_2$), the transit midpoint ($t_0$); and two coefficients of a
quadratic function of time to account for the longer-timescale stellar
flux variation near the time of the transit.  We adopt the usual
$\chi^2$ likelihood function and obtained the maximum-likelihood
solution using the Levenberg-Marquardt algorithm as implemented in the
{\tt Python} package {\tt lmfit} \citep{lmfit}.

After each transit is fitted individually we ``rectify'' the light
curves, dividing the data by the best-fitting quadratic function
of time.  Then we fit all the rectified light curves together with a
reduced set of parameters: $P_{\rm orb}$, $R_p/R_\star$, $a/R_\star$,
$b$, $t_0$, $u_1$ and $u_2$.

Next we allow for the possibility of TTVs, and for transit depth
variations caused by untransited active regions.  When the untransited
portions of the star are relatively faint, the loss of light due to
the planet increases, and vice versa.  We account for these
effects by fitting the individual light curves again but with only two
adjustable parameters: the time of transit ($t_0$), and an additive
constant ($\Delta F$) to account for the overall changes in flux of
the star.  We hold the other parameters fixed at the values determined
in the preceding joint fit of all the light curves.  The model for the
observed transit light curve is then
\begin{equation}
  F_{\text{calc}} = \frac{F_{\text{tra}} + \Delta F} {1 + \Delta F},
\end{equation}
where $F_{\text{tra}}$ is the transit model calculated by {\tt Batman}
and the denominator sets $F_{\text{calc}} = 1$ outside of transits,
the same normalization that was adopted for the data.
To prevent overfitting we require $|\Delta F|$ to be smaller
than the observed peak-to-peak variation of the relative
flux across the entire light curve. After this final fit to each light
curve, we record the residual flux $F_{\text{res}} \equiv
F_{\text{obs}}- F_{\text{calc}}$, and transform the time stamps into
transit longitude $\Phi$ using Equation~\ref{eqn:longitude}.

\subsection{Searching for Transit Chord Correlations}

\begin{figure*}
\begin{center}
  \includegraphics[width = 2.2\columnwidth]{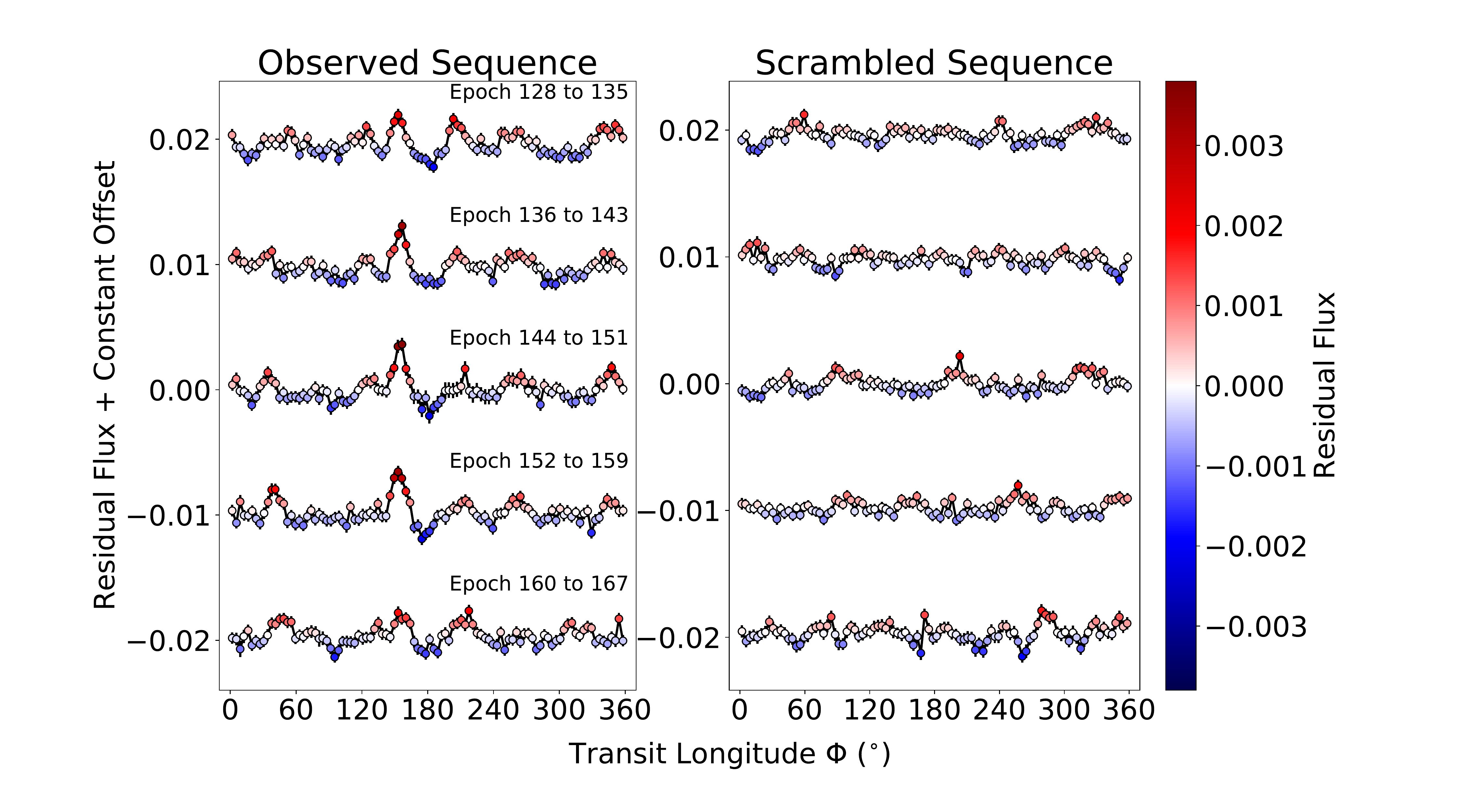}
  \caption{ {\bf Residual fluxes for Kepler-17b.} The colored circles
    show the binned residual flux as a function of transit longitude.
    As in Figure~\ref{Kepler17_tapestry}, positive residuals are red
    and negative residuals are blue.  The left panel shows data from
    five consecutive groups of transits.  Each group consists of eight
    transits, which together provide complete longitude coverage from 0 to
    360$^{\circ}$. A coherent pattern recurs from one group to the
    next. This is indicative of a low-obliquity orbit, as illustrated
    in Figure \ref{sky_model}. The right panel shows the same dataset
    after scrambling the epoch numbers of the transits. No recurring
    pattern is seen.}
  \label{Kepler17_residuals}
\end{center}
\end{figure*}

We are now in the position to seek statistical evidence for the
recurring pattern in the residuals that one would expect for a
low-obliquity system.  First we group together the residuals from a
certain number $N_{\rm tra}$ of consecutive transits.  The reason for
grouping is two-fold: (1) each transit only probes the visible side of
the photosphere, and we need at least a few transits to obtain
complete longitude coverage; (2) grouping transits together enhances
the SNR.  The case $N_{\rm tra}$ = 1 corresponds to no grouping.  To
illustrate the method, Figure~\ref{Kepler17_residuals} shows the data
for Kepler-17, a young and active G star with a 1.5-day hot Jupiter.
The system was found to have a low obliquity through an earlier
application of the spot-crossing method \citep{Desert2011}.  Shown are
the residuals for five neighboring groups of transits, each of which
is composed of five consecutive transits.  All five of these groups
show a very similar pattern as a function of longitude, as we would
expect for a low-obliquity star.

To quantify the significance of the correlations, we perform the
following steps. First we bin the residual flux uniformly in transit
longitude.  We compute normalized residuals,
\begin{equation}
r_{i, j} \equiv  \frac{F_{\text{res}}(i,j)}{\sigma_{\text{res}}(j)},
\end{equation}
where $F_{\text{res}}(i,j)$ is the median of the residual fluxes in the
$i$th group of transits and $j$th bin in transit longitude, and
$\sigma_{\text{res}}(i,j)$ is the standard deviation of the residuals of all the data in the $i$th group that contribute to the $j$th bin.
Finally, we define the transit chord correlation (TCC) as the average
of the dot
products of the residuals between neighboring groups:
\begin{equation}
  {\rm TCC} \equiv \frac{1}{N_{\text{gr}}-1} \sum_{i}^{N_{\text{gr}}-1}
  \frac{1}{N_{\Phi}} \sum_{j}^{N_{\Phi}}{ r_{i,j} \times r_{i+1, j} }
\end{equation}
where $N_{\text{gr}}$ is the total number of groups observed, and
$N_{\Phi}$ is the number of longitude bins.  We selected $N_{\Phi}$
such that each bin contains at least five data points;
typically $N_{\Phi} \approx 100$. In some cases, gaps in the data
resulted in an empty bin, in which case we set
$F_{\text{res}}(i,j)=0$.

Note that the dot products of the residuals are only computed between
neighboring groups. This is because each neighboring pair is only
separated by a relatively short timescale (a few orbital periods). On
longer timescales, we expect the correlation between different groups
to dwindle because there might have been enough time for the active
regions to undergo substantial evolution or migration. We will see
later that Kepler-17 and many other systems do show evidence for
evolution of the active regions.

\begin{figure}
\begin{center}
\includegraphics[width = 1.0\columnwidth]{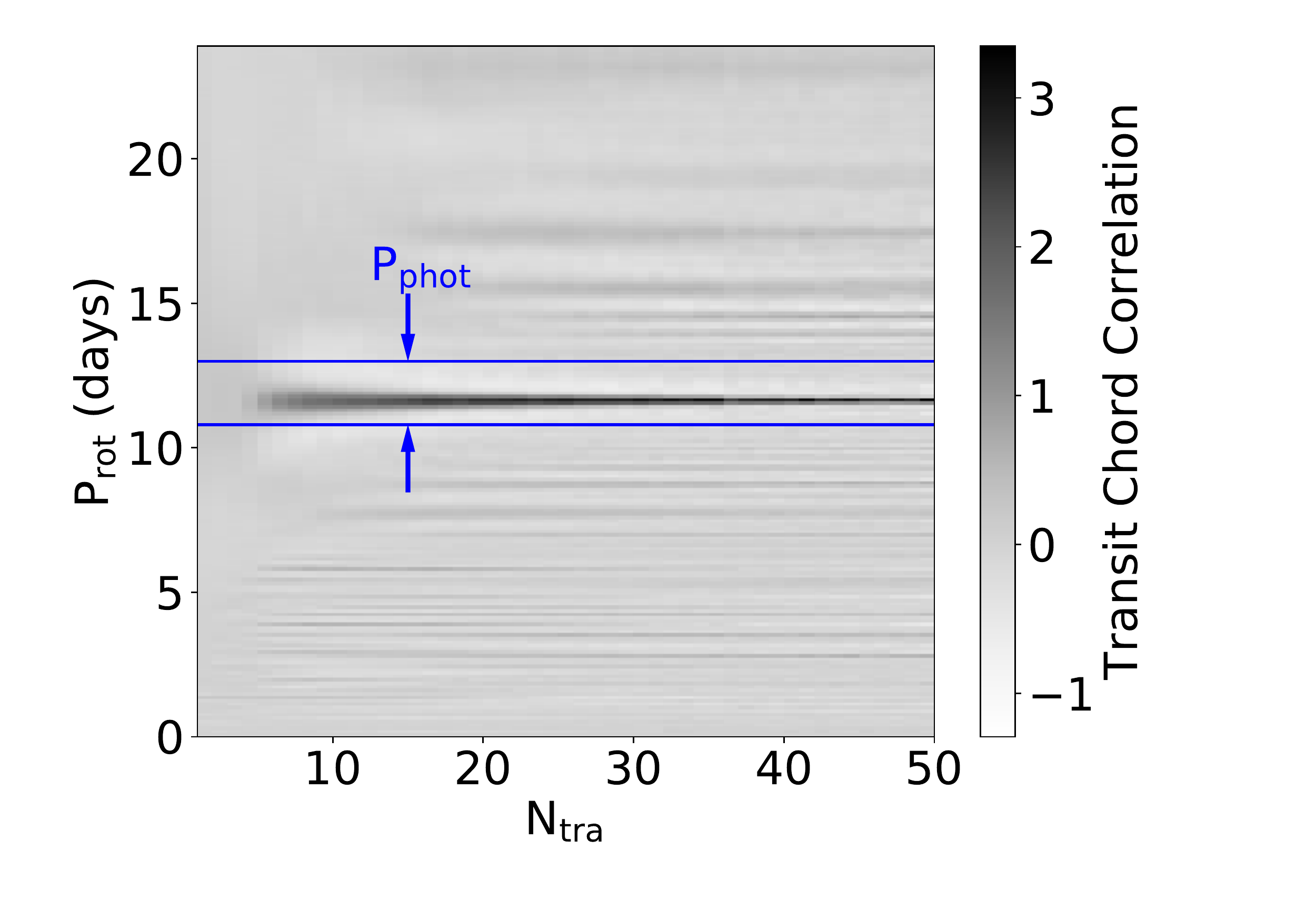}
\caption{ {\bf Transit Chord Correlation (TCC) for Kepler-17b}, as a
  function of $P_{\text{rot}}$, the rotation period that is assumed
  when computing transit longitudes, and $N_{\rm tra}$, the number of
  transits per group. The blue lines show the 1-$\sigma$ confidence
  interval in the measurement of $P_{\text{phot}}$, the stellar
  rotation period based on the photometric periodicity in the
  out-of-transit light curve.  The strongest TCC occurs at
  $P_{\text{tcc}} = 11.8\pm 0.7$ days which agrees well with
  $P_{\text{phot}} = 11.9\pm 1.1$~days.  This agreement is strong
  support for the claim that the recurring pattern we see in Figure
  \ref{Kepler17_residuals} is due to active regions on a low-obliquity
  host star.}
\label{Kepler17_grid}
\end{center}
\end{figure}

We compute the TCC on a 2-d grid of possible choices for the stellar
rotation period $P_{\rm rot}$ and the number $N_{\rm tra}$ of
neighboring transits that are included in each group.  The stellar
rotation period that gives rise to the strongest correlation is
denoted as $P_{\text{tcc}}$.  For a star with a low obliquity, the
strongest TCC should be observed when we calculate the transit
longitude using the independently measured stellar rotation period,
i.e., for $P_{\rm tcc} = P_{\rm phot}$, within the uncertainties in
the observations and the degree of differential rotation.
Figure~\ref{Kepler17_grid} shows the TCC grid for Kepler-17.  In this
case, $P_{\text{tcc}} = 11.8\pm0.7$~days agrees very well with
$P_{\text{phot}} = 11.9\pm 1.1$~days, which is good evidence that this
star has a low obliquity.

\begin{figure}
\begin{center}
\includegraphics[width = 0.9\columnwidth]{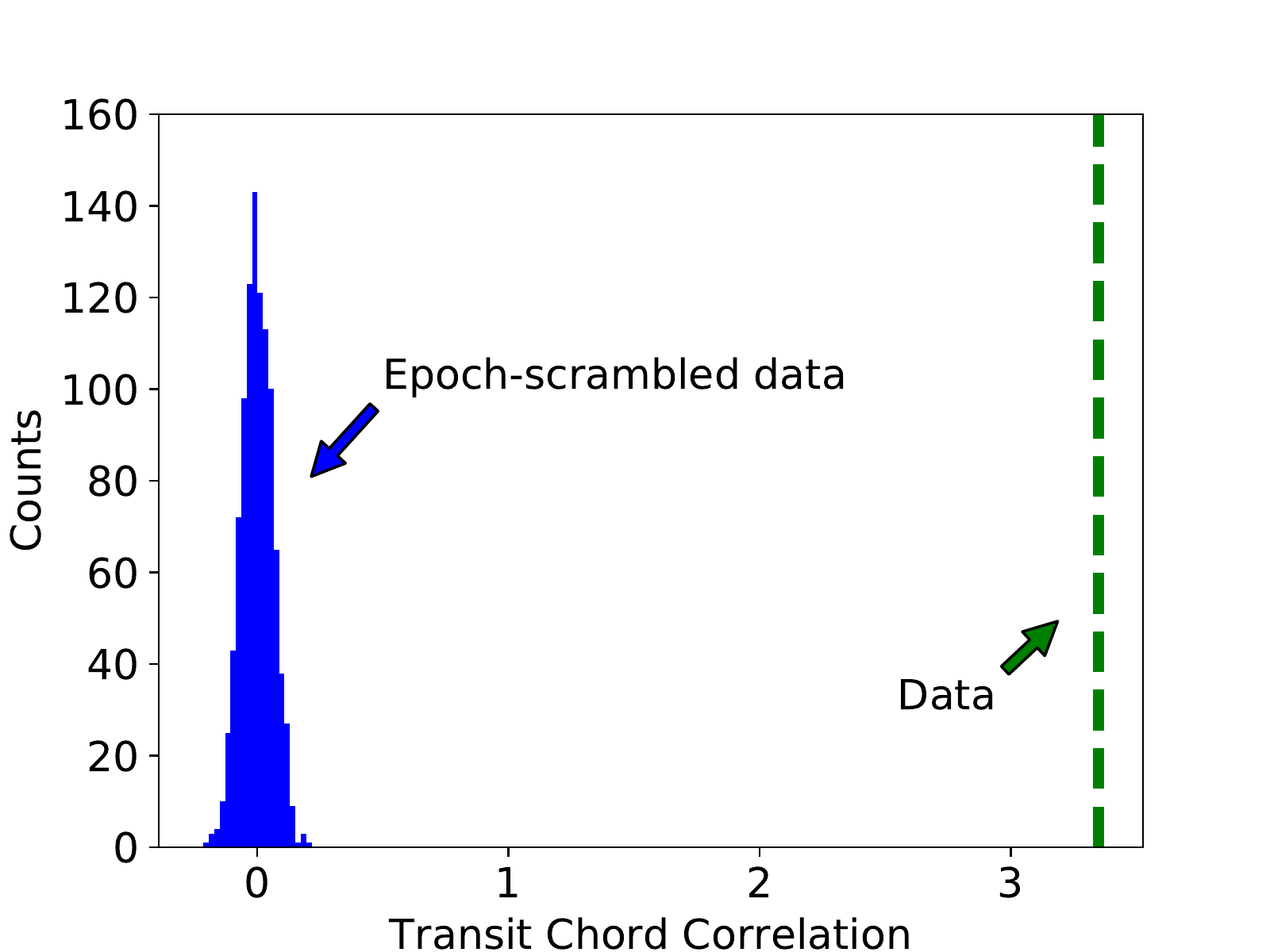}
\caption{ {\bf Statistical significance of the TCC for Kepler-17b.}  To
  quantify the statistical significance of the recurring pattern seen
  in Figure \ref{Kepler17_residuals} and \ref{Kepler17_tapestry}, a
  scrambling test is conducted. We recompute the TCC after randomizing
  the epoch numbers of the individual transits. The resultant
  distribution is plotted as the blue histogram. The TCC computed with
  the correct epoch sequence (shown in green) is much stronger than
  when computed for the scrambled sequences.}
\label{Scrambled}
\end{center}
\end{figure}

We also need to establish the threshold value of TCC to be considered
statistically significant.  We do so with a Monte Carlo procedure.  In
each of $10^3$ trials, we scramble the transit epochs, i.e., we
randomize the order of the individual transits without affecting the
individual light curves. In this way, we remove any correlations
between neighboring transits due to spot-crossing events, while
preserving the structure of the residuals (the ``red noise'') within
each transit light curve.  For each trial we recalculate the TCC
using the values of $P_{\rm tcc}$ and $N_{\rm tra}$ that produced the
strongest correlation in the real dataset.  We find the resulting
distribution of TCC to be nearly Gaussian with a well defined
standard deviation, as one would expect from the central limit theorem.
The TCC distribution for Kepler-17 is shown in Figure~\ref{Scrambled}.
The statistical significance of the TCC of the real data is quantified
by the number of standard deviations ($N_\sigma$) away from the median
of the Monte Carlo distribution.

The final question is, what threshold should be set on $N_{\sigma}$
for a statistically robust detection of correlation?  We found that
statistical fluctuations alone sometimes produce a TCC with
$\text{N}_{\sigma} =$~2--3 for a random value of $P_{tcc}$ (not
necessarily close to $P_{phot}$). This is because when searching for
the strongest TCC, we step through a grid of different stellar
rotation periods.  At each grid point, the transit longitudes and
hence the TCC is recomputed with the new rotation period. Typically,
we test a rotation period grid with about 200 different periods. It is
often the case that a random rotation period among the 200 periods
tested produce a 2 or 3-$\sigma$ outlier just by chance. We therefore
set a threshold of low-obliquity detection at $\text{N}_{\sigma}>4$.
Even more importantly, we guard against spurious detections by
requiring that $P_{\rm tcc}$ must agree with $P_{\text{phot}}$.

\subsection{The problem of spin-orbit commensurabilities}

When the ratio of the stellar rotation period and the planetary
orbital period is close to the ratio of small integers, i.e., $P_{\rm
  rot}/P_{\rm orb} = n/m$, then the planet and the star return to the
same sky-projected configuration every $n$ orbital periods or $m$
rotation periods.  Such a spin-orbit commensurability may be the
result of a physical process that drove a system into a spin-orbit
resonance, or may simply occur by chance.  Upon returning to the same
sky-projected configuration, the planet will occult the same set of
active regions.  The TCC will therefore be strong regardless of the
stellar obliquity.  Fortunately, these cases for which the TCC is
blind to obliquity can be easily recognized because $P_{\rm phot}$ and
$P_{\rm orb}$ are both known in advance.  Moreover, in such cases, the
TCC will usually be strongest at $P_{\rm tcc} = mP_{\rm rot} = nP_{\rm
  orb}$, rather than $P_{\rm rot}$.

In practice, our code raises an alarm whenever $P_{\rm phot}/P_{\rm
  orb}$ falls within 5\% of a ratio of small numbers and
$P_{\text{tcc}}$ also falls within 5\% of $nP_{\rm orb}$.  In these
cases, we carry out a visual inspection of the light curve, which can
easily distinguish the low-obliquity from the high-obliquity cases: in
the low-obliquity case spot/faculae crossing events recur in all
neighboring transits; whereas in the high-obliquity case spot/faculae
crossing events only recur after $nP_{\rm orb}$.

In summary, we declare a statistically significant detection of low
stellar obliquity if the system satisfies: (1) $P_{\text{tcc}}$ agrees
with $P_{\text{phot}}$ to within the uncertainties; (2)
$N_{\sigma}>4$; and (3) the orbital and rotation periods are not in
the ratio of small integers. For such systems we can place an upper
bound on the allowed obliquity using a simple geometric
argument. Between consecutive transits, the active regions move
through an angle of $2\pi/P_{\text{rot}} \times P_{\text{orb}}$. The
angular displacement of the active region between consecutive transits
is therefore
\begin{equation}
\Delta\theta = ~{\rm mod} (\frac{2\pi P_{\text{orb}}}{P_{\text{rot}}}, 2\pi).
\end{equation}
In a well-aligned system, the active regions move along the transit
chord, along the $x$ direction.
However a non-zero obliquity introduces a vertical motion, in the $y$ direction.
In order to observe a recurring pattern, the active regions must remain
within the transit chord for at least two transits.  For this to happen
the vertical motion of the active regions must be smaller than
the width of the transit chord,
\begin{equation}
\label{eqn:upperlimit}
\Delta\theta \times \text{sin}(\Psi) \lesssim \frac{R_p}{R_\star}.
\end{equation}
For giant planets around Sun-like stars this leads to a typical upper
limit of 10$^\circ$, though it is only approximate because we have
neglected the effects of the stellar inclination toward or away from
the observer.  Also, in cases where the active regions are larger than
the planet radius, we should replace $R_p/R_\star$ by the size of the
active regions relative to the star.

To allow a visual inspection of all the data we introduce the
``transit tapestry'', shown in the left panel of
Figure~\ref{Kepler17_tapestry} for Kepler-17.  This is a  heat map in
which the color scale represents the residual flux, the horizontal
dimension is transit longitude, and the vertical dimension is the
transit group number (essentially, the time).  If a star has a low
obliquity, then long-lived and stationary active regions will appear
as vertical features in the image.  Evolution of the active regions
will cause the features to vary slowly in the vertical direction, and
migration in longitude will cause them to vary in the horizontal
direction.  These effects produce a tapestry-like pattern of coherent
features, which displays the properties of the active regions such as
their sizes, relative intensities, lifetimes and migration patterns.
If the obliquity is large or the active regions are weaker then the
image will lose its artistic appeal and look like random static, as in
the right panel of Figure~\ref{Kepler17_tapestry}, which was generated
after scrambling the transit epochs.

\subsection{Retrograde Orbits}

Although our emphasis in this description has been on the detection of
low-obliquity orbits, it is worth noting that the same method can just
as easily be used to identify planets on nearly perfectly retrograde
orbits ($\Psi = 180^{\circ}$).  This is because the transit chord of a
retrograde-orbiting planet is also parallel to the lines of latitude
on the stellar photosphere.  In such a geometry, active regions on the
star can be occulted multiple times in neighboring transits and
produce strong correlations in the residual flux.  To search for
retrograde systems, we simply flip the sign of the second term in
Equation~\ref{eqn:longitude} and proceed again with all the other
steps in the analysis.

In fact it is not strictly necessary to perform two completely
different analyses, one for prograde and the other for retrograde
systems.  Even if the positive sign is always retained in
Equation~\ref{eqn:longitude}, a retrograde system will distinguish
itself by producing a strong TCC for a value of $P_{\rm tcc}$ that is
not equal to the stellar rotation period, but rather at the period
given by
\begin{equation}
\label{eqn:retro}
\frac{1}{P'_{\rm tcc}}  =\frac{1}{P_{\rm orb}} - \frac{1}{P_{\rm rot}}.
\end{equation}
This is because in between two transits, the active regions on a
retrograde star with rotation period $P'_{\rm tcc}$ will rotate
through the same angle as a prograde star with rotation period $P_{\rm
  rot}$.  In general $P'_{\rm tcc}$ differs significantly from $P_{\rm
  rot}$.  The only exception is the case $P_{\rm rot}$ = 2$P_{\rm
  orb}$, which is already handled in the test for spin-orbit
commensurabilities.

\section{Sample Selection}

We tried to assemble a collection of all the transit datasets for
which there seemed to be a reasonable chance of success for the TCC
method.  The requirement for high SNR and a large number of
consecutive transits restricts us to data from the space missions {\it
  CoRoT}, {\it Kepler}, and {\it K2}.

We included all 29 confirmed {\it CoRoT} systems in our sample.  We
downloaded the light curves from the {\it CoRoT} N2 public
archive\footnote{\url{http://idoc-CoRoT.ias.u-psud.fr}}. We used the
monochromatic flux labelled ``whiteflux'' in the FITS header.  We only
retained data points with a Quality Flag of 0, i.e., those not
affected by various known problems.

Analyzing all of the thousands of {\it Kepler} Objects of Interest
(KOIs) would have required too much computation time.  Instead, we
selected a sample of confirmed planets, planetary candidates, and
eclipsing binaries with high SNR and for which there was some
indication of spot-crossing anomalies.  The SNR was calculated as the
ratio between the transit or eclipse depth and the tabulated 6-hour
CDPP (combined differential photometric precision).  We identified the
360 KOIs with a single-transit SNR~$>$~20.  Evidence for anomalies is
based on the ratio of the standard deviation of the photometric
residuals during the transit $\sigma_{\rm in}$, and the standard
deviation of the out-of-transit data $\sigma_{\rm out}$.  Stellar
activity should cause this ratio to be significantly larger than
unity. We retained the high SNR systems that also have $\sigma_{\rm
  in}/\sigma_{\rm out} >1.7$.  A list of such systems had been
compiled by \citet{Sanchis-Ojeda_thesis}.  We also included any {\it
  Kepler} systems for which the stellar obliquity had been previously
measured using spot-crossing anomalies or any other method.  We
downloaded the light curves from the MAST website using the {\tt
  Python} package {\tt kplr} \footnote{\url{http://dfm.io/kplr}}.
Whenever possible we used the short-cadence light curves, with
one-minute time sampling. We used the light curves based on SAP
(simple aperture photometry), and we only kept data points with a
Quality Flag of 0.

Our sample also includes nine stars with confirmed planets that were
observed in the short-cadence mode during the {\it K2} mission
\citep{Howell2014}.  These systems are HATS-9, HATS-11, Qatar-2,
WASP-47, WASP-55, WASP-75, WASP-85, WASP-107 and WASP-118.  To produce
the light curves, we downloaded the pixel files from the Mikulski
Archive for Space Telescopes website (MAST).  To remove the spurious
intensity fluctuations caused by the uncontrolled rolling motion of
the {\it Kepler} spacecraft, we used the photometry pipeline described
by \citet{Dai2017qatar2}, which decorrelates the flux variations
against the measured coordinates of the center of light on the
detector.

The resultant sample consists mainly of FGK dwarf stars, except for
Kepler-13A (an A star) and Kepler-45 (an M dwarf).  Because of the SNR
requirement, the planets in our sample are almost all hot Jupiters.
Three-quarters of the planets are larger than $8~R_{\oplus}$ and have
orbital periods shorter than 10 days.  We believe the resulting sample
is fairly exhaustive, in the sense that it includes all of the data
currently available for which the TCC method has a significant chance
of revealing the obliquity.  However, the sample is strongly biased
toward close-in giant planets around active stars.  It is not
``complete'' in any other sense, i.e., it is not amenable to any
simple astrophysical description.  Caution is therefore needed when
trying to interpret the fraction of systems that are found to be
aligned, retrograde, or indeterminate.

\section{Results}
\label{Results}

Tables~\ref{tab:planet} and \ref{tab:eb} summarize the results for the
transiting planets and eclipsing binaries respectively.  For each
system we report the orbital period ($P_{\text{orb}}$); the scaled
semi-major axis ($a/R_{\star}$); the mass and radius of the planet and
the host star ($M_p$, $R_p$, $M_\star$, and $R_\star$); the effective
temperature of the host star ($T_{\text{eff}}$); the rotation period
measured from the quasiperiodic modulation in the light curve
($P_{\text{phot}}$); the rotation period that gives the strongest TCC
($P_{\text{tcc}}$); the statistical significance of the correlation
based on the scrambling test ($N_{\sigma}$); the ratio between the
rotation and orbital periods ($P_{\text{tcc}}/P_{\text{orb}}$); and
any previous measurement of the sky-projected obliquity that has
appeared in the literature $\lambda_{\text{lit}}$.  We have arranged
the table such that systems that show statistically significant TCC
(those with a low stellar obliquity) appear in front.  For those systems
we report an upper limit on the true obliquity using
Equation~\ref{eqn:upperlimit}.  The other systems are arranged in
descending order of $N_{\sigma}$.

To check on the validity of the TCC method, we examined the results
for all the systems for which stellar obliquities had been measured
with other methods.  For the low-obliquity systems (those that satisfy
Equation~\ref{eqn:upperlimit}), we may detect a strong TCC with
$P_{\text{tcc}}=P_{\text{phot}}$.  However, a null result could also
be obtained, if the host star is not sufficiently active, the planet's
transit chord misses the active latitudes, or the active regions
undergo rapid evolution.  Therefore we began by focusing on the
low-obliquity systems where previous studies had revealed strong
stellar activity through the clear identification of individual
spot-crossing events: Kepler-17 \citep{Desert2011}, CoRoT-2
\citep{Nutzman2011} and Qatar-2 \citep{Mocnik2016,
  Dai2017qatar2}. These are hot Jupiters around active G or K stars,
with orbital periods less than two days.  For all of these systems,
individual spot-crossing anomalies could be visually identified in the
light curves.  The recurrence of the spot-crossing events led previous
authors to conclude that all three systems are aligned to within
$\approx$$10^{\circ}$. Additionally, for the case of Qatar-2b,
\citet{Esposito2017} found $\lambda = 0 \pm 10^{\circ}$ based on the
Rossiter-McLaughlin effect.  Unsurprisingly, these three systems show
the strongest TCC among all the systems in our sample, and $P_{\rm
  tcc}$ agrees well with $P_{\rm phot}$.  Thus the TCC method easily
confirms the low stellar obliquities for these systems.

\citet{Holczer2015} reported five systems with prograde obliquities
(Kepler-17b, Kepler-71b, KOI-883.01, KIC 7767559b and Kepler-762b),
based on the observed correlation between the apparent TTV and the
local time derivative of the out-of-transit flux. This method was
described in the Section 1.  The TCC method shows that the orbits of
all five systems are not only prograde ($\Psi < 90^{\circ}$) but also
aligned to within about $\Psi \lesssim 10^{\circ}$.

We now turn to the eight systems for which other methods unveiled
a high stellar obliquity (see Table \ref{tab:planet} for the measured
obliquities and references).  For CoRoT-3b, CoRoT-19b, Kepler-13Ab,
Kepler-420b, HAT-P-7b, and WASP-107b, we did not detect a
statistically significant TCC.  For those cases in which the stellar
rotation period could be measured from the light curve,
$P_{\text{phot}}$ did not agree with $P_{\text{tcc}}$.  This is as
expected, for high-obliquity systems.  However, our method did return
a statistically significant TCC for the HAT-P-11b and Kepler-63b
systems, which are known to be misaligned
\citep{Sanchis-Ojeda2011Hat,Sanchis-Ojeda2013}.  We found $N_{\sigma}
= 6.0$ for HAT-P-11b and 5.8 for Kepler-63b.  These ``false
positives'' are examples of the problem of spin-orbit commensurability
described in the previous section.  \citet{Beky2014Hat} reported a 6:1
spin-orbit period ratio for HAT-P-11b, and \citet{Sanchis-Ojeda2013}
reported a nearly 4:7 ratio for Kepler-63b. The strongest TCC for
Kepler-63 was detected at $P_{\text{tcc}} = 37.7$ days (4$\times
P_{\text{orb}}$) rather than $P_{\rm phot}$ = 5.401$\pm 0.014$ days.

After completing the search for well-aligned systems, we used the same
sample to search for perfectly retrograde systems. We flipped the sign
of the second term in Equation \ref{eqn:longitude} before repeating
the TCC analysis.  We did not find any cases of a statistically
significant TCC.  Hence, although we could have detected perfectly
retrograde orbits just as easily as we detected prograde orbits, we
did not find any evidence for perfectly retrograde orbits in our
sample.  In the following sections we discuss some of the individual
systems in greater detail.

\subsection{Kepler-17}

Kepler-17b is a hot Jupiter orbiting a G star every 1.5 days.
\citet{Desert2011} identified and modeled a sequence of spot-crossing
events in neighboring transits. The recurrence of the spot-crossing
events led \citet{Desert2011} to conclude that the Kepler-17b has an
obliquity < 15$^{\circ}$. Our analysis revealed a strong TCC
($N_{\sigma}$ = 59.6) at $P_{\text{tcc}} = 11.8 \pm 0.7$ days. The
{\it Kepler} light curve shows rotational modulation in the
out-of-transit light curve which offers an independent check on the
stellar rotation period: $P_{\text{phot}} = 11.9 \pm 1.1$ days. These
observations together suggest a low stellar obliquity of Kepler-17b;
and we place an upper bound $\Psi \lesssim$ 10$^{\circ}$ with
Equation~\ref{eqn:upperlimit}.

\begin{figure*}
\begin{center}
\includegraphics[width = 1.8\columnwidth]{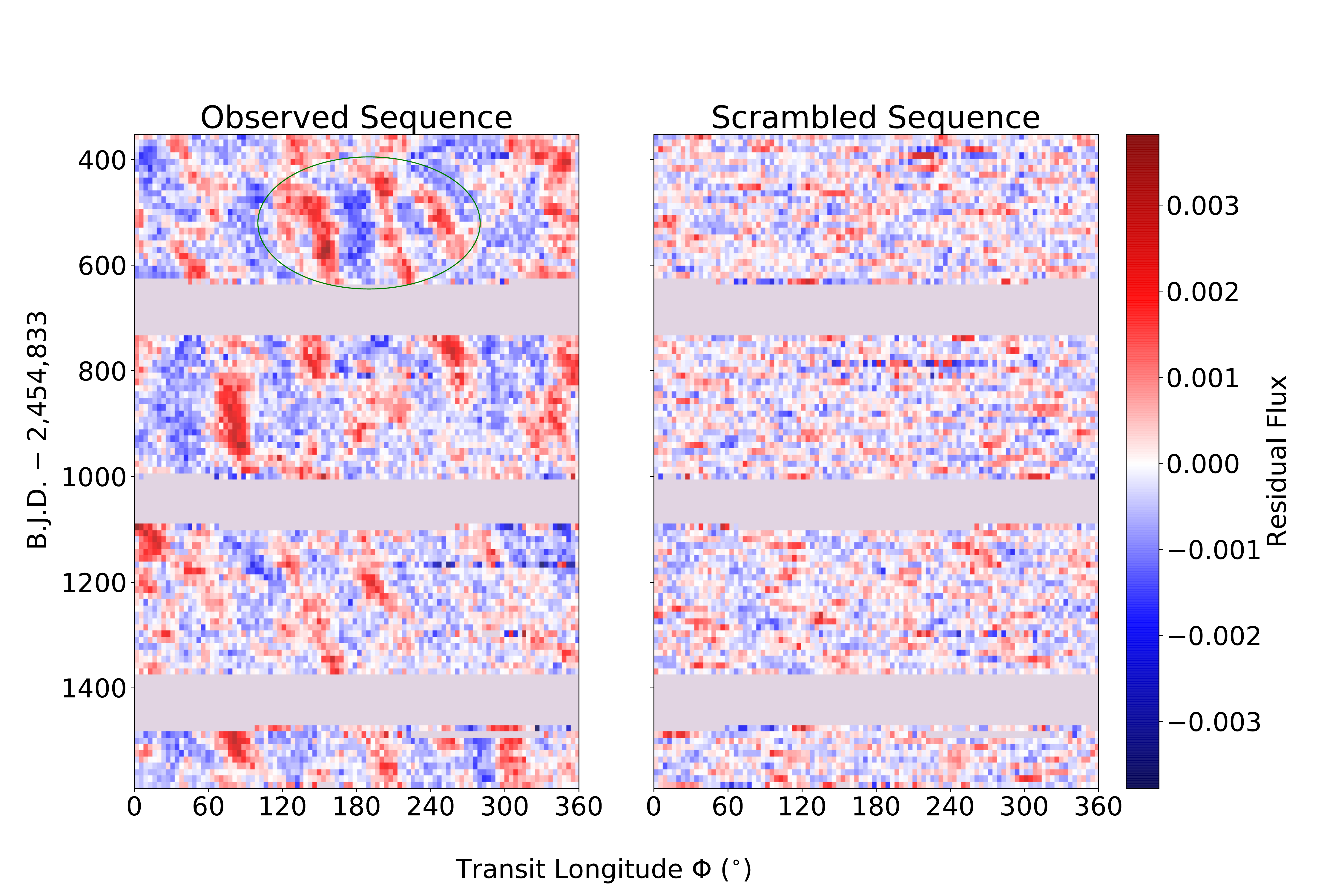}
\caption{ {\bf Transit Tapestry of Kepler-17b.} The residual flux of
  Kepler-17b as a function of transit longitude and time.  Red
  indicates positive residuals, and blue indicates negative residuals.
  A portion of these data (BJD~540--600) are shown in Figure
  \ref{Kepler17_residuals}.  The coherent and generally vertical
  patterns imply a low stellar obliquity.  These patterns allow for
  estimates of some key properties of the active regions, including
  size (5--20$^{\circ}$ in longitude), lifetime (100--200 days), and
  relative intensity (10--20\% dimmer than average photosphere).  Some
  active regions are observed to split into two smaller active regions
  (see, e.g., the green ellipse). We interpret this as the result of
  emerging magnetic flux tubes. For comparison, the right panel shows
  the transit tapestry for a scrambled dataset, in which the transit
  epochs were randomized.}
\label{Kepler17_tapestry}
\end{center}
\end{figure*}

Figure~\ref{Kepler17_tapestry} shows the transit tapestry for
Kepler-17.  It displays clear clusterings of positive and negative
residuals in transit longitude.  We interpret the observed clustering
as photometric signatures of active regions along the transit chord.
These patterns can be used to infer the properties of the host star's
magnetic activity.  At any particular time there are 1--4 active
regions present on the transit chord.  Each of these regions in the
tapestry spans 20--30$^{\circ}$ in transit longitude.  However, the
nonzero size of the the planet (about 16$^{\circ}$ in longitude) acts
to widen the apparent size of the active regions.  After accounting
for the size of the planet, the active regions span 5--20$^{\circ}$ in
true longitude on the star.  The active regions lasted for a 100--200
days before either disappearing or leaving the latitudes probed by the
planet. The intensity of the active regions changed gradually over
time; they did not burst into existence with maximum intensity.

One point of interest is whether the active regions remain stationary
in longitude or undergo longitudinal migration.  However, there is a
degeneracy between any constant rate of longitudinal migration, and
the rotation period used to calculate transit longitudes.
Specifically, if $P_{\rm tcc}$ is smaller than the actual stellar
rotation period, all the photometric features in the transit tapestry
would appear to undergo prograde longitudinal migration in the transit
tapestry.  The TCC method, by design, maximizes the correlation of
residual flux. As a result, it always selects the stellar rotation
period such that most photometric features remain fixed in stellar
longitude. Therefore, the transit tapestry can only reveal relative
migration rates between different active regions but not the absolute
rate of migration.  In the case of Kepler-17, there is indeed some
indication of relative migration. The pattern enclosed by the green
ellipse in Figure~\ref{Kepler17_tapestry} suggests that an active
region split apart into two smaller regions which separated
longitudinally over time.  This is reminiscent of the emerging
magnetic flux tubes on the Sun which produce bipolar magnetic regions
\citep{Spruit}.  A flux tube initially gives rise to one footpoint on
the host star when it just reaches the stellar surface. As the flux
tube emerges further, it gives rise to two footprints that spread
apart in longitude.

From now on, we will define the contrast of an active region as the
relative ratio between the brightness of the active regions to the
average photosphere, in the observing bandpass.  Specifically for
Kepler-17, the photometric features of the active regions had
amplitudes of about 0.002--0.004 in relative flux.  By comparing this
to the transit depth of about 0.02, the active regions on average are
seen to be roughly 10-20\% fainter than the average photosphere in the
broad optical {\it Kepler} bandpass (approximately 420--900~nm). We
note that the finite size of the transiting planet introduces a
convolution effect that not only broadens the photometric signature of
the active regions, but also reduces the contrast.  The contrast
calculated here and subsequently is the product of the relative
brightness of individual magnetic features that constitute the active
regions and the area filling factor within the shadow of the
planet.  Because of this degeneracy, we refrain from converting the
contrast into an effective temperature (which would have allowed for
more direct comparisons with sunspots).

Figure~\ref{Kepler17_tapestry} shows that the number, size, and
relative intensity of active regions decreased towards the latter part
of the {\it Kepler} mission. This decline might have been part of a
magnetic activity cycle, or it might represent the latitudinal
migration of the active regions away from the transit chord.  In
principle, the degree of differential rotation could be studied by
comparing the rotation period on the transit chord $P_{\text{tcc}}$
with the photometrically-derived rotation period $P_{\rm phot}$, which
is based on active regions across the entire star.  However, this is
difficult in practice because there is little information on how the
active regions are distributed latitudinally outside the transit
chord. Given the impact parameter of the transit $b=0.27$
\citep{Desert2011} and the size of the planet $R_p/R_\star
\approx$0.13, the stellar latitude probed by the planet is $16\pm
8^{\circ}$ north or south of the stellar equator.  At least, our
analysis can show that this latitudinal range is magnetically active.

\subsection{CoRoT-2}

CoRoT-2b is a hot Jupiter orbiting a G star every 1.7 days
\citep{Alonso2008}. The star rotates rapidly, with a period of about
4.5 days, and has strong magnetic activity which manifested as clear
spot-crossing events in the {\it CoRoT} light curve. By modeling the
recurrence of spot-crossing events, \citet{Nutzman2011} found the
sky-projected obliquity to be $\lambda = 4.7 \pm 12.3^{\circ}$.  Our
TCC method tells a consistent story. We find a strong TCC $N_{\sigma}$
= 22.6 at $P_{\text{tcc}} = 4.5 \pm 0.3$ days consistent with
$P_{\text{phot}}$ of $4.530 \pm 0.068$ days.  We conclude that CoRoT-2
has a low stellar obliquity

\begin{figure*}
\begin{center}
\includegraphics[width = 1.5\columnwidth]{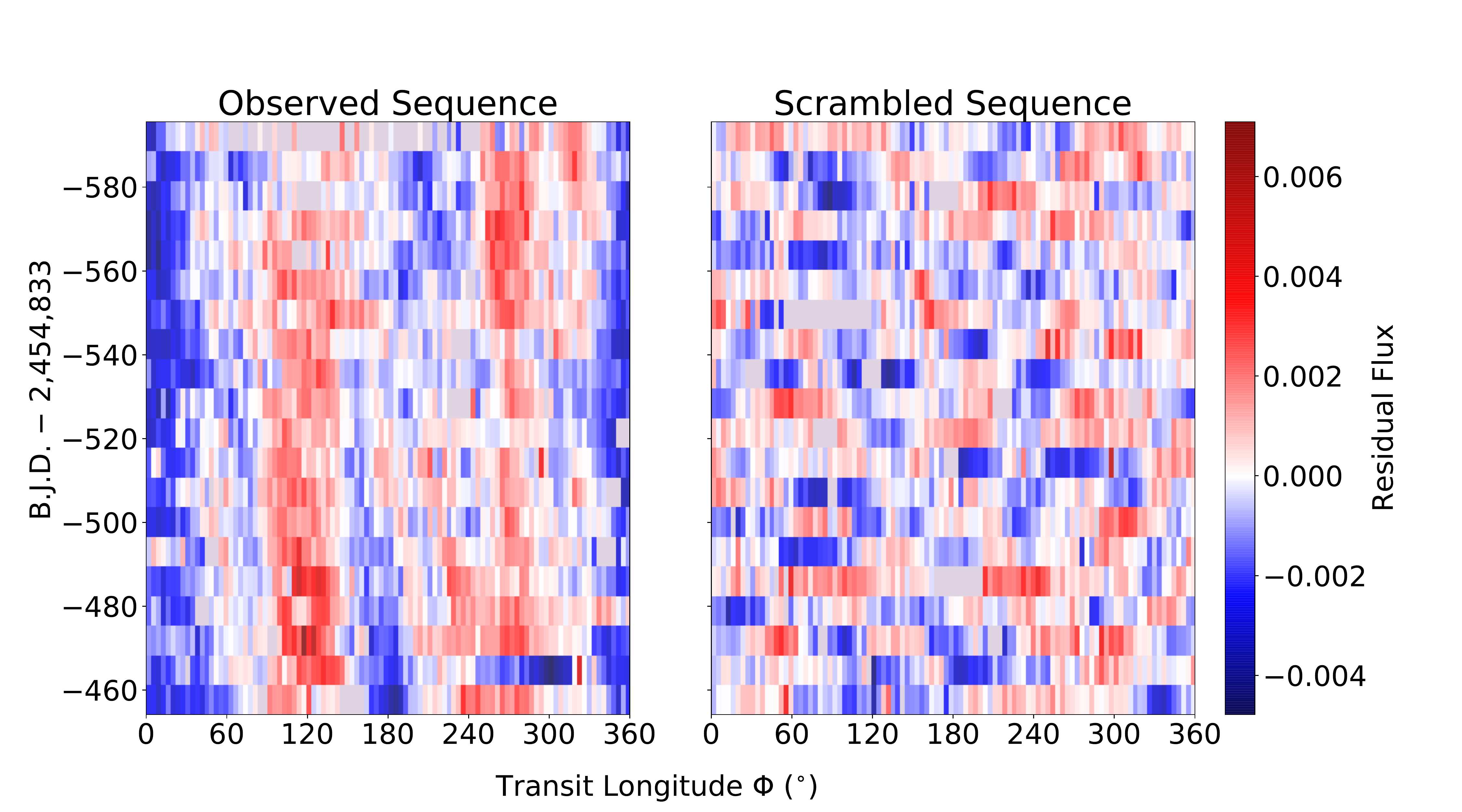}
\caption{ {\bf Transit Tapestry of CoRoT-2b.} Same as Figure
  \ref{Kepler17_tapestry} but for CoRoT-2.  The TCC method makes a strong case
  for a low stellar obliquity. In contrast to Kepler-17, there
  appear to be two active regions along the stellar latitude probed
  by the planet. They have similar intensity (7-17\% dimmer than
  average photosphere) and size (10-20$^{\circ}$ in longitude); and
  are located rather symmetrically on opposite hemispheres. The active
  regions persisted throughout the $\approx$150 days of {\it CoRoT}
  observations. The transit chord of CoRoT-2b correspond to a stellar
  latitude of $13\pm 10^{\circ}$.}
\label{Corot2_tapestry}
\end{center}
\end{figure*}

Figure~\ref {Corot2_tapestry} shows the transit tapestry.  The
clustering of residual flux indicates that there were two active
regions present on the stellar latitude probed by the planet. They had
similar sizes and intensities and were separated by about
$180^{\circ}$ in longitude. These two regions span about
30--40$^{\circ}$ in the transit tapestry.  After accounting for the
blurring effect of the planet, which extends about 20$^{\circ}$ in
longitude, we conclude the active regions are about 10-20$^{\circ}$
wide.  Both active regions persisted throughout the $\approx$150 days
of {\it CoRoT} observations, remaining nearly stationary in longitude.
With Figure~\ref {Corot2_tapestry}, we can estimate that active
regions produce flux variation of about 0.002--0.005 in relative flux.
Taking the ratio with the transit depth (about 3\%), we find the
active regions to be roughly 7-17\% dimmer than the rest of the
photosphere in the {\it CoRoT} bandpass.  Again, using the impact
parameter of the transit $b\approx 0.22$ \citep{Gillon2010b} and the
size of the planet $R_p/R_\star \approx 0.13$, the stellar latitude
probed by the planet is likely $13\pm 10^{\circ}$ in either the
northern or southern hemisphere.

\citet{Lanza2009} and \citet{Huber2010corot} also studied the magnetic
activity of CoRoT-2. \citet{Lanza2009} used a maximum entropy method
to determine the longitudinal distribution of active regions using the
rotational modulation in the out-of-transit light curve, while
\citet{Huber2010corot} modeled the transit light curve by dividing the
eclipsed and noneclipsed parts of photosphere into a number of
longitudinal intervals with different intensities. Both of these works
confirmed the presence of active regions on CoRoT-2.

\subsection{Qatar-2}

Qatar-2b is a hot Jupiter discovered by the Qatar Exoplanet Survey
\citep{Bryan2012}.  The planet orbits a K dwarf every 1.3 days.
Recent {\it K2} observations in the short-cadence mode unveiled dozens
of spot-crossing anomalies.  By modeling the spot-crossing anomalies,
both \citet{Mocnik2016} and \citet{Dai2017qatar2} showed that the
system has a low stellar obliquity.  \citet{Esposito2017}
independently confirmed the low stellar obliquity $\lambda = 0 \pm
10^{\circ}$ by observing the Rossiter-McLaughlin effect.  As expected,
we detect a strong TCC $N_{\sigma}$ = 18.0 in the {\it K2} light curve
at $P_{\text{tcc}} = 18.2 \pm 0.4$ days which agrees with
$P_{\text{phot}}$ = $18.5 \pm 1.5$ days.  We place an upper bound
$\Psi \lesssim 11^{\circ}$.

\begin{figure*}
\begin{center}
\includegraphics[width = 1.5\columnwidth]{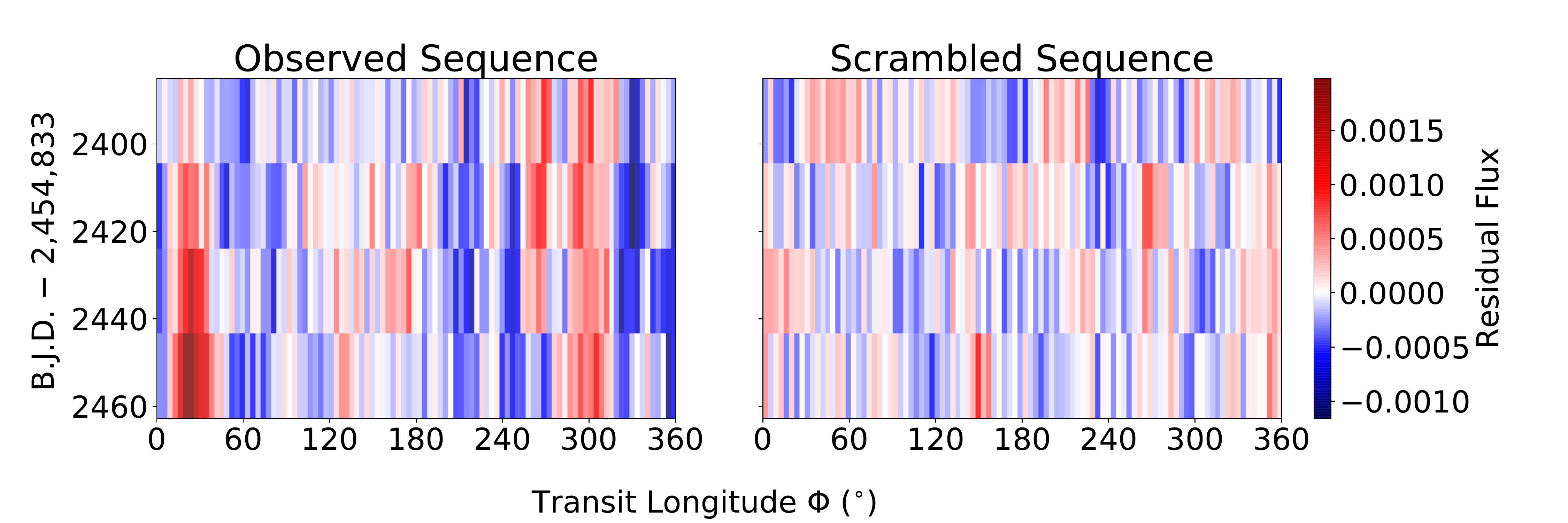}
\caption{ {\bf Transit Tapestry of Qatar-2b.} Same as Figure
  \ref{Kepler17_tapestry} but for Qatar-2.  TCC analysis reveals a low
  stellar obliquity.  One of the active regions was
  located at 300$^{\circ}$ in longitude. Its intensity remained
  relatively constant during the 80 days of {\it K2} observation. The
  other active region was located near 20$^{\circ}$ in longitude. Its
  size and intensity underwent significant increase during the {\it
    K2} campaign.}
\label{Qatar2_tapestry}
\end{center}
\end{figure*}

Figure \ref{Qatar2_tapestry} shows the tapestry plot for
Qatar-2. Although {\it K2} observations only spanned 80 days, the
transit tapestry still reveals two active regions along the transit
chord. One of them was located at 300$^{\circ}$ in longitude and was
about 15--25$^{\circ}$ wide (after accounting for the broadening due
to the nonzero size of the planet). Its intensity remained relatively
constant throughout the 80 days of {\it K2} observations. The other
active region was located near 20$^{\circ}$ in longitude. Its size and
intensity underwent a significant increase during the {\it K2}
observation.  This might have been caused by the emergence of a
magnetic flux tube, or latitudinal migration of an active region into
the transit chord.  These active regions maintained a constant
separation in longitude.  The active regions on average are roughly
3-7\% dimmer than the photosphere.  Given the impact parameter of the
transit $b \approx 0.03$ \citep{Dai2017qatar2} and $R_p/R_\star=
0.15$, the transits probe the region close to the stellar equator
($2\pm9^{\circ}$).

\subsection{Kepler-71}

Kepler-71b is a 3.9-day hot Jupiter around a G star
\citep{Howell2010}. With a V magnitude of 15.4, Kepler-71 is too faint
for precise radial-velocity follow-up. \citet{Holczer2015} detected a
strong correlation between the detected TTV and the time derivative of
the stellar flux, which suggested a prograde orbit and the presence of
starspots. Our TCC analysis shows that the orbit of Kepler-71b is not
only prograde but also well-aligned.  We detect a strong TCC
($N_{\sigma}$ = 10.7; $P_{\text{tcc}} = 19.7 \pm 0.8$ days;
$P_{\text{phot}}$ = $19.87 \pm 0.18$ days). The upper bound on the
obliquity is $\Psi$ $\lesssim$ 6$^{\circ}$.

\begin{figure*}
\begin{center}
\includegraphics[width = 1.5\columnwidth]{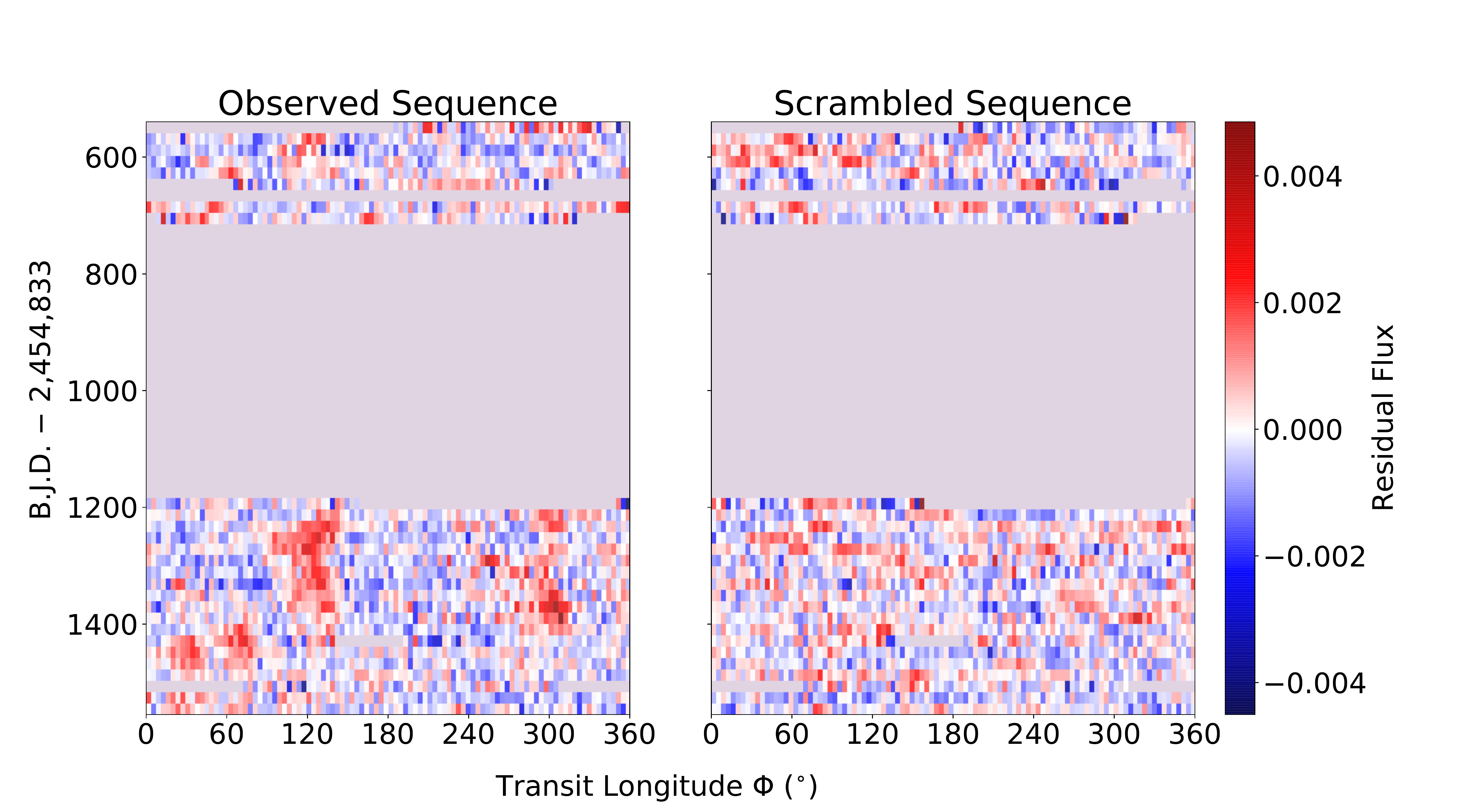}

\caption{ {\bf Transit Tapestry of Kepler-71b.} Same as Figure
  \ref{Kepler17_tapestry} but for Kepler-71.  TCC detects a strong
  correlation consistent with a low stellar obliquity.  Two
  large-scale active regions emerged in the later part of the {\it
    Kepler} mission. The active regions were separated by about
  180$^{\circ}$. They each spanned about 5-15$^{\circ}$in
  longitude. They have similar lifetime ($\sim$ 150 days) and relative
  intensity (10-20\% dimmer than average photosphere). Given its
  impact parameter, the planet probes the equatorial region ($2 \pm
  8^{\circ}$ in latitude).}
\label{Kepler71_tapestry}
\end{center}
\end{figure*}

We made a transit tapestry for Kepler-71 (Figure
\ref{Kepler71_tapestry}).  Two active regions are most discernible in
the later part of the {\it Kepler} mission (BJD-2454833 > 1200).
These two active regions were separated by about 180$^{\circ}$.  They
each spanned about 5-15$^{\circ}$ in longitude after accounting for
the convolution effect of the planet (16$^{\circ}$ in longitude). They
lasted for about 150 days; and on average were roughly 10-20\% dimmer
than the photosphere. We estimate the latitude probed by the planet
using the impact parameter of the transit \citep[0.04,][]{Howell2010}
and the size of the planet ($R_p/R_\star \approx$ 0.14). The result
shows that the equatorial region of the photosphere ($2 \pm
8^{\circ}$) is magnetically active.

\subsection{KOI-883.01}

KOI-883.01 is a planetary candidate discovered by the {\it Kepler}
mission. The transit light curve is consistent with a hot Jupiter
orbiting a K star every 2.7 days. As was the case with Kepler-71,
\citet{Holczer2015} detected a strong correlation between TTV and
local flux variation and inferred a prograde orbit for the system. We
show that the system is not only prograde but also well-aligned. The
$N_{\sigma}$ is as high as 9.2 while $P_{\text{tcc}} = 9.1 \pm 0.2$
days and $P_{\text{phot}} = 9.11 \pm 0.11$ days agree well. We place
an upper bound $\Psi$ $\lesssim$ 4$^{\circ}$.

\begin{figure*}
\begin{center}
\includegraphics[width = 1.5\columnwidth]{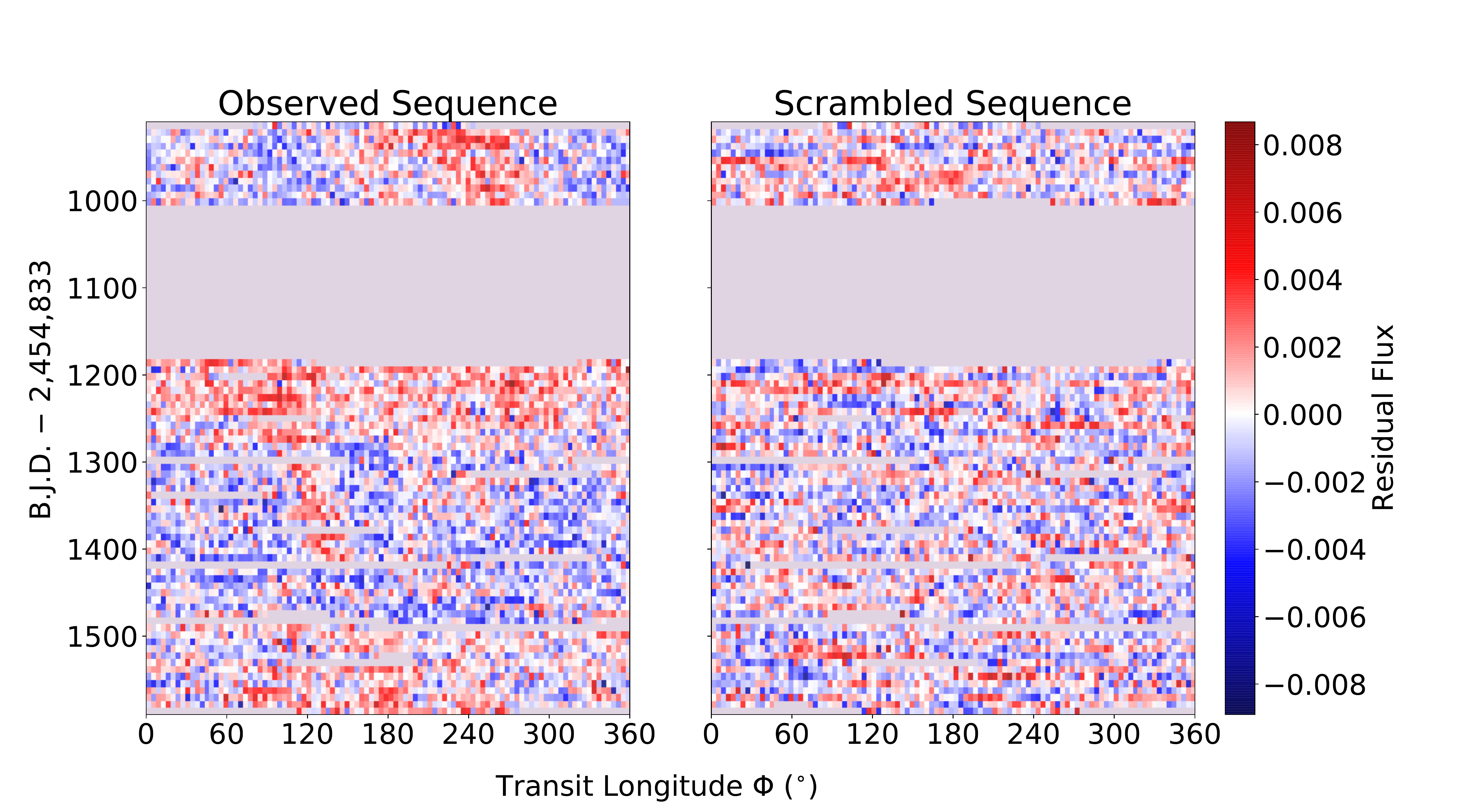}
\caption{ {\bf Transit Tapestry of KOI-883.01} Same as Figure
  \ref{Kepler17_tapestry} but for KOI-883.  TCC analysis revealed a
  low stellar obliquity.  An active region was located at
  240$^{\circ}$ in longitude during the first 100 days of {\it Kepler}
  observation. The second active region emerged near 120$^{\circ}$ in
  longitude with similar width (20-30$^{\circ}$ in longitude) and
  intensity (10-20\% dimmer than the average photosphere) after the
  data gap of about 200 days.}
\label{KOI-883.01_tapestry}
\end{center}
\end{figure*}

The transit tapestry is shown in Figure \ref{KOI-883.01_tapestry}. The
photometric signatures of active regions are more diffuse than for the
systems described earlier.  In the first $\approx$100 days of {\it
  Kepler} observations, an active region was located near
240$^{\circ}$ in longitude and was 20--30$^{\circ}$ wide (after
accounting for the 18$^\circ$ angular size of the planet).  After a
gap in the data, another active region emerged at 120$^{\circ}$ in
longitude with a similar size and intensity (10-20\% dimmer than the
average photosphere).  Both active regions lasted for at least 100--200
days, and perhaps even longer if the active region near 240$^{\circ}$
persisted through the data gap.  The impact parameter of the transit
is about 0.01 and $R_p/R_\star$ is about 0.18
(ExoFoP \footnote{\url{https://exofop.ipac.caltech.edu}}). The transit
chord overlaps with the equatorial region ($0 \pm 10^{\circ}$) which
appears to be magnetically active.

\subsection{Kepler-45}

Kepler-45b is a 2.5-day hot Jupiter orbiting a M dwarf. The system was
confirmed by a combination of radial velocity monitoring, adaptive
optics imaging, and near-infrared spectroscopy \citep{Johnson2012}.
It is one of the three hot Jupiters around M dwarfs that have been
reported to date.  The other two are HATS-6b \citep{Hartman2015} and
NGTS-1b \citep{Bayliss2017}.  No constraint on the stellar obliquity
of Kepler-45b has been published yet.

Application of the TCC method gave $N_{\sigma}$ = 6.3 at
$P_{\text{tcc}} = 16.7 \pm 0.8$ days. The $P_{\text{tcc}}$ coincides
with the measured $P_{\text{phot}}$ = $15.8 \pm 0.2$ days, to within
the uncertainties.  We conclude that Kepler-45b is very likely on a
well-aligned orbit, with an upper bound on $\Psi$ $\lesssim$
11$^{\circ}$.  To our knowledge, this is only the second case of a
stellar obliquity measured for an M dwarf.  The first such report was
for GJ~436b, a 2.6-day Neptune-mass planet \citep[][$\lambda =
  72^{+33}_{-24}~^{\circ}$]{Bourrier2017}.  Enlarging the number of
such measurements would be interesting.  Being fully convective, some
M-dwarfs are structurally very different from solar-type stars for
which most of the obliquity measurements have been obtained.  The
excitation and damping of stellar obliquities may operate in different
ways for M dwarfs than for solar-type stars. Given that other, more
traditional obliquity measurements tend to fail for M dwarfs, the TCC
method may be particularly promising for constraining the stellar
obliquities of M-dwarf planet hosts. We will return to this point in
Section \ref{discussion}.

\begin{figure*}
\begin{center}
\includegraphics[width = 1.5\columnwidth]{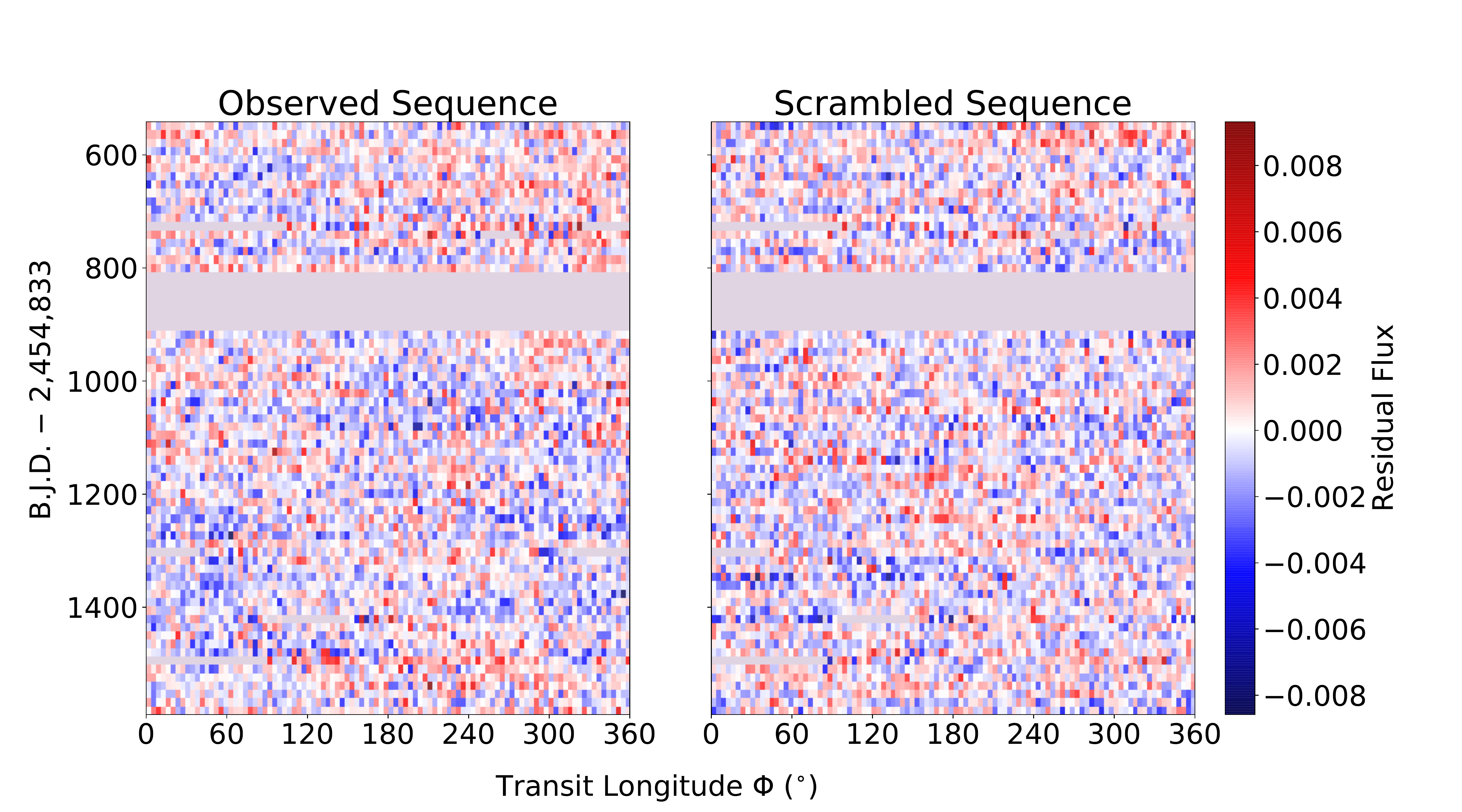}

\caption{ {\bf Transit Tapestry of Kepler-45b} Same as Figure
  \ref{Kepler17_tapestry} but for Kepler-45. Our TCC analysis
  suggests a low stellar obliquity. However, any presence of
  magnetically active regions is not visually obvious. This may be
  attributed to the low SNR of the light curve. Another possibility is
  that magnetically activity of the M dwarfs is dominated by bright
  chromospheric plages/networks instead of dark photospheric
  spots. The chromospheric features may have much larger areas, less
  well-defined shapes and shorter lifetimes than the spots. As a
  result, their photometric features in the transit light curves are
  harder to discern visually.}
\label{Kepler-45_tapestry}
\end{center}
\end{figure*}

Kepler-45b has an impact parameter of about 0.6 \citep{Johnson2012}
and a $R_p/R_\star$ of 0.18. Together this suggests that the transit
chord probes the stellar latitude of $37 \pm 16^{\circ}$. The transit
tapestry, shown in Figure~\ref{Kepler-45_tapestry}, does not show any
visually compelling patterns.  This is because we have organized this
section in order of decreasing correlation strength.  At this point,
the correlations may not be visually obvious, even though our Monte
Carlo procedure shows that they are statistically significant when
summed over the entire dataset.  Alternatively, it may be the case
that active regions on M-dwarfs preferentially take the form of broad
chromospheric plages and networks rather than dark photospheric spots
on sun-like stars. The resultant photometric features in the transit
light curves would be more spread out, more rapidly evolving and hence
harder to discern visually, compared to the sun-like stars described
earlier. The increased chromospheric activity of M dwarfs was observed
in Ca II H and K lines \citep{Isaacson2010}. Moreover, the difference
in magnetic behavior between M dwarfs and solar-type stars is also
theoretically motivated: fully convective M dwarfs lack the
tachocline, the strong shearing zone between the radiative and
convective layer of the sun, which is thought to be important for the
operation of sun-like dynamos and the formation of sunspots
\citep{Charbonneau2014}.

\subsection{TrES-2}

TrES-2 is a G star hosting a 2.5-day hot Jupiter
\citep{O'Donovan2006}. Rossiter-McLaughlin observations have shown
that TrES-2b has a low stellar obliquity $\lambda = -9 \pm 12^{\circ}$
\citep{Winn2008}, although the confidence in that measurement was
lower than usual because of the star's relatively low rotation
velocity.  We find a strong TCC $N_{\sigma}$ = 5.8 in the residual
flux at $P_{\text{tcc}} = 29.9 \pm 1.4$ days. This period is in
agreement with the independently measured photometric period,
$P_{\text{phot}} = 28.35 \pm 0.34$ days.  We conclude that TrES-2
likely has a low stellar obliquity with an upper bound $\Psi$
$\lesssim$ 10$^{\circ}$.

\begin{figure*}
\begin{center}
\includegraphics[width = 1.5\columnwidth]{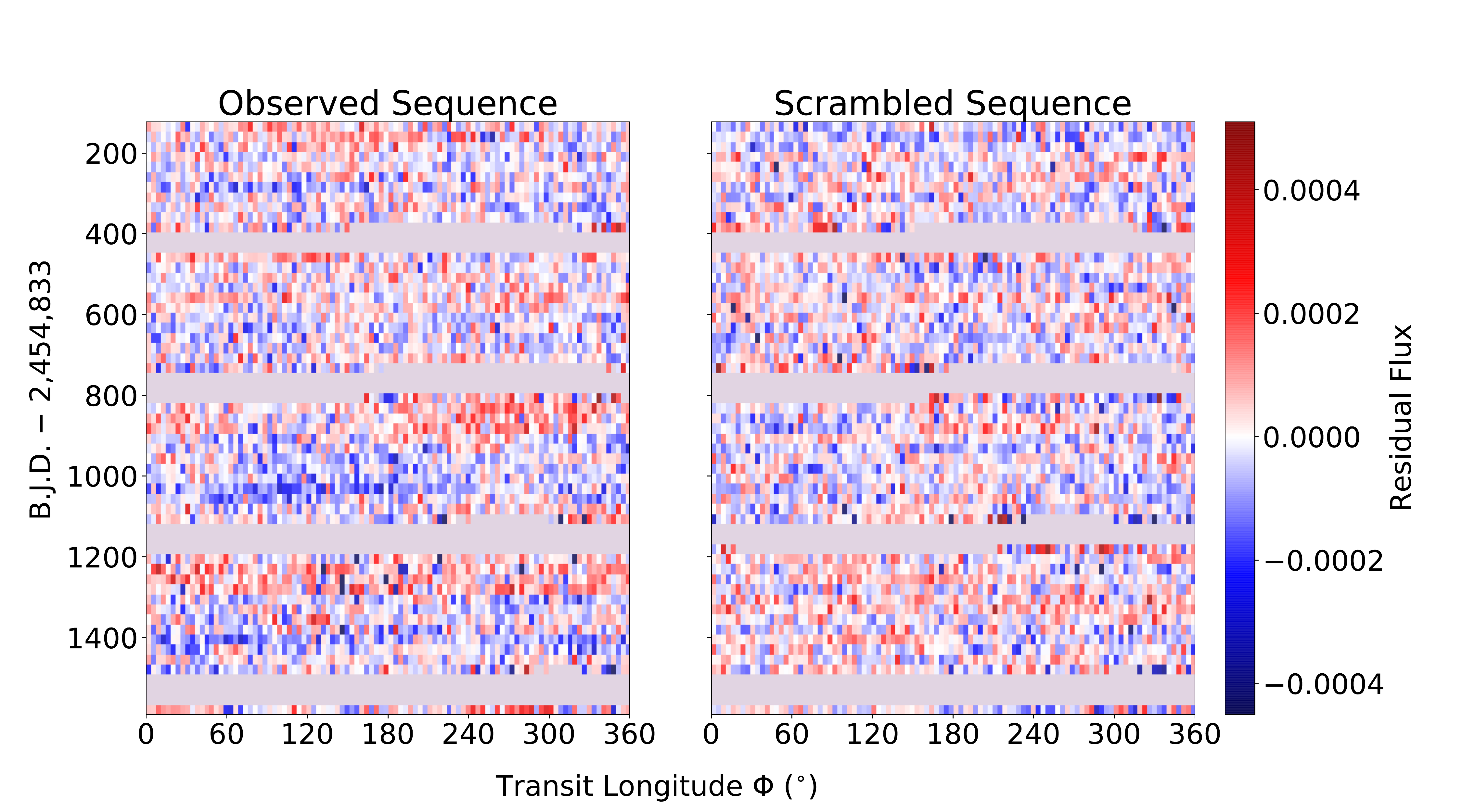}
\caption{ {\bf Transit Tapestry of TrES-2b} Same as Figure
  \ref{Kepler17_tapestry} but for TrES-2. TCC makes a good case for a
  low stellar obliquity of TrES-2. With a slow stellar rotation of
  about 29 days, TrES-2 may be dominated by faculae and active
  networks. The blue patch with transit latitude between 60 and
  240$^{\circ}$ near BJD-2454833 = 1000 may be caused by a bright,
  extended active region on TrES-2.}
\label{TrES-2_tapestry}
\end{center}
\end{figure*}

We show the transit tapestry in Figure~\ref{TrES-2_tapestry}.  We can
see correlations between neighboring groups, but not any distinct,
long-lasting features of the type that appeared in some of the systems
described earlier. According to \citet{Montet2017}, the photometric
activity of solar-type stars with rotation periods greater than 25
days are more likely to be dominated by bright patches (faculae) than
by dark spots.  These bright patches and the active network, at least
those on the Sun, are more extended and evolve more quickly than spots
\citep{Foukal1998,Pontieu,Shapiro2016}.  Given the slow stellar
rotation period of TrES-2, we may be seeing the more extended and less
persistent photometric features of the faculae or active networks. In
particular, the blue patch spanning about $180^{\circ}$ near
BJD-2454833 = 1000 may be associated with a bright, more extended
active region.

\subsection{Kepler-762}

\begin{figure*}
\begin{center}
\includegraphics[width = 1.5\columnwidth]{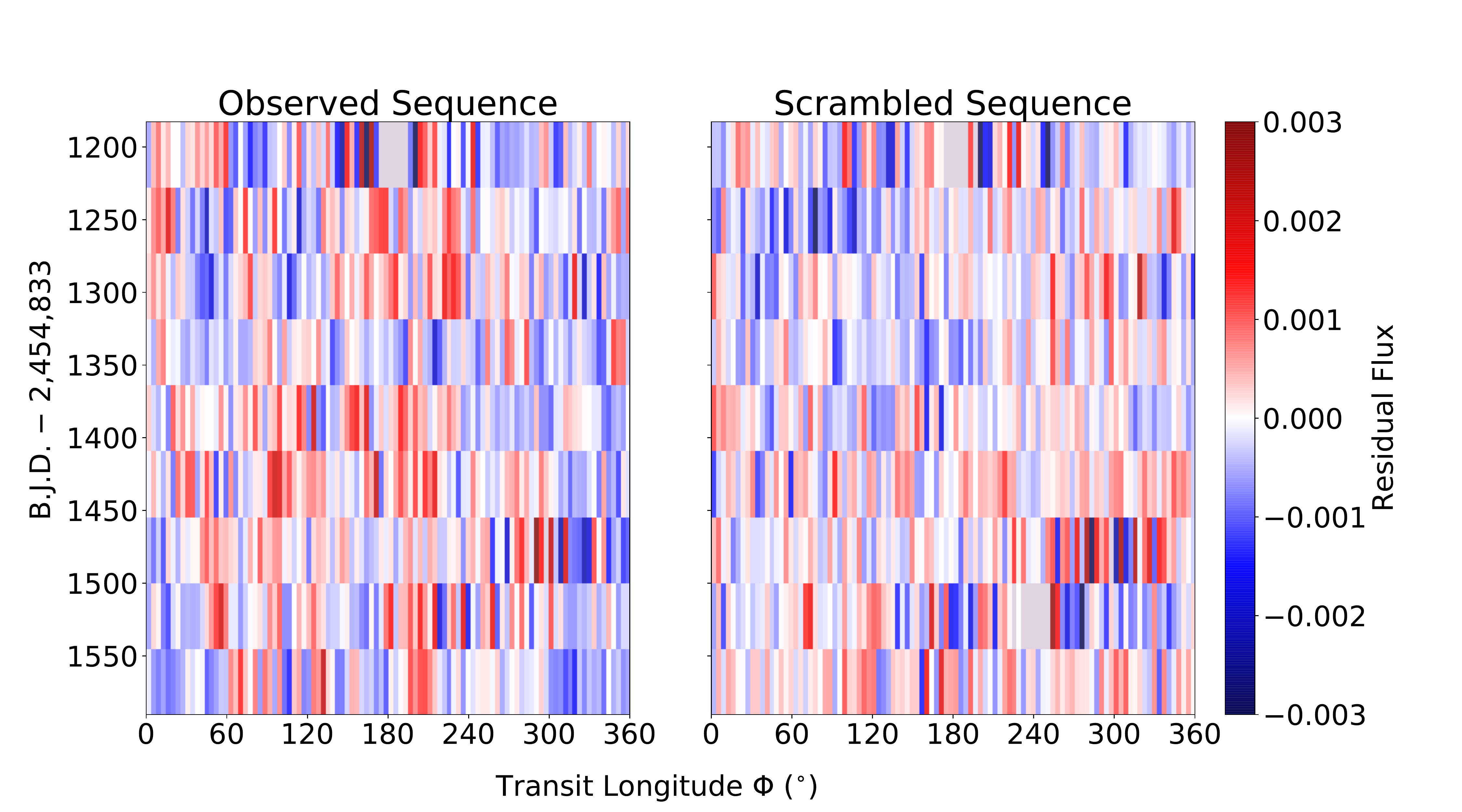}
\caption{ {\bf Transit Tapestry of Kepler-762b} Same as Figure
  \ref{Kepler17_tapestry} but for Kepler-762. A strong correlation
  was detected by the TCC method suggesting a low stellar
  obliquity. The presence of active regions is not visually obvious.}
\label{Kepler-762_tapestry}
\end{center}
\end{figure*}

Kepler-762b is a 3.8-day hot Jupiter around a G star.  The planet went
from being a ``candidate'' to being ``validated'' through the
statistical analysis of \citet{Morton2016}.  \citet{Holczer2015}
reported a prograde orbit for this system given the observed strong
correlation between TTV and local flux variation.  Our TCC analysis
shows a good agreement between $P_{\text{tcc}} = 4.0 \pm 0.1$ days and
$P_{\text{phot}}$ = $4.045 \pm 0.025$ days.  The statistical strength
of the correlation is $N_{\sigma}$ = 5.2. We thus argue that the orbit
of Kepler-762 is not only prograde but also well-aligned ($\Psi$
$\lesssim$16$^{\circ}$).  The transit tapestry,
Figure~\ref{Kepler-762_tapestry}, is not impressive to the eye, even
though the TCC is strong enough for a confident conclusion.  Given the
impact parameter of the transit of about 0.05 (ExoFop) and
$R_p/R_\star$ of 0.10, the transit chord probes the stellar latitude
of $3 \pm 7^{\circ}$.
  
\subsection{Kepler-423}

\begin{figure*}
\begin{center}
\includegraphics[width = 1.5\columnwidth]{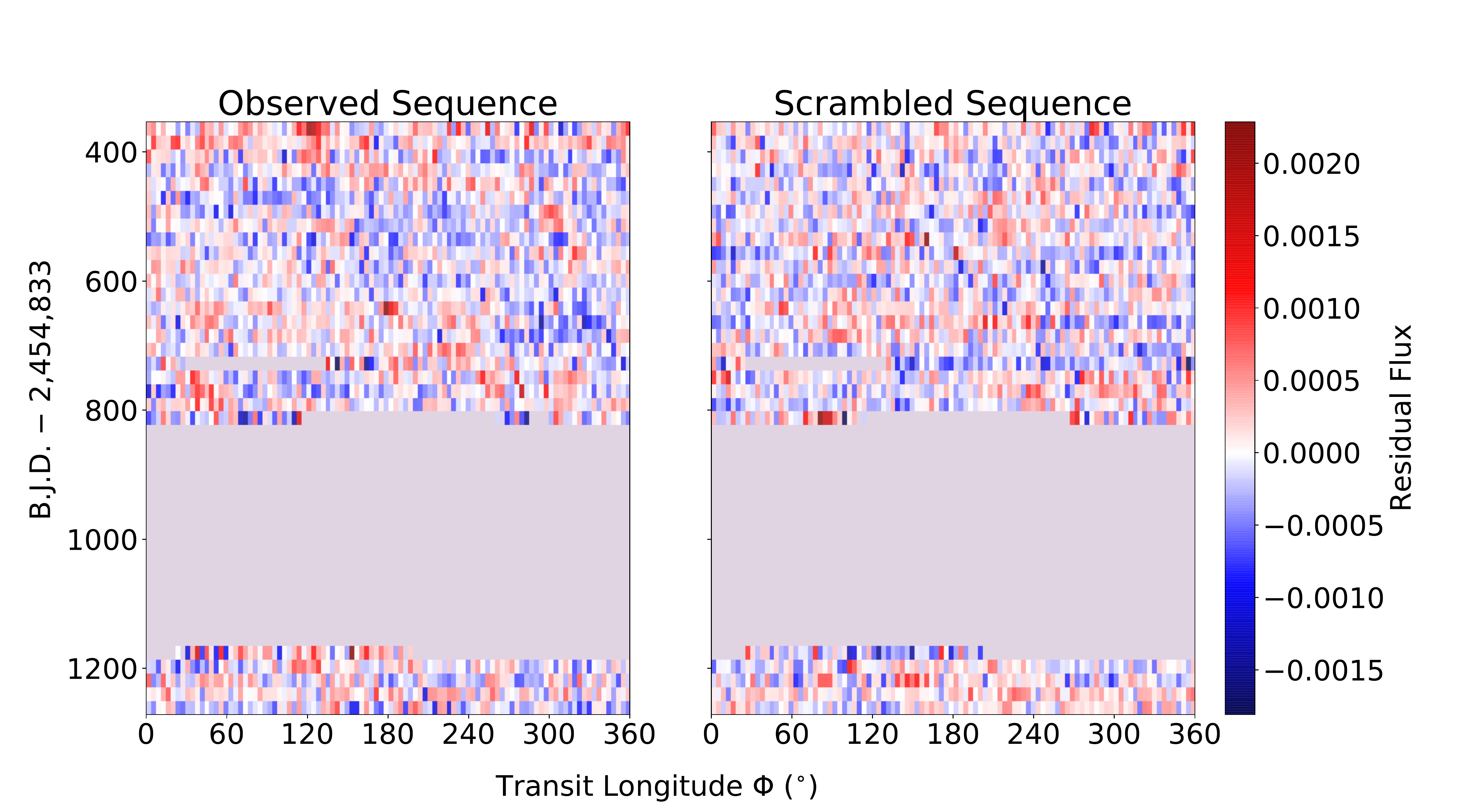}
\caption{ {\bf Transit Tapestry of Kepler-423b} Same as Figure
  \ref{Kepler17_tapestry} but for Kepler-423b.  TCC analysis suggests
  a low stellar obliquity. Although there are hints of group-group
  correlation, we do not see prominent, long-lasting active regions.
  Similar to TrES-2, Kepler-423 may be faculae-dominated given its
  slow stellar rotation period (22 days). In contrast to the
  spot-dominated stars, Kepler-423 may not have persistent, localized
  features in the transit light curves.}
\label{Kepler-423_tapestry}
\end{center}
\end{figure*}

Kepler-423b is a 2.7-day hot Jupiter orbiting a G star
\citep{Gandolfi2015}.  The maximum TCC has $N_{\sigma} = 4.7$ at
$P_{\text{tcc}} = 23.0 \pm 0.9$ days. The stellar rotation period was
also independently measured to be $P_{\text{phot}}$ = $22.047 \pm
0.121$ days based on the out-of-transit light curve. Therefore
Kepler-762 likely has a low stellar obliquity ($\Psi$ $\lesssim$
10$^{\circ}$).  In the transit tapestry
(Figure~\ref{Kepler-423_tapestry}), there are hints of group-group
correlations, however no large-scale, long-lasting active regions can
be discerned by eye. As we said with regard to TrES-2, Kepler-423 may
be also faculae-dominated given its slow stellar rotation period.
 
\subsection{WASP-85}
\begin{figure*}
\begin{center}
\includegraphics[width = 2.0\columnwidth]{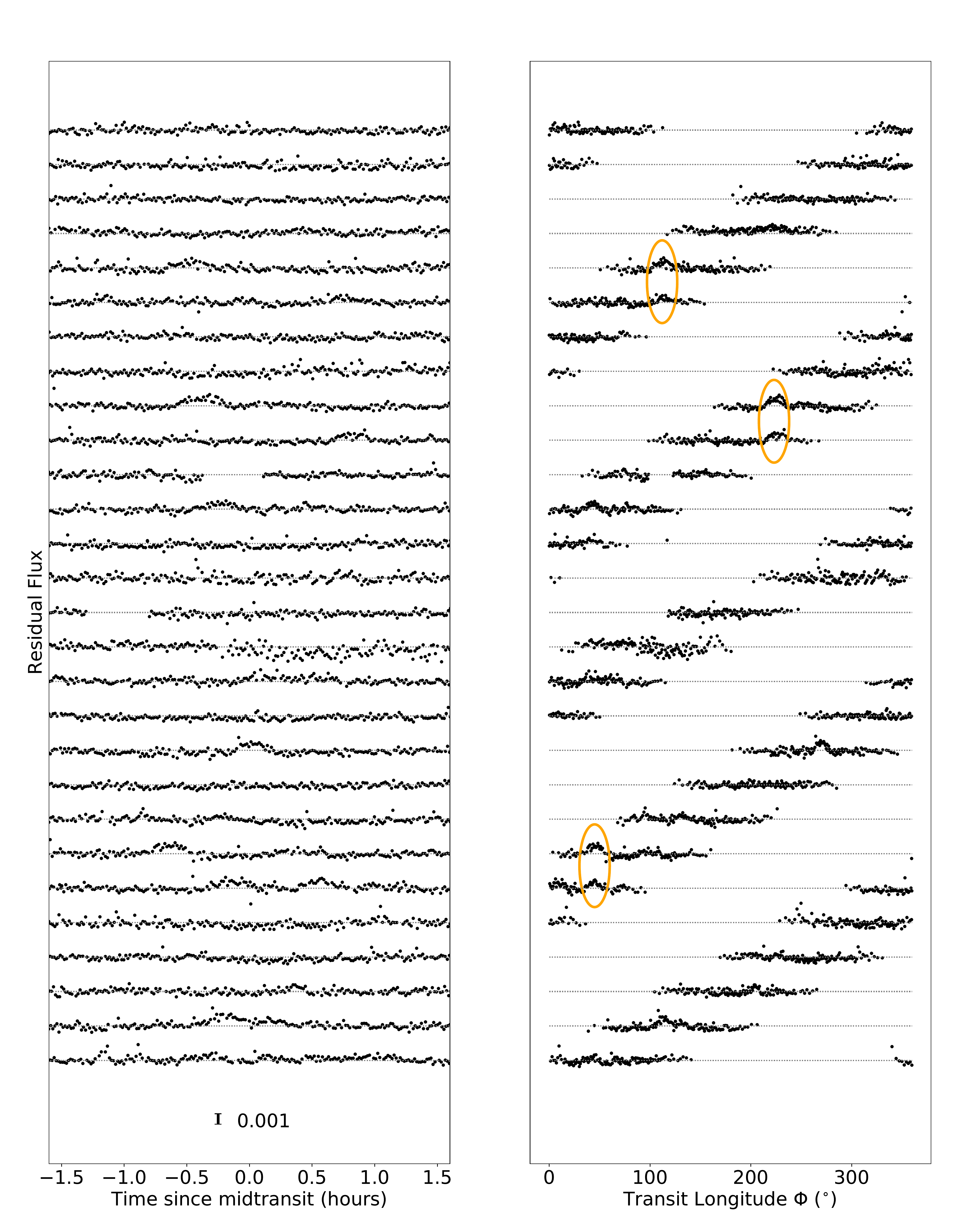}
\caption{ {\bf {\it K2} light curve of WASP-85} ---Left: the residual
  flux (the observed transit light curve minus the best-fitting
  transit mode) as a function of time. ---Right: the residual flux as
  a function of transit longitude $\Phi$ calculated with
  $P_{\text{tcc}} = 15.2$ days. As shown with the orange ellipses,
  spot-crossing anomalies recur at fixed transit longitudes from one
  transit to the next unless the spot-crossing anomalies happen close
  to the ingress/egress of the transits. During the ingress/egress,
  geometrical foreshortening and limb darkening both weaken the signal
  of spot-crossing anomalies.}
\label{201862715_resdiual}
\end{center}
\end{figure*}

WASP-85Ab is a 2.7-day hot Jupiter around a G star in a visual binary
system \citep{Brown2015}. The companion is a K star at an angular
separation of about 1.5 arcseconds ($\sim$190~AU).
\citet{Mocnik2016wasp85} analyzed the spot-crossing anomalies
identified in the {\it K2} short-cadence data for WASP-85A.  They
concluded that the recurrence of the spot-crossing events not only
suggests a low stellar obliquity ($\lambda$ $\lesssim$ 14$^{\circ}$),
but also constrains the stellar rotation period to be $15.1 \pm 0.6$
days. On the other hand, they measured the stellar rotation period to
be $13.6 \pm 1.6$ days based on the rotational modulation seen in the
out-of-transit light curve.

We detect a TCC of $N_{\sigma} = 4.5$ at $P_{\text{tcc}} = 15.2 \pm 0.3$ days.
This is consistent with the value of $15.1 \pm 0.6$ days
reported by \citet{Mocnik2016wasp85}.  We measure a $P_{\text{phot}}$
= $13.28 \pm 0.41$ days which was also consistent with the $13.6 \pm
1.6$ days reported by \citet{Mocnik2016wasp85}. Since the $N_{\sigma}$
of 4.5 is not as strong as the previous cases and the $P_{\text{tcc}}$
differ substantially from $P_{\text{phot}}$, we examined WASP-85 more
closely.

The left panel of Figure \ref{201862715_resdiual} shows the residual
flux of {\it K2} short-cadence observation of WASP-85.  After
transforming time into transit longitude using $P_{\text{tcc}} = 15.2$
days, the spot-crossing anomalies clearly recur at fixed transit
longitude from one transit to the next, except in cases where
spot-crossing events happened near the ingress or egress.  Geometrical
foreshortening and limb darkening effect are most severe during the
ingress/egress; and both effects tend to weaken the signal of
spot-crossing anomalies. Therefore the spot-crossing anomalies are
expected to be suppressed here. The recurrence of spot-crossing events
are compelling enough that we agree with \citet{Mocnik2016wasp85} that
WASP-85Ab likely has a low stellar obliquity. The fact that
$P_{\text{tcc}} = 15.2 \pm 0.3$ days disagrees with $P_{\text{phot}}$
= $13.28 \pm 0.41$ days may be a sign differential
rotation. Alternatively, this may be the result of the short lifetime
of starspots. The rotational modulation in the {\it K2} light curve
underwent significant evolution during the $\approx$80 days of {\it
  K2} observation suggesting a short spot lifetime. Furthermore,
Figure~\ref{201862715_resdiual} shows that many of crossed spots only
persisted for days before disappearing.

The spot-crossing anomalies are about 0.0004-0.0008 in relative flux
while the transit depth is about 2\%. If the active regions uniformly
fills the shadow of the planet, their relative intensity is hence
2-4\% dimmer than the averaged photosphere. The spot-crossing events
last about $\sim 15$ min or $\sim 20^{\circ}$ in longitude.  After
accounting for the blurring effect of the planet ($R_p/R_\star$ of
0.13), the spots are smaller than $5^{\circ}$. For WASP-85, we are
likely seeing smaller, shorter-lived, individual starspots rather
than the more stable and extended active regions on Kepler-17.
Curiously, on the Sun, the lifetime of a sunspot is proportional to
its size \citep{Gnevyshev,Waldmeier}. With this ``GW rule'' in mind,
it is not surprising that the spots on WASP-85 have shorter lifetimes
compared to the extended active regions on Kepler-17.  We estimate the
latitude probed by the transit chord as $3 \pm 7^{\circ}$ using the
the impact parameter of the transit of about 0.05
\citep{Mocnik2016wasp85} and $R_p/R_\star$ of 0.13.

\begin{figure*}
\begin{center}
\includegraphics[width = 1.5\columnwidth]{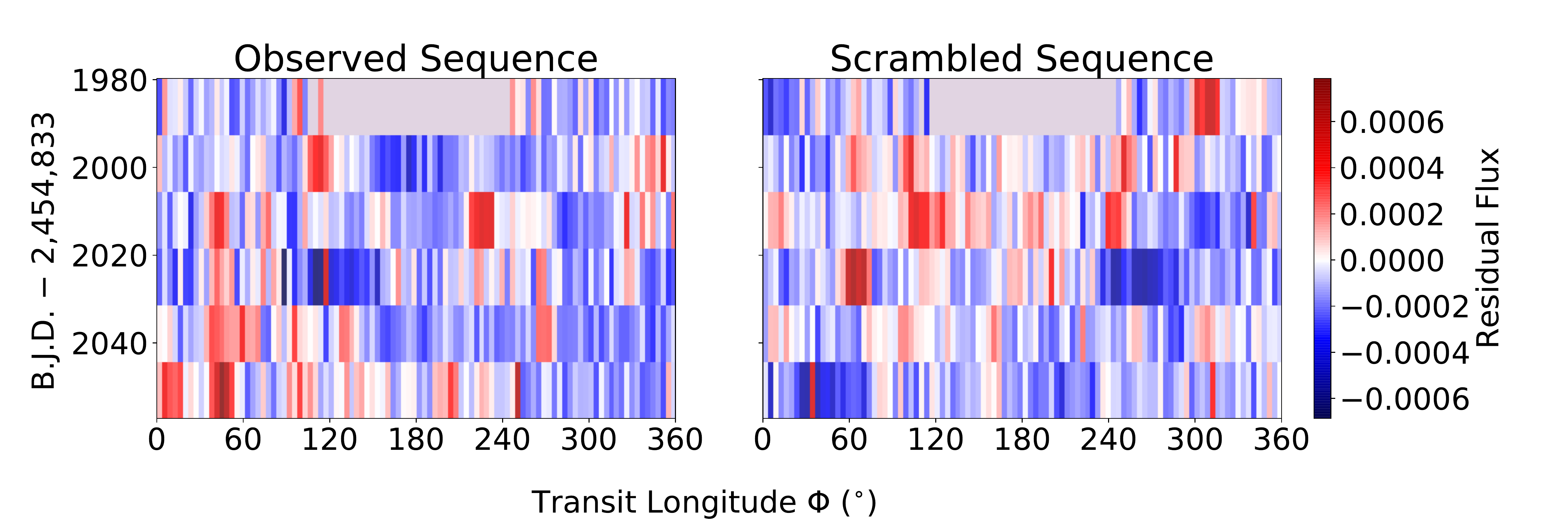}
\caption{ {\bf Transit Tapestry of WASP-85b.} Same as Figure
  \ref{Kepler17_tapestry} but for WASP-85. TCC of $N_{\sigma}$ = 4.5
  is suggestive of a low obliquity.  Close examination of
  Figure~\ref{201862715_resdiual} showed that spot-crossing events did
  recur in neighboring transits. We agree with
  \citet{Mocnik2016wasp85} that WASP-85Ab likely has a low stellar
  obliquity. We are likely seeing the crossing of individual spots
  instead of extended active regions. This would explain the shorter
  lifetimes ($\sim 10$ days), weaker photometric signal and smaller
  size ($\sim 5^{\circ}$) of the anomalies seen in the transit light
  curves.}
\label{WASP-85_tapestry}
\end{center}
\end{figure*}

\subsection{HAT-P-11}

HAT-P-11b is a 4.9-day super-Neptune around a K dwarf.  With a
relatively large scaled orbital distance of $a/R_\star$ = 15.6, it is
more akin to the ``warm Jupiters'' than hot Jupiters.
Rossiter-McLaughlin observations revealed $\lambda = 103^{+23}_{-10}$
degrees.  \citet{Sanchis-Ojeda2011Hat} confirmed that the orbit is
nearly perpendicular to the stellar equator by analyzing the
spot-crossing anomalies seen in the {\it Kepler} light curve.  They
showed that spot-crossing anomalies cluster near two specific phases
of the transit.  They attributed this phenomenon to the presence of
two active latitudes on the photosphere. As the planet transits the
host star on a perpendicular orbit, the two symmetric active latitudes
on both hemisphere will be occulted sequentially. The photometric
signatures of the active regions are two persistent brightening
features in the residual flux that remain fixed relative to the
mid-transit time.  Recent work by \citet{Morris} mapped out the
distribution of starspots on HAT-P-11 by explicitly modeling
individual spot-crossing events. They drew attention to many
similarities between sunspots and the starspots on HAT-P-11 such as
size and latitudinal distribution. \citet{Morris_activity} showed that
HAT-P-11 is chromospherically more active than other planet hosts with
similar properties.

\citet{Beky2014Hat} reported that the ratio of spin period to orbital
period is very nearly 6 to 1.  Because of this commensurability, every
six transits the planet and star return to the same configuration. As
a result, the transiting planet can revisit the same active regions
despite the high stellar obliquity. This results in a repeating
pattern in the residual flux. Consequently, our TCC method detects a
strong correlation ($N_{\sigma}$ = 6.0) even though the system is
known to have a high obliquity.  Visual inspection of the {\it Kepler}
light curve shows that spot-crossing events only recur at fixed
transit longitude after 6 consecutive transits. The close spin-orbit
commensurability distinguishes HAT-P-11 from the low-obliquity systems
discussed earlier.
 
\subsection{Kepler-63}

Kepler-63b is a sub-Saturn orbiting a G dwarf every 9.4
days. \citet{Sanchis-Ojeda2013} showed that the system has a high
stellar obliquity ($\Psi$ = 104$^{+22}_{-14}~^{\circ}$) using both
Rossiter-McLaughlin observations and spot-crossing events in the {\it
  Kepler} light curve. Similar to HAT-P-11, Kepler-63 displays an
apparent spin-orbit resonance (4:7 in this case). Similar to HAT-P-11,
our TCC method finds a relatively strong signal ($N_{\sigma}$ = 5.8)
even though the system has a high stellar obliquity. The strongest TCC
is detected at $P_{\text{tcc}}$ = 37.7 $\pm$ 0.8 days
(c.f. $P_{\text{phot}}$ = 5.041 $\pm$ 0.014 days) which coincide with
the least common multiple between the stellar rotation period and the
orbital period ($4 P_{\text{orb}}$ or $7 P_{\text{phot}}$) i.e. the
period at which the system return to the same configuration. This is
indicative of a spin-orbit commensurability rather than a
low-obliquity orbit as we explained earlier.

\subsection{Kepler-25}

The two transiting planets b and c of Kepler-25, an F star, were
confirmed by \citet{Steffen2012} via transit timing
variations. \citet{Albrecht2013} performed Rossiter-McLaughlin
measurement during the transits of planet c which is a sub-Saturn on a
12.7-day orbit. The Keck/HIRES data they obtained showed that
Kepler-25c has a low stellar obliquity of $\lambda = -0.5 \pm
5.7^{\circ}$. We applied the TCC method to the short-cadence {\it
  Kepler} light curve of Kepler-25. No statistically significant
correlation was detected. The strongest TCC has $N_{\sigma}= 3.2$ at
$P_{\text{tcc}} = 31 \pm 1$ days. We could not measure the stellar
rotation period independently in the out-of-transit light curve. Here
and elsewhere in similar cases, we use the measured rotational
broadening $v \rm{sin} i_\star$ and the stellar radius to estimate the
stellar rotation period. For Kepler-25, \citet{Albrecht2013} reported
$v \rm{sin} i_\star = 9.5\pm0.5$~km~s$^{-1}$ and a stellar radius of
$1.36\pm 0.13 R_\odot$. Assuming an edge-on geometry ($i_\star$ =
90$^{\circ}$), the stellar rotation period should be close to 7
days. The non-detection of correlation in the residual flux may be
attributed to several possibilities. The SNR of any photometric
signature of active regions is expected to be smaller since the
transit depth is only about 0.1\% (compared to the 1-4\% for the
strong TCC cases described earlier). The lack of rotational modulation
in the out-of-transit light curve suggests that the host star
($T_{\text{eff}} \approx 6200\,K$) may not be magnetically
active. Finally, the impact parameter of the planet is high ($b =
0.881\pm 0.004$); the transit chord might have missed any active
latitudes that are closer to the stellar equator.

\subsection{CoRoT-11}

CoRoT-11b is a 3.0-day hot Jupiter orbiting an F star, first reported
by \citep{Gandolfi2010}.  Later, \citet{Gandolfi2012} showed that the
system has a low stellar obliquity ($\lambda = 0.1 \pm 2.6^{\circ}$)
based on the Rossiter-McLaughlin effect.  Our TCC analysis could not
detect a statistically significant correlation ($N_{\sigma}= 3.0$;
$P_{\text{tcc}} = 30 \pm 4$ days; $P_{\text{phot}}$
unconstrained). Assuming an edge-on geometry
($i_\star = 90^{\circ}$),
the stellar rotation period should be close to 2 days
\citep[$v\rm{sin} i_\star = 40\pm 5$~km~s$^{-1}$ and $R_\star = 1.37\pm 0.03 R_\odot$,
  from][]{Gandolfi2012}.
Similar to Kepler-25, the CoRoT-11 star may
be too massive and hot ($T_{\text{eff}} \approx 6400\,K$) to be
magnetically active. In addition the impact parameter is high ($b =
0.8108\pm 0.0077$) and the transit chord might miss the active
latitudes.

\subsection{CoRoT-19}

First reported by \citet{Guenther2012}, CoRoT-19b is a hot Jupiter
orbiting a F star every 3.9 days.  \citet{Guenther2012} also measured
the Rossiter-McLaughlin effect and found the sky-projected stellar
obliquity to be $\lambda = -52^{+27}_{-22}~^{\circ}$. Our TCC analysis
did not detect a statistically significant correlation ($N_{\sigma} =
3.0$; $P_{\text{tcc}} = 10 \pm 1$ days; $P_{\text{phot}}$
unconstrained). The $v\rm{sin} i_\star$ argument gives a stellar rotation
period close to 14 days \citep[$v\rm{sin} i_\star = 6\pm 1$~km~s$^{-1}$
  and $R_\star = 1.65\pm 0.04 R_\odot$, from][]{Guenther2012}. The
non-detection of correlation is consistent with the high obliquity
reported by \citet{Guenther2012}. However, it might also be because
the host star is too massive and hot ($T_{\text{eff}} \approx
6100\,K$) to be magnetically active.

\subsection{WASP-47}

WASP-47b is 4.2-day hot Jupiter around a G star
\citep{Hellier2012}. Observations by {\it K2} revealed two additional
transiting planets in the system, thus making WASP-47b the first hot
Jupiter known to have close-in planetary companions
\citep{Becker2015}. \citet{Sanchis-Ojeda2015} measured the
Rossiter-McLaughlin effect induced by the hot Jupiter WASP-47b,
finding $\lambda = 0 \pm 24 ^{\circ}$.

We do not find a statistically significant correlation ($N_{\sigma} =
3.0$; $P_{\text{tcc}} = 6.3 \pm 0.4$ days; $P_{\text{phot}}$
unconstrained). The stellar rotation period may be close to 37 days
based on the combination of $v\rm{sin} i_\star =
1.80^{+0.24}_{-0.16}$~km~s$^{-1}$ and $R_\star = 1.16 \pm
0.26~R_\odot$ \citep{Hellier2012,Sanchis-Ojeda2015}. WASP-47 may be
magnetically quiet, which is consistent with the lack of rotational
modulation in the {\it K2} light curve.

\subsection{Kepler-13}

Kepler-13Ab is 1.8-day hot Jupiter around an A star.
\citet{Barnes2011} reported a misaligned orbit based on the asymmetric
transit profile induced by gravity darkening. Later Doppler tomography
\citep{Johnson2014} and further analysis of gravity darkening
\citep{Masuda2015} showed that the system has a high obliquity
($\lambda = 58.6 \pm 2.0 ^{\circ}$).

Before applying our TCC method, we removed the best-fitting gravity
darkening model \citep{Masuda2015}, rather than the usual transit
model, in order to obtain the residual flux time series.  No
statistically significant correlation was detected. The strongest TCC
has $N_{\sigma} = 2.7$ at $P_{\text{tcc}} = 3.12 \pm 0.10$ days.
However, \citet{Masuda2015} reported a stellar rotation period close
to one day. The lack of correlation is expected, given the high
obliquity and the spectral type of the host star. \citet{Szabo2014}
reported the detection of a 3:5 spin-orbit resonance based on a
pattern seen in the residual flux. We did not detect a correlation at
the stellar rotation period implied by the 3:5 spin-orbit resonance
using the best-fitting gravity-darkened model from \citep{Masuda2015}.

\subsection{CoRoT-3}

CoRoT-3b is a 4.3-day brown dwarf ($21.66 \pm 1.00~M_{\text{Jup}}$)
transiting an F star \citep{Deleuil2008}. \citet{Triaud2009} performed
Rossiter-McLaughlin observations which gave a constraint on the
obliquity ($\lambda = 37.6^{+10.0}_{-22.3}~^{\circ}$).  We did not find
a strong TCC; the maximum correlation has $N_{\sigma} = 2.7$ and
$P_{\text{tcc}} = 24.3 \pm 0.5$ days. The photometric rotation period
is unconstrained.  The $v\rm{sin} i_\star$ argument gives a stellar
rotation period close to 3.0 days \citep[$v\rm{sin} i_\star =
  17\pm1$~km~s$^{-1}$ and $R_\star = 1.56 \pm 0.09 R_\odot$,
  from][]{Deleuil2008}.  The non-detection of correlation is
consistent with a mildly oblique orbit.  Alternatively, it may be due
to the magnetically inactive host star as indicated by the lack of
rotational modulation in the {\it CoRoT} light curve.

\subsection{CoRoT-18}

CoRoT-18b is a 1.9-day hot Jupiter around a G star
\citep{Hebrard2010}.  Previous Rossiter-McLaughlin observations
revealed a low stellar obliquity ($\lambda = -10 \pm 20 ^{\circ}$).
We find the strongest TCC of $N_{\sigma} = 2.6$ at $P_{\text{tcc}} =
15.7 \pm 0.7$ days. On the other hand, the rotational modulation in
the out-of-transit light curve suggests a rotation period of
$P_{\text{phot}} = 5.4 \pm 0.4$ days. We cannot confirm the low
stellar obliquity using our correlation method due to the lack of
significant correlation and the disagreement between $P_{\text{tcc}}$
and $P_{\text{phot}}$. This may partly be attributed to the low SNR
light curve of this 15th~mag star.

\subsection{Kepler-448}

Kepler-448b is a 17.9-day warm Jupiter around a rapidly rotating F
star \citep[$v\rm{sin} i_\star = 60.0\pm0.9$~km~s$^{-1}$][]{Bourrier2015}.
The {\it Kepler} light curve shows a 7.5-hour periodic modulation,
presumably the stellar rotation period.  The sky-projected obliquity
has been determined through Doppler tomography to be $\lambda =
12.6^{+3.0}_{-2.9}~^{\circ}$.  We do not detect any statistically
significant correlation ($N_{\sigma} = 2.7$; $P_{\text{tcc}} = 21.0
\pm 0.6$~days; $P_{\text{tcc}}$ unconstrained). If the stellar
rotation period is indeed 7.5 hours, the fast rotation has a similar
timescale as the transit duration. Any photometric anomalies caused by
active regions would likely be smeared out by the fast rotation.

\subsection{Kepler-420}

Kepler-420b is a 86.6-day warm Jupiter around a G dwarf.
\citet{Santerne2014} used radial velocity data to show that the orbit
has a high eccentricity of $0.772 \pm 0.045$ and possibly a high
stellar obliquity $\lambda = 75^{+32}_{-46}~^{\circ}$.  There is no
strong TCC ($N_{\sigma} = 2.5$; $P_{\text{tcc}} = 11.7 \pm 0.7$ days;
$P_{\text{phot}}$ unconstrained).  If the obliquity is low then the
rotation period is close to 13.5~days, based on the reported values of
$v\rm{sin} i_\star = 4.6 \pm 0.2$~km~s$^{-1}$ and $R_\star = 1.13 \pm 0.14
R_\odot$ \citep{Santerne2014}. The non-detection of correlations in
the residual flux is consistent with an oblique orbit, although the
host star may simply lack strong surface magnetic activity.

\subsection{WASP-107}

WASP-107b is a warm Jupiter ($a/R_\star = 18.2$) around a K dwarf star
\citep{Anderson2017}. \citet{Dai2017} analyzed the short-cadence {\it
  K2} light curve of WASP-107. They inferred that WASP-107 has a high
stellar obliquity, with $\Psi = 40-140^{\circ}$, because the observed
spot-crossing anomalies did not recur in neighboring transits. The
lack of recurrence also implied that the stellar rotation period and
the planet's orbital period cannot be in an exact spin-orbit resonance
as previously suggested by \citet{Anderson2017}. If the period ratio
were exactly commensurate, spot-crossing anomalies would recur
regardless of the stellar obliquity. The high obliquity of this system
has now been confirmed by Rossiter-McLaughlin measurement which
indicates a nearly perpendicular orbit (A.~Triaud, private
communication).

We applied the TCC method to the {\it K2} light curve of WASP-107. The
results show a weak signal, with $N_{\sigma} = 2.5$ at $P_{\text{tcc}}
= 10.0 \pm 0.5 $ days. On the other hand, the rotational modulation in
the out-of-transit light curve suggests a rotation period of
$P_{\text{phot}} = 17.1 \pm 1.0$~days. The lack of significant
correlation and the disagreement between $P_{\text{tcc}}$ and
$P_{\text{phot}}$ are consistent with the oblique orbit.

\subsection{Kepler-8}

Kepler-8b is a 3.5-day hot Jupiter around an F star
\citet{Jenkins2010}.  \citet{Albrecht2012} found $\lambda = 5 \pm
7^{\circ}$ based on observations of the Rossiter-McLaughlin effect.
The strongest TCC analysis has $N_{\sigma} =2.2$ at $P_{\text{tcc}} =
27.1 \pm 0.3$ days. The rotational modulation in the out-of-transit
light curve suggests a rotation period of $P_{\text{phot}} = 28.64 \pm
0.32$~days. The weak correlation in the residual flux may be
attributed to the magnetically inactive host star ($T_{\text{eff}}$ =
6200\,K). Moreover, the impact parameter of the planet is high $b =
0.724\pm 0.020$.

\subsection{HAT-P-7}

HAT-P-7b is a hot Jupiter with an orbital period of 2.2 days and an
F-type host star \citep{Pal2008}.  \citet{Albrecht2012} observed the
Rossiter-McLaughlin effect and found a high obliquity, with $\lambda =
155 \pm 37~^{\circ}$. \citet{Masuda2015} measured the true obliquity,
$\Psi = 101 \pm 2~^{\circ}$, based on the observable manifestations of
gravity darkening effects in the {\it Kepler} transit light curves.
As we did with Kepler-13, we removed the best-fitting
gravity-darkening model to isolate any anomalies in the residual flux
time series.  No statistically significant TCC was detected. The
strongest signal has $N_{\sigma}= 2.0$ and $P_{\text{tcc}} = 0.8 \pm
0.2$ days, which conflicts with the rotation period of 1.5--2.1 days
estimated by \citep{Masuda2015}.

\subsection{WASP-75}

WASP-75b is another hot Jupiter around an F star.  The orbital period
is 2.5 days.  The system is distinguished by an unusually large
transit impact parameter, $b =0.887\pm0.008$.  The {\it K2} light
curve reveals a stellar rotation period of $P_{\rm phot} = 13.7\pm1.1$
days.  \citet{Gomez} reported the projected rotation velocity ($v~sin
i_\star = 4.3\pm 0.8$~km~s$^{-1}$) and stellar radius ($1.26 \pm
0.04~R_\odot$), leading to a weak constraint on the stellar
inclination: $\sin i_\star > 0.36$ with 95\% confidence.  We do not
see a strong correlation in our TCC analysis. The strongest TCC has
$N_{\sigma} = 1.9$ at $P_{\text{tcc}} = 22 \pm 3$ days.

\subsection{KIC~6307537}

KIC~6307537 is an eclipsing binary system discovered with {\it Kepler}
data.  The star listed in the Kepler Eclipsing Binary Catalogue is a K
dwarf.  It eclipses the companion star every 29.7 days, causing the
system brightness to fade by 7\% for about 23 hours.  When the
companion occults the catalogued star, the system fades by 17\%.  This
indicates that the companion is likely an evolved star, with a larger
radius and lower effective temperature than the K dwarf.  \citet{Prsa}
classified this system as an eclipsing binary of the Algol type
(detached). The {\it Kepler} light curve also shows a clear rotational
modulation of $78\pm 3$ days which we attribute to the slower rotation
of the evolved star. From the separation between the primary and
secondary eclipse, \citet{vanEylenEB} obtained a constraint of the
eccentricity: $e\cos\omega = 0.0042 \pm 0.0008$.

We apply the TCC method to the 7\% eclipses of KIC~6307537, i.e., when
the evolved star is being eclipsed.  We detect a strong correlation
($N_{\sigma} = 11.4$) in the residual flux at $P_{\text{tcc}} = 80.4
\pm 1.6$ days. This is consistent with the $P_{\text{phot}} = 78 \pm
3$~days.  We interpret the strong TCC as evidence for a low obliquity
of the evolved star.  We place an upper bound of $\Psi \lesssim
6^{\circ}$. The agreement between $P_{\text{tcc}}$ and
$P_{\text{phot}}$ not only supports the low-obliquity interpretation
but also suggests that the rotational modulation in the out-of-eclipse
light curve originates from the evolved star.

\begin{figure*}
\begin{center}
\includegraphics[width = 1.5\columnwidth]{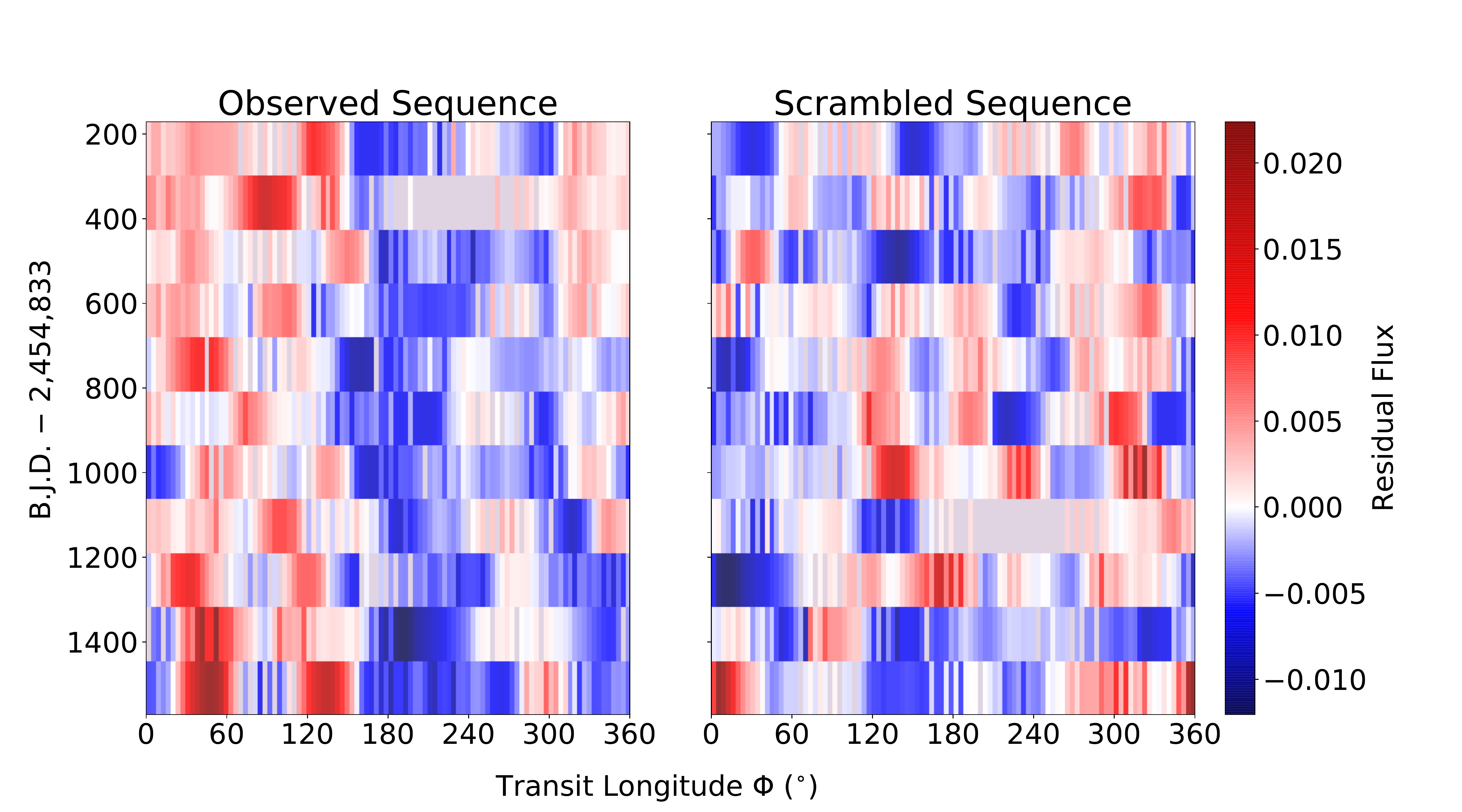}
\caption{ {\bf Eclipse Tapestry of KIC~6307537.} Same as Figure
  \ref{Kepler17_tapestry} but for the eclipsing binary system
  KIC~6307537. We find a strong TCC consistent with a low stellar
  obliquity for the larger star in the system. There are two active
  regions near transit longitudes of 40$^{\circ}$ and
  120$^{\circ}$. The lifetime of these active regions are at least
  $10^3$ days, longer than those inferrred for the dwarf stars
  described earlier.}
\label{6307537_tapestry}
\end{center}
\end{figure*}

Figure~\ref{6307537_tapestry} shows the eclipse tapestry. The tapestry
suggests there are two active regions near longitudes of 40$^{\circ}$
and 120$^{\circ}$ each spanning about 5-10$^{\circ}$ in longitude. The
active regions appear wider in Figure~\ref{6307537_tapestry} because
of the finite radius ratio. With $R_1/R_2 \approx 0.24$, the
photometric features due to the active regions are broadened by about
30$^{\circ}$ in longitude. These photometric features in the residual
flux had amplitudes of about 1-2\% in relative flux. Comparing with
the eclipse depth of about 7\%, the active regions on average were
roughly 15-30\% dimmer than the photosphere. The active regions seem
to have lasted the entire {\it Kepler} campaign indicating a lifetime
of at least $10^3$ days.

\subsection{KIC~5193386}

KIC~5193386 is another {\it Kepler} eclipsing binary system.  One set
of eclipses occurs every 21.4 days, during which the total light
decreases by 8\% for about 22 hours.  The other set of eclipses is
deeper (24\%) and flat-bottomed.  The secondary star (the star being
eclipsed during the 8\% fading events) is likely evolved, with a
larger radius and lower effective temperature than its companion.  We
measure a rotational modulation of $26.0 \pm 0.8$~days in the {\it
  Kepler} light curve. From the separation between the primary and
secondary eclipse, \citet{vanEylenEB} constrained the eccentricity:
$e\cos\omega = -0.0022 \pm 0.0015$.

\begin{figure*}
\begin{center}
\includegraphics[width = 1.5\columnwidth]{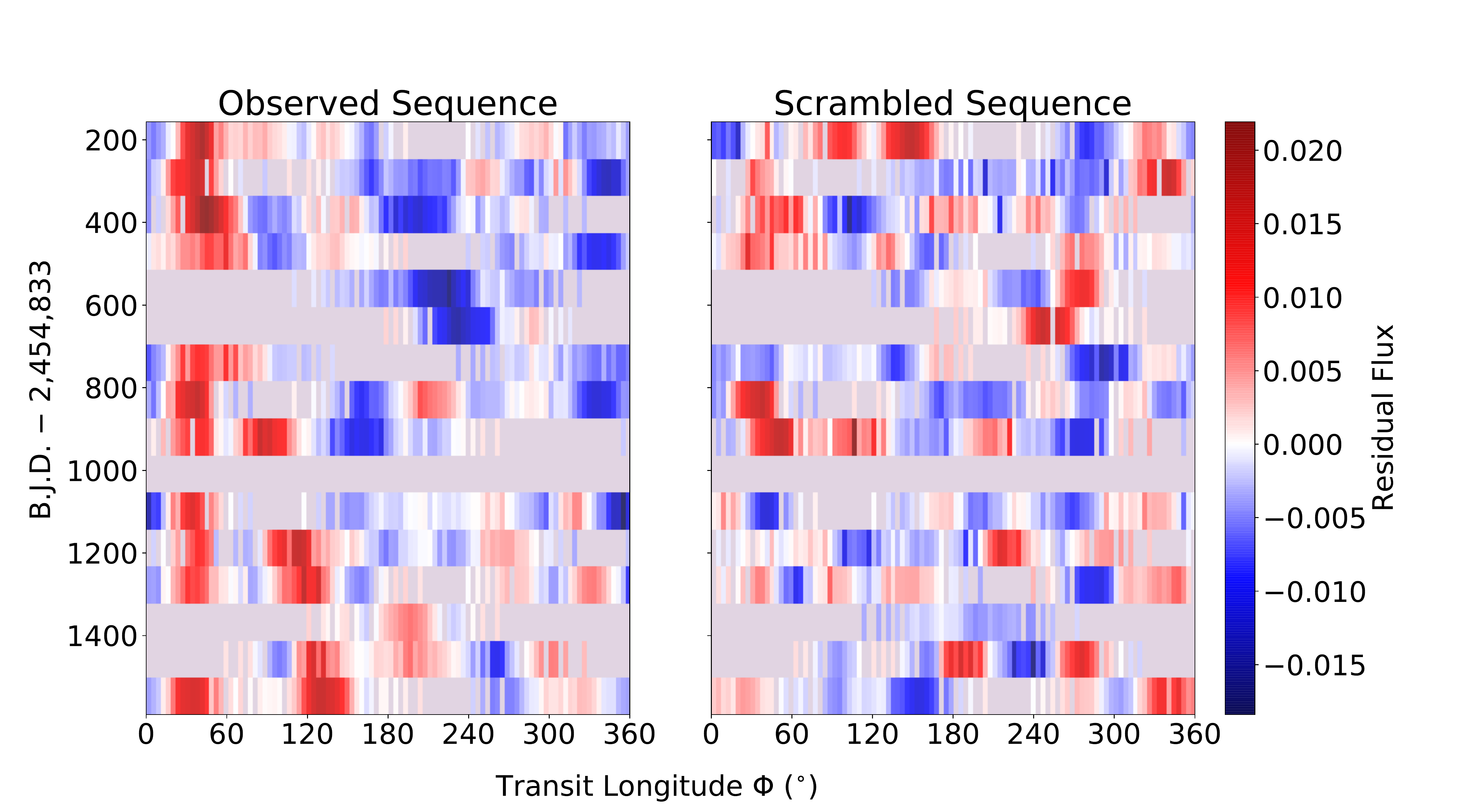}

\caption{ {\bf Eclipse Tapestry of KIC~5193386.} Same as Figure
  \ref{Kepler17_tapestry} but for the eclipsing binary system
  KIC~5193386.  The strong TCC detected suggests a low obliquity for
  the larger, likely evolved star of the system. One prominent active
  region was located at transit longitude $ 30 ^{\circ}$ at the start
  of the {\it Kepler} observation. It then split into two distinct
  regions that separated longitudinally from each other. This may be
  the result of an emerging magnetic flux tube.}
\label{5193386_tapestry}
\end{center}
\end{figure*}

By analyzing the 8\% eclipses we find a strong TCC, with
$N_{\sigma}=9.5$ and $P_{\text{tcc}} = 25.8 \pm 0.3$ days, indicating
a low obliquity ($\Psi \lesssim 14^{\circ}$).
Figure~\ref{5193386_tapestry} shows the tapestry.  During the early
stages of {\it Kepler} observations, one prominent active region was
located at a transit longitude of 30$^{\circ}$. It gradually split
into two distinct active regions that separated longitudinally from
each other. Again, we interpret this phenomenon as the emergence of a
magnetic flux tube and the subsequent separation of the two footprints
of the tube on the stellar photosphere.  The active regions are
consistent of being about 10$^{\circ}$ wide and are 10--25\% fainter
than the photosphere.

\subsection{KIC~6603756}

KIC~6603756 is an eclipsing binary system in the {\it Kepler} catalog
of \citet{Prsa}, who classified this system as an eclipsing binary of
the Algol type (detached). Only one set of eclipses was detected.
These 4\% eclipses last for about 6 hours and repeat every 5.2
days. The {\it Kepler} light curve shows a clear rotational modulation
of 6.128 $\pm$ 0.054 days.

\begin{figure*}
\begin{center}
\includegraphics[width = 1.5\columnwidth]{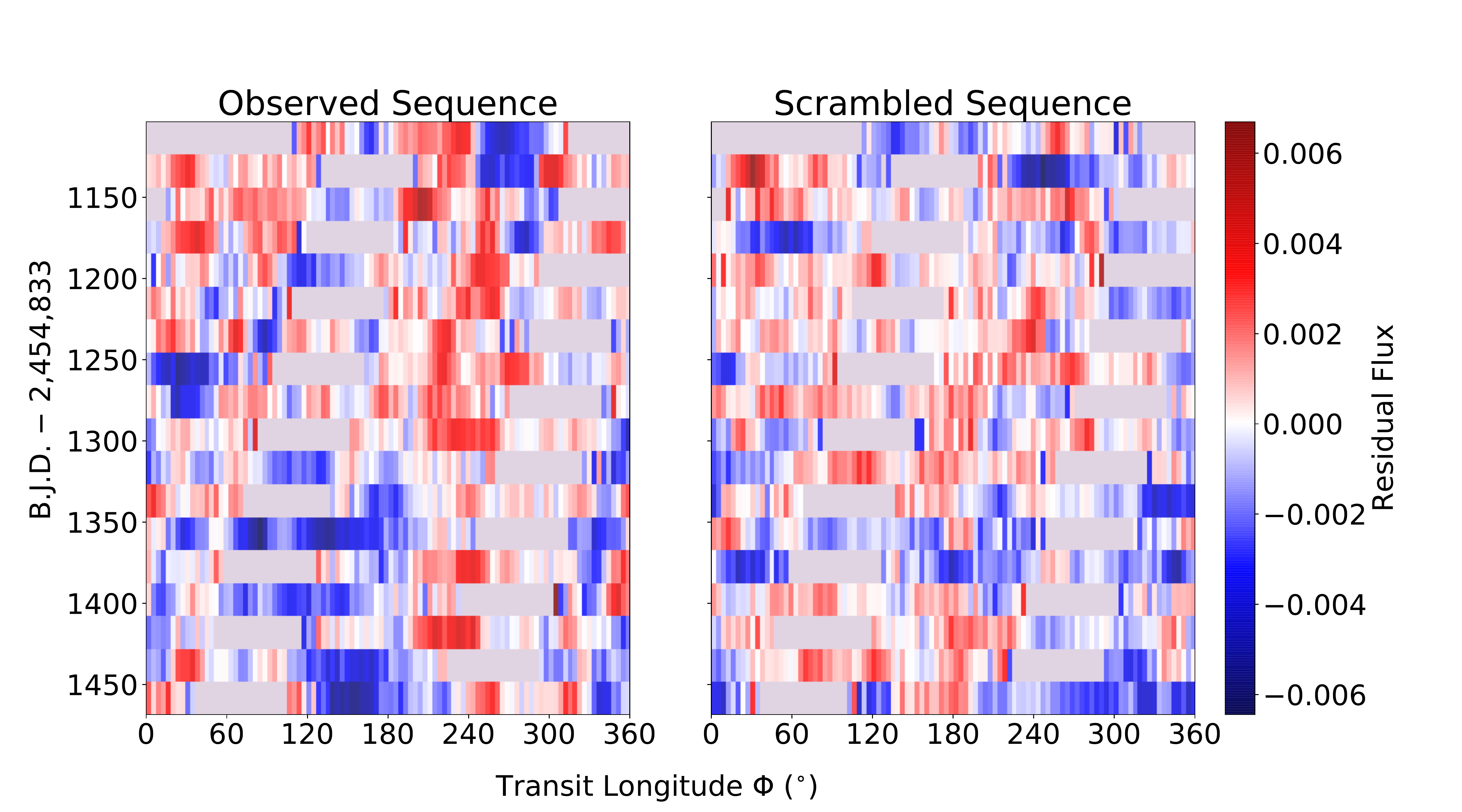}

\caption{ {\bf Eclipse Tapestry of KIC~6603756.} Same as Figure
  \ref{Kepler17_tapestry} but for the eclipsing binary system
  KIC~6603756. TCC analysis revealed a low-obliquity for this
  system. Active regions seem to occur preferentially near transit
  longitude of 230$^{\circ}$ throughout the 300 days of {\it Kepler}
  observations.}
\label{6603756_tapestry}
\end{center}
\end{figure*}

Our TCC analysis gives a strong correlation, $N_{\sigma} = 9.2$, at
$P_{\text{tcc}} = 6.0 \pm 0.1$ days. We interpret the strong TCC as a
sign of a low obliquity for eclipsed star and place an upper bound of
$\Psi\lesssim 12^{\circ}$.  Figure \ref{6603756_tapestry} shows the
eclipse tapestry.  Active regions seemed to occur preferentially near
a transit longitude of 230$^{\circ}$ throughout the 300 days of {\it
  Kepler} observations. These photometric features in the residual
flux had amplitudes of about 0.002--0.006 in relative flux which, when
compared to the eclipse depth of 4\%, implies that the implying that
the active regions were on average 5--15\% dimmer than the
photosphere.

\subsection{KIC~5098444}

KIC~5098444 is a {\it Kepler} eclipsing binary with 2\% primary
eclipses and 0.4\% secondary eclipses (occultations), and an orbital
period of 26.9 days.  The {\it Kepler} light curve shows a clear
rotational modulation of 23.49 $\pm$ 0.19 days.  The primary eclipses,
analyzed here, have a duration of about 11 hours.  The TCC is strong,
with $N_{\sigma} = 7.9$ and $P_{\text{tcc}} = 23.40 \pm 0.24$~days.
We conclude the primary star has a low obliquity, with $\Psi\lesssim
8^{\circ}$. The agreement between $P_{\text{tcc}}$ and
$P_{\text{phot}}$ also shows that the rotational modulation in the
out-of-eclipse light curve originates from the primary.

\begin{figure*}
\begin{center}
\includegraphics[width = 1.5\columnwidth]{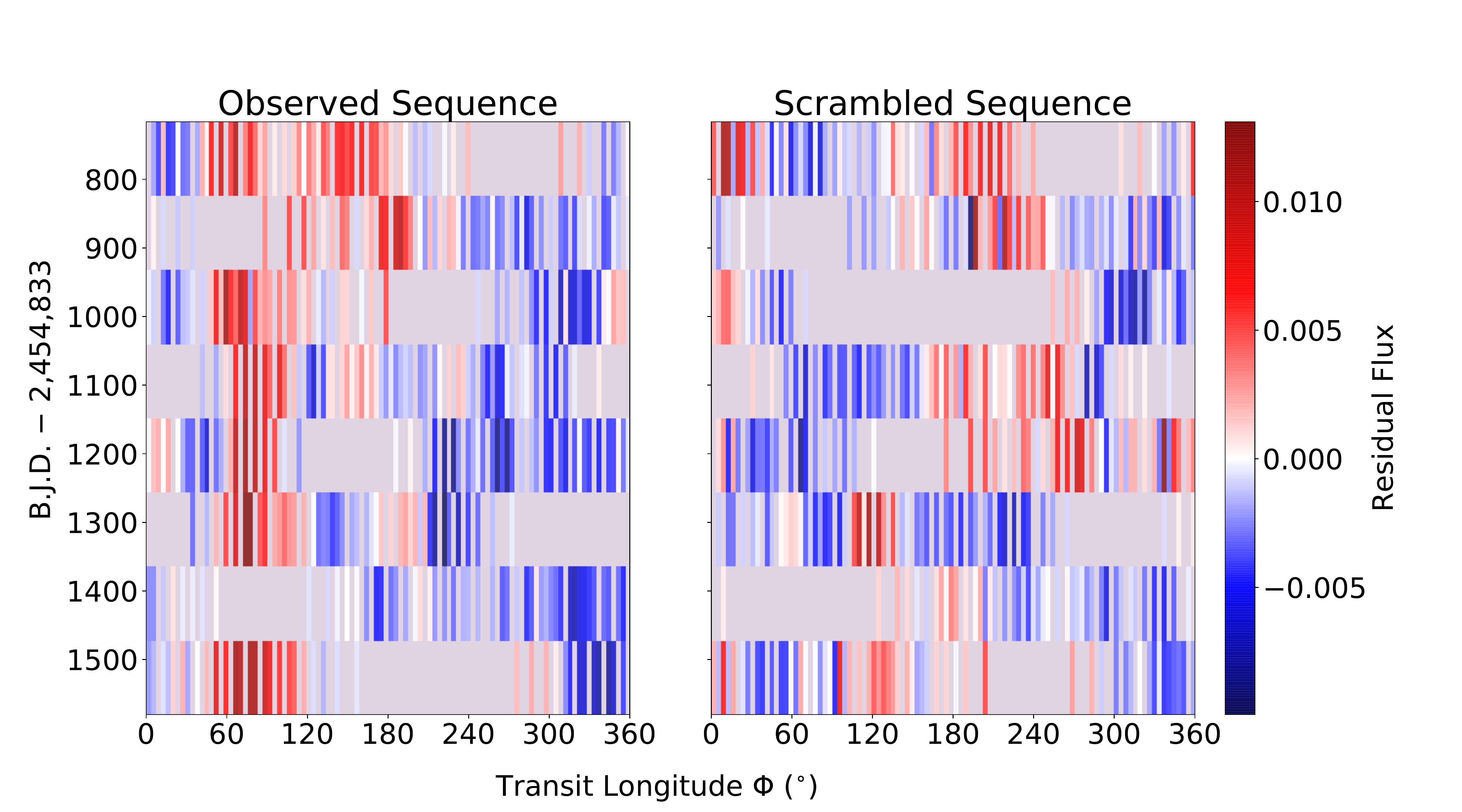}

\caption{ {\bf Eclipse Tapestry of KIC~5098444.} Same as Figure
  \ref{Kepler17_residuals}, but for the eclipsing binary system
  KIC~5098444. TCC analysis suggests a low stellar obliquity for the
  primary.  An active region (40$^{\circ}$ wide, 20-50\% dimmer than
  average photosphere) near transit longitude of 90$^{\circ}$
  persisted throughout the 800 days of {\it Kepler} observations. }
\label{5098444_tapestry}
\end{center}
\end{figure*}

Figure \ref{5098444_tapestry} is the eclipse tapestry. An active
region near transit longitude of 90$^{\circ}$ persisted throughout the
800 days of {\it Kepler} observations.  The flux anomalies have
amplitudes of 0.5--1\%.  Comparing with the eclipse depth of about
2\%, the active regions on average were roughly 20-50\% dimmer than
the photosphere. The active regions span about 40$^{\circ}$ in
longitude after accounting for the finite size of the secondary.

\subsection{KIC~7767559}

KIC~7767559 was originally listed as a planetary candidate, KOI-895.
This system was later classified as an eclipsing binary due to the
detection of significant secondary eclipses. The primary eclipse
occurs every 4.4 days with a depth of 1\% and a duration of 3.9
hours. The {\it Kepler} light curve shows a clear rotational
modulation of $P_{\text{phot}}$ = 5.02 $\pm$ 0.20 days.
\citet{Holczer2015} detected a strong correlation between TTV and
local flux variation, indicating a prograde orbit.

\begin{figure*}
\begin{center}
\includegraphics[width = 1.5\columnwidth]{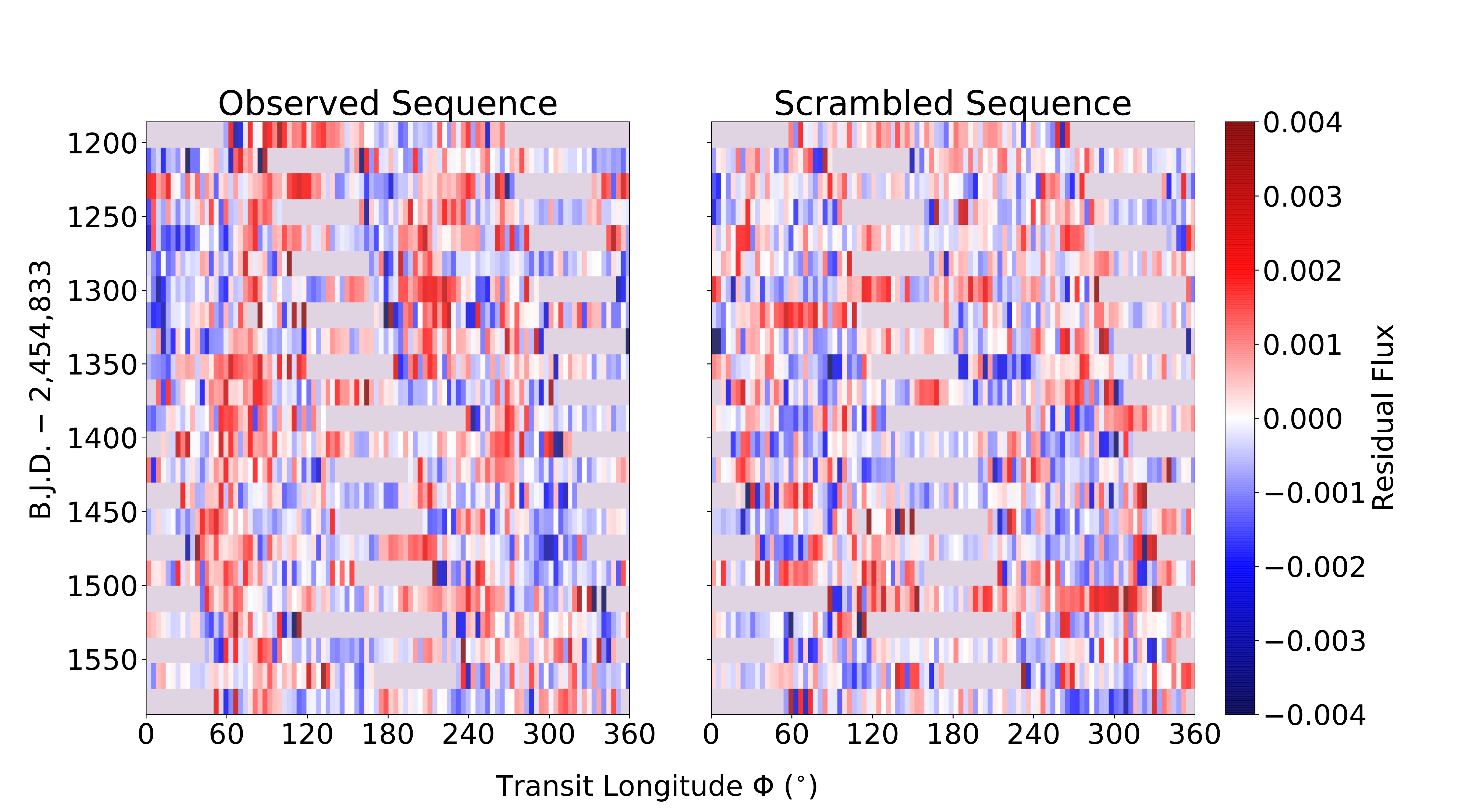}

\caption{ {\bf Eclipse Tapestry of KIC~7767559.} Same as Figure
  \ref{Kepler17_tapestry} but for the eclipsing binary system
  KIC~7767559. A strong TCC signal indicates a low obliquity for the
  primary star.}
\label{7767559_tapestry}
\end{center}
\end{figure*}

The 1\% primary eclipses show a strong TCC with $N_{\sigma} = 7.7$ at
$P_{\text{tcc}} = 5.1 \pm 0.1$ days. We place an upper bound on the
obliquity, $\Psi\lesssim 7^{\circ}$.  Figure \ref{7767559_tapestry}
shows the Eclipse Tapestry. Although there are signs of group-group
correlation, any large-scale, long-lasting active regions are not
visually obvious.

\subsection{KIC~5376836, 3128793, 5282049, 5282049}

\begin{figure*}
\begin{center}
\includegraphics[width = 1.5\columnwidth]{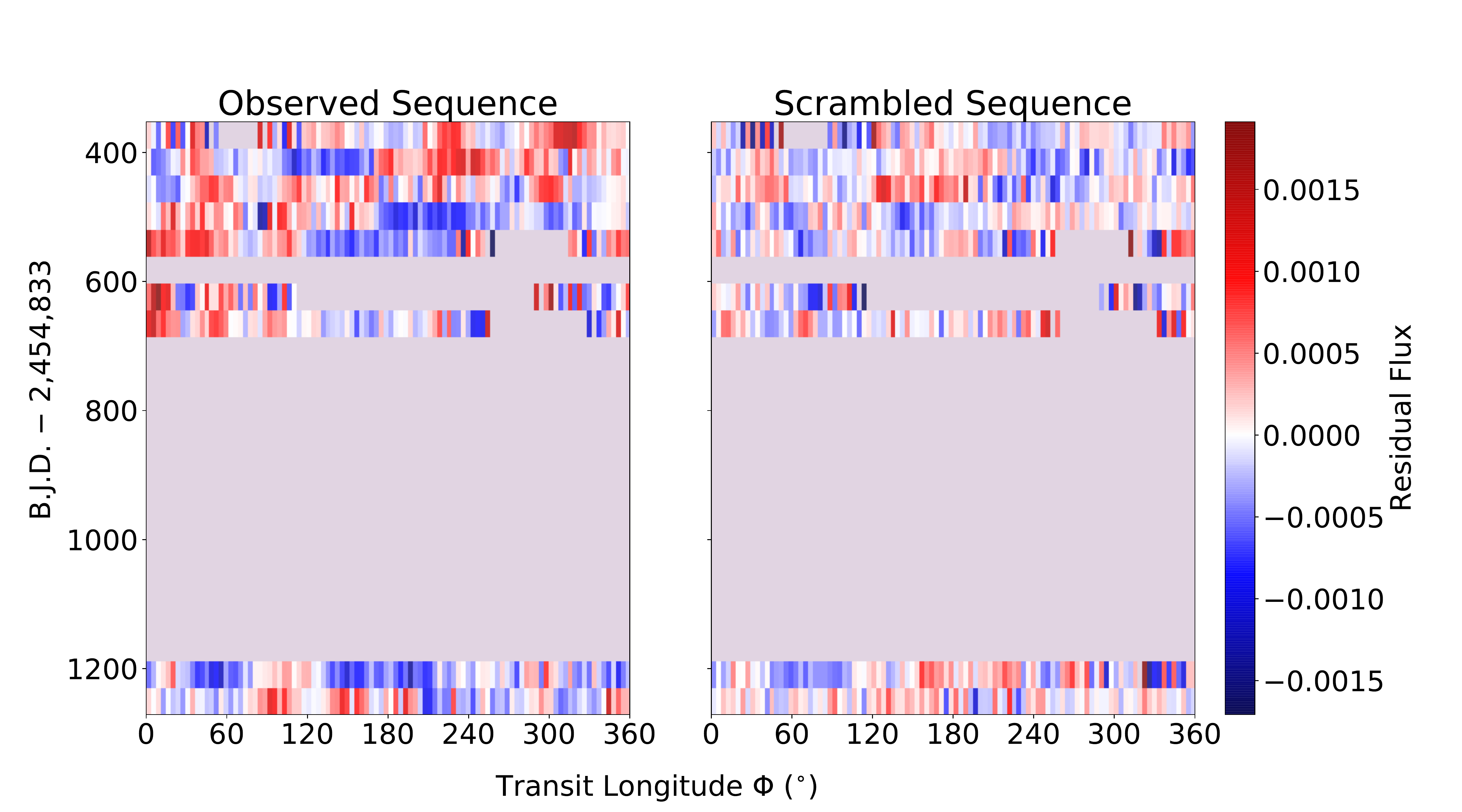}
\includegraphics[width = 1.5\columnwidth]{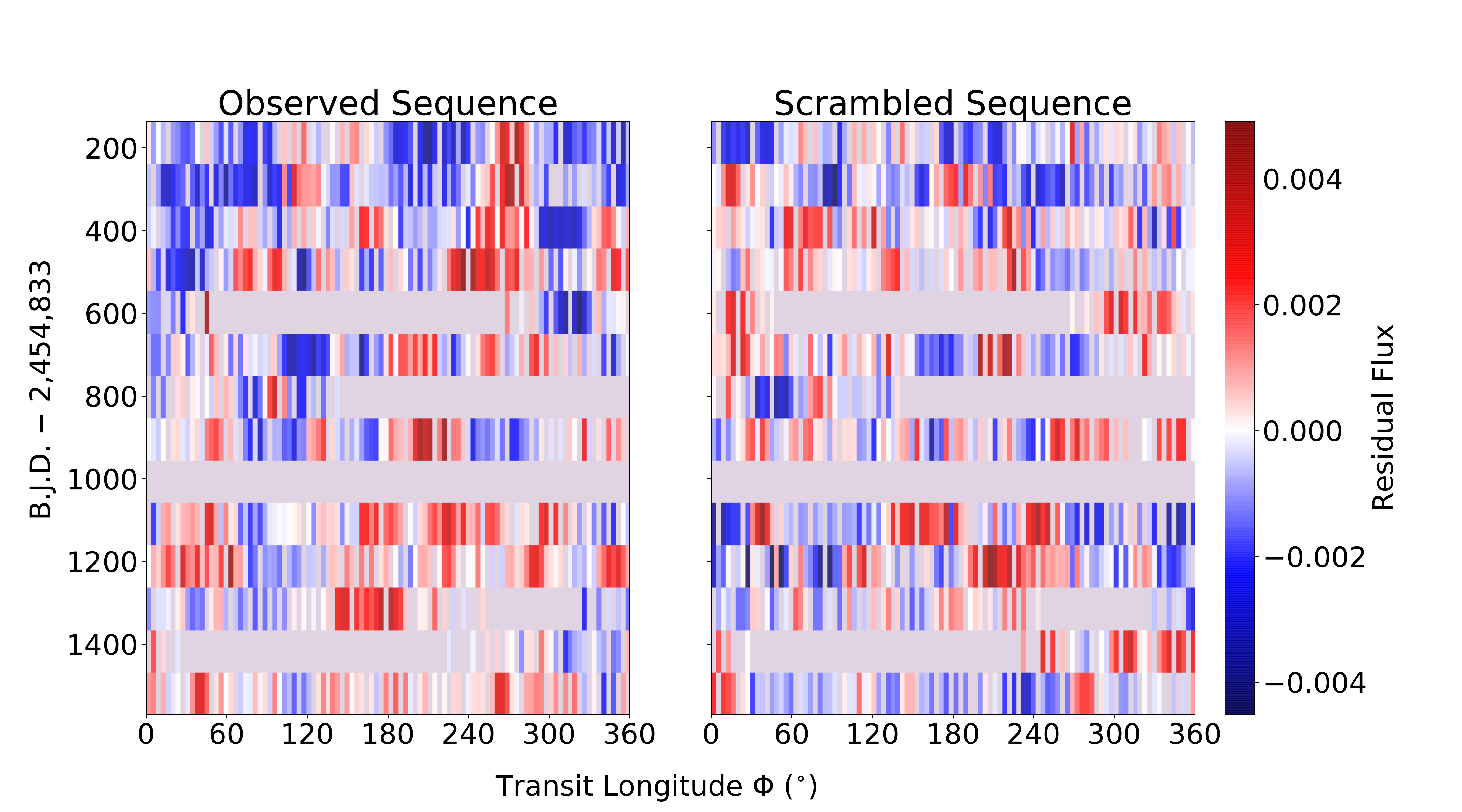}
\includegraphics[width = 1.5\columnwidth]{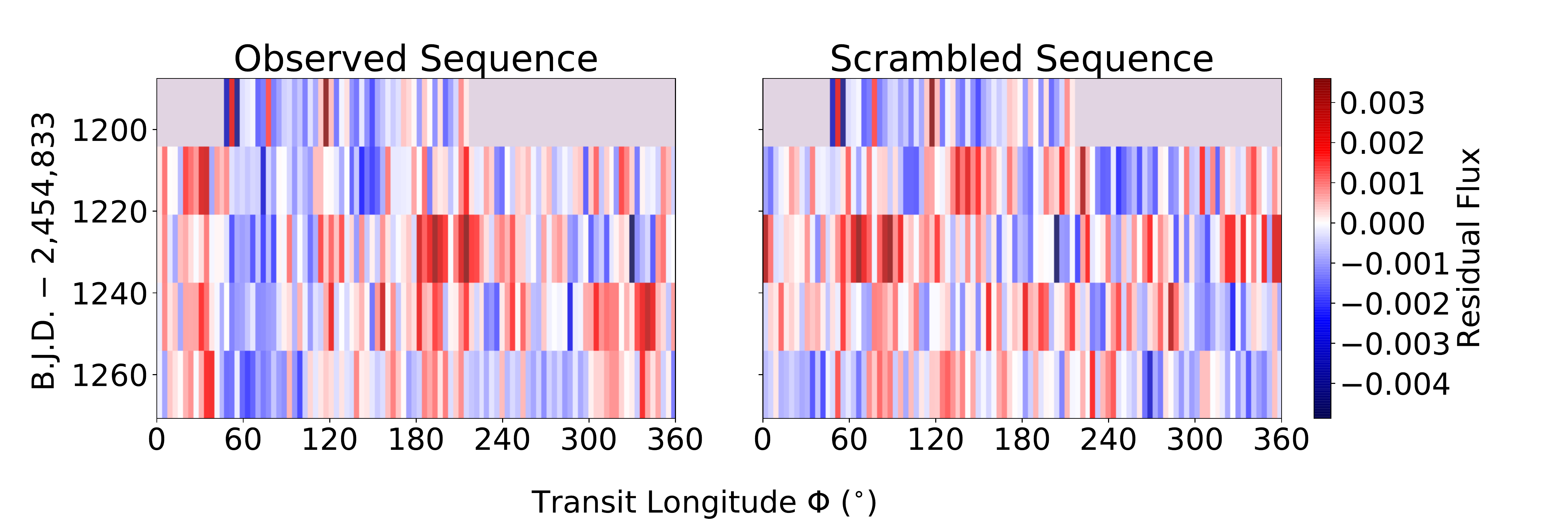}
\includegraphics[width = 1.5\columnwidth]{fig/5282049_tapestry.pdf}

\caption{ {\bf Eclipse Tapestry of KIC~5376836, 3128793, 5282049,
    5282049.} Same as Figure \ref{Kepler17_tapestry} but for these
  four eclipsing binary system from top to bottom. The strong TCC and
  the agreement between $P_{\text{tcc}}$ and $P_{\text{phot}}$ suggest
  low stellar obliquity for all four systems. However, the
  correlations are not obvious to the eye.}
\label{5376836_tapestry}
\end{center}
\end{figure*}

KIC~5376836, 3128793, 5282049, 5282049 are all eclipsing binary
systems discovered by {\it Kepler}.  In all cases our analysis
revealed strong TCC signals, with $N_{\sigma}$ = 5--7, and TCC periods
that agree with the independently measured photometric periods.  We
put an upper bounds on the obliquity $\Psi \lesssim$ 3 to 20$^{\circ}$
(See Table~\ref{tab:eb}).  Figure \ref{5376836_tapestry} shows the
eclipse tapestries for each system, all of which lack high-contrast,
long-lasting active regions.

\section{Discussion and Conclusion}
\label{discussion}
\subsection{The TCC method}

As any other method to obtain information on stellar obliquities, the
TCC method has strengths and weaknesses.  One of the advantages of the
TCC method is that it does not make any assumption about the size,
shape and intensity distribution of the active regions.  In contrast,
the traditional spot-tracking methods
\citep[e.g.][]{Sanchis-Ojeda2011WASP, Tregloan-Reed, Beky2014} often
assume circular and uniform starspots for ease of modeling.  The TCC
method looks for recurrence in the residual flux regardless of the
shape of the recurring pattern. Therefore it can handle spots of
arbitrary shape and intensity distribution and even bright active
regions such as plages and faculae.

We have created transit tapestries to allow the properties of the
active regions on transit chord to be tracked and visualized in a
model-independent manner.  The TCC statistic combines all the data
together, and can therefore be effective even in systems with lower
SNR for which no features can be discerned in the transit tapestry. In
contrast, the traditional spot-tracking method is based on the visual
identification of individual spot-crossing events.  This is more
subjective and is only possible in the systems for which the data have
the highest SNR.  The TCC method is largely automated and does not
require additional follow-up observations, facilitating the
application to a large sample of systems.

One of the limitations of the TCC method is that it requires light
curves with a high temporal sampling rate. This is because the
durations of spot-crossing events are similar to the brief durations
of the ingress and egress phases of the transit.  Continuous
monitoring of the system is also very important for the TCC method.
Continuous monitoring enables an independent check on the stellar
rotation period $P_{\text{phot}}$ from the rotational modulation in
the out-of-transit light curve.  Moreover, active regions have a
limited lifetime, and may drift in longitude and latitude.  The
recurring pattern in residual flux caused by active regions may
quickly change or disappear after a few transits.  This is why it is
important to observe many pairs of neighboring transits, separated by
a relatively short amount of time.

In the TCC method, the detection of a strong correlation requires a
confluence of factors.  The host star must be magnetically active,
with stable active regions.  Unless the system is close to a
spin-orbit commensurability, the stellar obliquity $\Psi$ must be
nearly 0$^\circ$ or 180$^\circ$, such that the planet repeatedly
transits the same active regions.  In addition, the impact parameter
of a transiting planet must be such that the transit chord overlaps
with the active latitudes on the host star.  On the other hand, when
no strong TCC is observed, the system is not guaranteed to have a high
obliquity, as the failure of any of the preceding conditions could
explain the lack of correlation. Therefore the TCC method is best at
picking out stars with obliquities that are very low or are close to
exactly retrograde.  In contrast, the Rossiter-McLaughlin effect can
lead to tight constraints for just about any angle, although it is
only sensitive to the sky-projected obliquity $\lambda$.

\subsection{Detected low-obliquity systems}

We applied the TCC method to selected {\it CoRoT}, {\it Kepler} and
{\it K2} transiting planets and planetary candidates.  We found 10
cases in which the star has a low obliquity (see Table
\ref{tab:planet}).  Among these, five were not previously known to
have a low obliquity: Kepler-71b, KOI-883.01, Kepler-45b, Kepler-762b
and Kepler-423b. For the other five systems, we confirmed the low
stellar obliquities reported previously. Notably, all of these
low-obliquity detections are for stars with giant planets
($R_p>0.9~R_{\text{Jup}}$), with scaled semi-major axis $a/R_{\star}
<12$.  This is not too surprising; the requirements for a relatively
high SNR and many consecutive transits are most easily met for
close-in giant planets.

All of these low-obliquity stars are G and K dwarfs, with the sole
exception of Kepler-45, which is an M dwarf.  The low obliquities of
the G and K dwarfs are consistent with the general trend that hot
Jupiters around relatively low-mass stars below the Kraft break tend
to have low obliquities \citep{Winn2010a}. The presence of a
convective outer layer in these low-mass stars may be also crucial for
the generation of stellar magnetic activity. In contrast, several
previously reported low-obliquity systems (Kepler-25c, CoRoT-11b and
Kepler-8b) with more massive host stars did not show strong TCC.  This
may be ascribed to the lack of magnetic activity; these stars have
effective temperatures exceeding the Kraft break, and do not have
thick outer convective zones.

Kepler-45 is only the second M dwarf for which the stellar obliquity
has been measured, the other one being GJ\,436 \citep{Bourrier2017}.
As noted in the introduction, the distribution of stellar obliquities
has been observed to depend on the properties of the host star.
\citet{Winn2010a} found that the stars above and below the
\citet{Kraft} break have different obliquity distributions.  Stars
cooler than about 6250\,K have thick convective envelopes, while
hotter stars have radiative envelopes.  The differing obliquity
distributions may be related to the differing magnetic fields, tidal
dissipation rates, or rotational histories of the stars on either side
of this boundary.  Performing obliquity measurements on a wider range
of stellar types may help to clarify the situation.  This is
especially true for M dwarfs which may be completely convective and
for which obliquity measurements have been very limited.
Rossiter-McLaughlin observations have not been very successful for M
dwarfs, because of their higher stellar variability, slower stellar
rotation, and faint optical magnitudes, all of which hinder the
acquisition and interpretation of high-resolution spectra.  The TCC
method may be particularly useful for M dwarfs because high levels of
activity and slow rotation are actually favorable for the technique,
and because high-resolution spectroscopy is not needed.

As we mentioned previously, the TCC method is capable of identifying
both well-aligned and perfectly retrograde systems.  However, among
the 64 planetary systems and 24 eclipsing binaries in our sample, none
were found with a strong TCC corresponding to a retrograde orbit.  In
particular, all 10 of the planetary systems that had a strong TCC were
found to have very well-aligned orbits, even though retrograde orbits
would have been equally easy to detect.  All 10 of the systems feature
cool stars, below the Kraft break.  Thus the results for these 10
systems is further evidence that cool stars with hot Jupiters mainly
have prograde orbits rather than retrograde orbits.

\subsection{Constraints on stellar magnetic activity}

The properties of sunspots have been tracked since the 18th century.
It is well known that sunspots preferentially emerge at latitudes of
about $30^{\circ}$ north or south of the equator at the start of a
magnetic cycle \citep{Hathaway}. These active latitudes gradually
drift towards the equator as the magnetic cycle continues, giving rise
to the ``butterfly diagram'' when spot latitudes are plotted against
time.  It has also been noted that sunspots tend to cluster at certain
active longitudes \citep[see e.g.][]{Bumba1965}. Other solar magnetic
phenomena such as flares, coronal mass ejections, and X-ray emission
are also associated with active longitudes
\citep{Zhang2008,Gyenge2017,Zhang2007}.  Typically there are two
active longitudes on the Sun separated by 180$^{\circ}$, each of which
has a width of order 20$^{\circ}$ \citep{BU2003}.  At solar maximum
there can be as many as four active longitudes, and at solar minimum
active longitudes may disappear \citep{deToma2000}.  The lifetime of
active longitudes has been found to be seven rotation cycles
\citep{deToma2000} and perhaps even as long as a century
\citep{BU2003}. While the exact physical mechanism responsible for
persistently active longitudes is still unknown, they do not seem to
be a unique property of our Sun. \citet{Lanza2009} and \citet{GA2011}
found evidence for active longitudes on other stars too.

The transit tapestries of several low-obliquity systems with high SNR
and strong magnetic activity --- Kepler-17b, CoRoT-2b, Qatar-2b,
Kepler-71b and KOI-883.01 --- allow us to track the size, contrast,
lifetime and latitudinal distribution of the active regions along the
transit chords of these planet-hosting stars.  These are the
most magnetically active among the systems we have studied.  We
summarize the properties of the active regions of these systems in
Table \ref{active_regions}.

We note the following trends.  In all cases the Rossby number --- the
ratio between the rotation period and the convective overturn time ---
is smaller than that of the Sun.  The Rossby number is routinely used
for expressing thresholds and scaling relations in the study of
stellar magnetic activity and magnetic braking \citep[see,
  e.g.,][]{vanSaders}.  The strong magnetic activity that is
characteristic of our sample is linked to their fast rotation.  All
five stars are below the Kraft break.  The typical contrast of the
active regions is on the order of 10\% in the wide optical bandpasses
of the {\it Kepler} and {\it CoRoT} data.  The typical lifetime of the
active regions is on the order of 100 days, although we note that the
inferred lifetime of the active longitudes is degenerate with the rate
of latitudinal migration. If the active regions migrate away from the
transit chord, the active regions will no longer cause photometric
signatures in the transit light curve.

For all of the systems in Table \ref{active_regions}, there are
typically two active longitudes that are present on the same stellar
latitude simultaneously.  At a given time there may be no active
longitudes or as many as four, similar to the Sun.  Interestingly, the
size, contrast, and the lifetime of the different active regions on
the same stars are very similar to each other.  The latitude probed by
the planet is $\lesssim$20$^{\circ}$ in all cases, suggesting that
active regions tend to occur at these low latitudes, which is also
similar to the situation on the Sun.  If active regions had no
preferred latitudes, the systems in Table~\ref{active_regions} might
have shown a broader distribution of transit impact parameters.

\citet{Montet2017} noted a pattern in the origin of photometric
variability of Sun-like stars observed by {\it Kepler} ($T_{\rm eff}$
within 150\,K of the solar value and $\log g >4.2$).  The brightness
variations of stars with rotation periods shorter then 25 days are
consistent with those produced by dark spots, while slower rotators
are more consistent with bright spots (plages and faculae).  The most
magnetically active systems in our sample (Table \ref{active_regions})
are all consistent with being spot-dominated.  This is inferred from
the prevalence of positive residuals seen in the transit tapestries.
For the slower rotators in our sample, TrES-2 and Kepler-423, we did
not observe strong features in the transit tapestries, perhaps because
the active regions are faculae spanning a larger angular size.

\subsection{Eclipsing binaries}

We now turn to the eclipsing binaries in our sample.  We found eight
{\it Kepler} eclipsing binaries that show strong correlation and good
agreement between $P_{\text{phot}}$ and $P_{\text{tcc}}$ (see Table
\ref{tab:eb}).  It is more difficult to detect and interpret the TCC
signals for these systems.  Due to the lack of follow-up observations,
little is known about the properties of the stars in these
binaries. The stellar parameters in Table \ref{tab:eb} were estimated
from fitting stellar models and evolutionary tracks to archival
broadband photometry \citep{Prsa}. Moreover, only the properties of
the primary stars were reported.  Second, because of the larger radius
ratio in these binaries, the TCC method provides poor angular
resolution for probing the active regions.  Furthermore, some of the
eclipsing binaries were only observed in the long-cadence mode.  The
30-minute time averaging smears out the photometric patterns of active
regions.  Finally the orbital periods of some binaries are longer than
20 days.  This means we have a smaller number of eclipses to analyze,
and leaves the active regions more time in between eclipses for their
properties to change, weakening the TCC signal.

Nonetheless, we do see good evidence of active regions in the eclipse
tapestries of KIC~6307537, KIC~5193386 and KIC~5098444. Given the
shallow eclipse depths and long eclipse durations, one of the stars in
each of these binaries is likely evolved. This is further supported by
the deeper and flat-bottomed eclipse seen in KIC~6307537 and
KIC~5193386 which indicates the eclipsed star has a larger radius and
lower effective temperature than the eclipsing star.  The active
regions extend by about $10^{\circ}$ in longitude and are on the order
of 10\% fainter than the rest of the photosphere.  We did not detect
significant longitudinal migration except for KIC~5193386, where one
active region gradually split into two regions that drifted apart in
longitude.  These regions may be the two footprints of an emerging
magnetic flux tube.  The active regions on these evolved stars
persisted for longer than those on the planet-hosting dwarf stars, up
to at least a thousand days.

The low obliquities of these binaries could be a consequence of tidal
evolution, which tends to bring stars into a double-synchronous
aligned configuration.  However we note that the all eight of the
low-obliquity binary systems have not achieved a state of synchronized
rotation: $P_{\rm{orb}} \ne P_{\rm{rot}}$ (see Table
\ref{tab:eb}).  Thus the low-obliquity state may reflect the initial
conditions of formation, rather than the outcome of tidal evolution.

\subsection{Future prospects}

Crucial to the success of the TCC methods are light curves with high
SNR, high cadence, and continuous time coverage for many orbital and
rotation periods.  New opportunities for applying TCC will be provided
by the existing and upcoming space-based missions {\it TESS}
\citep{Ricker2014}, {\it CHEOPS} \citep{Cheops}, and {\it PLATO}.  The
      {\it TESS} mission, scheduled for launch in 2018, will provide a
      fresh sample of transiting planets which will be suitable
      targets for the TCC analysis.  The plan for the mission is to
      obtain data with 2-minute cadence for several hundred thousand
      preselected target stars.  This will be ideal for resolving the
      photometric signatures of active regions.  Most {\it TESS} stars
      will only be observed for one month during the two-year primary
      mission, but a small fraction of the stars (near the ecliptic
      poles) will be observed for as long as a year. Our best hopes
      are for {\it TESS} targets that are observed for at least a few
      months, similar in duration to the 80-day {\it K2} campaigns
      that have already yielded some strong detections with the TCC
      method.  Using the {\it TESS} mission simulations of
      \citet{Sullivan} and \citet{Bouma}, we have made a rough
      estimate that there will be a few dozen {\it TESS} planets with
      single-transit SNR~$>50$, amenable to TCC analysis.

\acknowledgements We are grateful to Ronald Gilliland for helping to
arrange short-cadence {\it Kepler} observations for many of the
targets analyzed here. Work by J.N.W.\ was supported through the {\it
  Kepler} Participating Scientist program and by NASA award
NNN12AA01C.

\bibliography{manuscript}

\clearpage
\begin{sidewaystable}

\fontsize{5}{5}\selectfont

\LongTables
\caption{List of Planetary Systems of Searched}
\begin{tabular}{lccccccccccccccll}
\hline
\hline
 \footnote{The columns are status of the planet, the orbital period $P_{\text{orb}}$, the scaled semi-major axis $a/R_{\star}$, the planetary mass $M_p$, the planetary radius $R_p$, the stellar mass $M_\star$, the stellar radius $R_\star$, the stellar effective temperature $T_{\text{eff}}$, the stellar rotation period measured from rotational modulation in the light curve $P_{\text{phot}}$, stellar rotation period measured from correlation in the residual flux $P_{\text{tcc}}$, the significance of correlation compared to the results of scrambling test $N_{\sigma}$, ratio between the stellar rotation period and orbital period $P_{\text{tcc}}$/$P_{\text{orb}}$,  the upper limit on the true obliquity $\Psi_{\text{Upper}}$, the obliquity constraint from literature $\lambda_{\text{lit}}$ and the references. The systems are sorted by the significance of correlation in the residual flux. Upper limit on the true obliquity is only calculated when a low stellar obliquity is detected.} & Status & $P_{\text{orb}}$ (days) & $a/R_{\star}$ & $M_p$ ($M_{\text{Jup}}$) & $R_p$ ($R_{\text{Jup}}$) & $M_\star$ ($M_{\odot}$) & $R_\star$ ($R_{\odot}$) & $T_{\text{eff}}$ (K) & $P_{\text{phot}}$ (days)& $P_{\text{tcc}}$ (days)& $N_{\sigma}$ & $P_{\text{tcc}}$/$P_{\text{orb}}$ & $\Psi_{\text{Upper}}$ ($^{\circ}$) & $\lambda_{\text{lit}}$($^{\circ}$) & Ref. \\
\hline
Kepler-17b & Confirmed & 1.4857108(2) & 5.48(2)  & 2.450(14) & 1.312(18) & 1.16(6) & 1.05(3) & 5781(85) & 11.9$\pm$1.1 & 11.8(7) & 59.6  & 7.94  & <10 & < 15 & \citet{Desert2011}\\
CoRoT-2b & Confirmed & 1.7429964(17) & 6.70(3) & 3.31(16) & 1.465(29) & 0.97(6) & 0.902(18) & 5575(66) & 4.530(68) & 4.5(3) & 22.6   & 2.58  & <4 & 4.7$\pm$12.3& \citet{Alonso2008}; \citet{Nutzman2011}\\
Qatar-2b & Confirmed & 1.3371182 (37)  & 6.52(10) & 2.487(86) & 1.144(35) & 0.740(37) & 0.713(18) & 4645(50) & 18.5 $\pm$1.9 & 18.2(4) & 18.0  & 13.61 & <11 &  0$\pm$10 & \citet{Esposito2017}\\
Kepler-71b & Confirmed & 3.905081(72) & 11.92(15) &  & 1.11(2) & 0.95(5) & 0.86(2) & 5591(105) & 19.87(18) & 19.7(8) & 10.7  & 5.04 & <6 & Prograde & \citet{Howell2010}; \citet{Holczer2015}  \\
KOI-883.01 & Candidate & 2.688899317(109) & 10.40 &  & 1.05(13) & 0.702(63) & 0.643(81) & 4809(151) & 9.11(11) & 9.1(2) & 9.2 & 3.38  & <4 & Prograde & ExoFOP; \citet{Holczer2015}  \\
Kepler-45b & Confirmed & 2.455239(5)  & 10.6 $\pm$ 1.0 & 0.51(9) & 0.96(11) & 0.59(6) & 0.55(11)  & 3820(90) & 15.8(2) & 16.7(8) & 6.3 & 6.80& <11 & & \citet{Johnson2012}\\
TrES-2b & Confirmed & 2.4706133738(187) & 7.98(17) & 1.253(52) & 1.189(25) & 0.98(62) & 1.000(36) & 5850(50) & 28.35(34) & 29.9$\pm$1.4 & 5.8 & 12.10 & <10  &9$\pm$12 & \citet{Winn2008}\\
Kepler-762b & Confirmed & 3.7705521(94) & 8.16 &  & 1.10(56) & 1.06(7) & 1.08(23) & 5944(124) & 4.045(25) & 4.0(1) & 5.2 & 1.06  & <16 & Prograde & ExoFOP; \citet{Holczer2015} \\
Kepler-423b & Confirmed & 2.68432850(7) & 8.106$^{+0.117}_{-0.259}$ & 0.595(81) & 1.192(52) & 0.85(4) & 0.95(4) & 5560(80) & 22.047(121) & 23.0(9) & 4.7  & 8.57  & <10 & & \citet{Gandolfi2015}\\
WASP-85b & Confirmed & 2.6556777(44) & 8.79(23) & 1.265(65) & 1.24(3)  & 1.04(7)  & 0.96(13) & 5685(65) & 13.28(41) & 15.2(3) & 4.5  & 5.69  &  <7 &  < 14 & \citet{Mocnik2016wasp85}\\
\hline
HAT-P-11b & Confirmed & 4.887802443(30) & 15.58$^{+0.17}_{-0.82}$ & 0.08243(90) & 0.42(13) & 0.81(3) & 0.75(2) & 4780(50) & 29.472(134) & 29.34(30) & 6.0  & 6.00 &  & $103^{+26}_{-10}$ & \citet{Winn2010hat}\\
Kepler-63b & Confirmed & 9.4341505(10) & 19.12(8) & <120 & 0.545(20) & 0.984(40) & 0.901(27) & 5576(50) & 5.401(14)  & 37.7(8) & 5.8  & 4.00 & &$-110^{+22}_{-14}$&\citet{Sanchis-Ojeda2013}\\
CoRoT-16b & Confirmed & 5.35227(20)  & 11.20$^{+1.21}_{-1.09}$ & 0.535(85) & 1.17(14) & 1.098(78) & 1.19(14) & 5650(100) &  & 4.0(4) & 3.9  & 0.75 &  &  & \citet{Ollivier2012} \\
WASP-55b & Confirmed & 4.465633(4) & 10.81(14) & 0.57(4) & 1.30(5) & 1.013(37) & 1.011(36) & 5900(100) &  & 27(2) & 3.5   & 6.05 &  & & \citet{Hellier2012} \\
CoRoT-17b & Confirmed & 3.7681(3) & 6.23(24) & 2.43(30) & 1.02(7) & 1.04(1) & 1.59(7) & 5740(80) &  & 17(2) & 3.5  & 4.51  &  &  & \citet{Csizmadia2011}\\
CoRoT-22b & Confirmed & 9.75598(11) & 17.30$^{+1.3}_{-0.57}$ & 12.2$^{+14}_{-8.8}$ & 0.435$_{-0.035}^{+0.015}$ & 1.099(49) & 1.136(90) & 5939(120) &  & 25(2) & 3.2  & 2.56  &  & &\citet{Moutou2014}\\
CoRoT-24b & Confirmed & 5.1134(6) & 10.9 $\pm$ 2.8 & <0.018 & 0.33(4) & 0.91(9) & 0.86(9) & 4950(150) &  & 14(1) & 3.2   & 2.74 & & &\citet{Alonso2014} \\
Kepler-25c\footnote{\label{note1}Systems whose stellar obliquity was previously reported to be low yet did not show strong TCC. These systems are likely magnetically inactive: the host stars are above the Kraft break; the light curve lacks rotational modulation. Alternatively the high impact parameter of the planets indicate that the transit chord might have missed the active latitude.} & Confirmed & 12.7203678(35) & 18.52(24) & 0.045(8) & 0.4598(54)  & 1.22(6)  & 1.36(13) & 6190(80) & 23.21(27) & 29.2(7) & 3.2   & 2.30 &  &0.5$\pm$5.7 & \citet{Albrecht2013}\\
CoRoT-13b & Confirmed & 4.03519(3) & 10.81(32) & 1.308(66)  & 0.885(14) & 1.09(2) & 1.01(3) & 5945(90) &  & 20.6(6) & 3.1   & 5.11 &  & &\citet{Cabrera2010}\\
CoRoT-11b$^{\rm \ref{note1}}$ & Confirmed & 2.994330(11) & 6.890(80) & 2.33(34) & 1.430(33)  & 1.27(5)  & 1.36(13) & 6440(120) &  & 30(4) & 3.0   & 10.02 &   &0.1$\pm$2.6 & \citet{Gandolfi2012}\\
CoRoT-19b & Confirmed & 3.89713(2) & 6.7(1) & 1.11(6) & 1.29(5) & 1.21(6)  & 1.65(4) & 6090(70) &  & 10(1) & 3.0   & 2.57 &  &  -52$^{+27}_{-22}$ &\citet{Guenther2012}\\
WASP-47b$^{\rm \ref{note1}}$ & Confirmed & 4.1591287(49) & 9.715(50) & 1.123(510) & 1.17(79)  & 1.11(49) & 1.16(26) & 5576(67) &  & 6.3(4) & 3.0   & 1.51 &   & 0$\pm$24 & \citet{Sanchis-Ojeda2015}\\
CoRoT-14b & Confirmed & 1.51214(13) & 4.78(28) & 7.6(6) & 1.09(7) & 1.13(9) & 1.21(8) & 6035(100) &  & 5.4(8) & 2.9  & 3.57 &  &  & \citet{Tingley2011} \\
CoRoT-8b & Confirmed & 6.21229(3) & 17.6(4) & 0.22(3) & 0.569(20) & 0.88(4)  & 0.77(2) & 5080(80)  &  & 25.3(0.3) & 2.8   & 4.07  &  &  & \citet{Borde2010}\\
CoRoT-25b & Confirmed & 4.86069(6) & 10.2$^{+1.1}_{-0.5}$ & 0.27(4) & 1.08 $_{-0.1}^{+0.3}$ & 1.09$_{-0.05}^{+0.11}$ & 1.19 $_{-0.03}^{+0.14}$ & 6040(90) &  & 31(1) & 2.8   & 6.38 &  &  & \citet{Almenara2013}\\
KOI-212.01 & Candidate & 5.6959023(15) & 12.2 &  & 0.58(23) & 1.013(135) & 0.94(38) & 6106(156) & 16.295(336) & 33(1) & 2.8   & 5.79 &  &  & ExoFOP\\
CoRoT-6b & Confirmed & 8.886593(4) & 17.94(33) & 2.96(34) & 1.166(35) & 1.055(55) & 1.025(26) & 6006(73) & 6.35(13) & 5.4(2) & 2.8  & 0.61 &  &  & \citet{Fridlund2010}\\
Kepler-13Ab  & Confirmed & 1.7633587(20)  & 3.16(8) & 6.6$\pm$1.5 & 1.335$_{-0.264}^{+0.484}$ & 1.49$_{-0.18}^{+0.22}$ & 1.80$_{-0.36}^{+0.65}$ & 8500(400) &  & 3.12(10) & 2.7   & 1.77  &  & 58.6$\pm$2.0 & \citet{Masuda2015} \\
CoRoT-3b &  Brown Dwarf & 4.256800(5) & 7.8(4) & 21.7 $\pm$ 1.0 & 1.01(7) & 1.41(8) & 1.44(8)  & 6740(140) &  & 24.3(5) & 2.7  & 5.71 &  &   -37.6$^{+22.3}_{-10.0}$ &\citet{Deleuil2008}; \citet{Triaud2009}\\
HATS-9b & Confirmed & 1.9153073(52) & 4.36$^{+0.10}_{-0.25}$ & 0.837(29) & 1.40(6) & 1.28(13) & 1.32(7) & 5366(70) &  & 13.6$\pm$1.1 & 2.7  & 7.10 &  &  & \citet{Brahm2015} \\
Kepler-471b & Confirmed & 5.01423457(59) & 8.3 &  &  0.94(69)&  1.314(257) & 1.271(927) & 6475(169) & 26.51(57) & 20.5(8) & 2.7   & 4.09 &  &  & ExoFOP\\
Kepler-494b & Confirmed & 8.0251182(23) & 11.5 &  & 0.637(118) & 1.10(7) & 1.26(23) & 5513(184) & 8.985(186)  & 30.1(4) & 2.6   & 3.75 &  &  & ExoFOP\\
CoRoT-12b & Confirmed & 2.828042(13) & 7.74(18) & 0.917(70) & 1.44(13) & 1.078(72) & 1.116(92) & 5675(80) &  & 9.8(8) & 2.6  & 3.47 &  &  & \citet{Gillon2010} \\
CoRoT-18b$^{\rm \ref{note1}}$ & Confirmed & 1.9000693(28) & 6.35(40) & 3.47(38) & 1.31(18) & 0.95(15) & 1.00(13) & 5440(100) & 5.4(4) & 15.7(7) & 2.6  & 8.26 &  &   -10$\pm$20 & \citet{Hebrard2010} \\
CoRoT-28b & Confirmed & 5.20851(38) & 7.29(16) & 0.484(87) & 0.955(66)  & 1.01(14) & 1.78(11) & 5150(100) &  & 41(3) & 2.6  & 7.87 &  &  & \citet{Cabrera2015} \\
Kepler-448b$^{\rm \ref{note1}}$ & Confirmed & 17.8552333(9) & 18.8(4) & < 10  & 1.43(13) & 1.452(93) & 1.63(15) & 6820(120) & 1.245(124)  & 21.0(6) & 2.6  & 1.18 &  &  12.6$^{+3.0}_{-2.9}$& \citet{Bourrier2015}\\
CoRoT-21b & Confirmed & 2.72474(14) & 4.60(26) & 2.26(31) & 1.30(14) & 1.29(9)  & 1.95(21) & 6200(100) &  & 16.7(5) & 2.5   & 6.13 &  &  & \citet{Patzold2012} \\
Kepler-420b & Confirmed & 86.647661(34) & 155.4$\pm$2.4 & 1.45(35) & 0.94(12) & 0.99(5)  & 1.13(14) & 5520(80) &  & 11.7(7) & 2.5   & 0.14 &  & 75$^{+32}_{-46}$ & \citet{Santerne2014}\\
Kepler-74b & Confirmed & 7.340718(1) & 11.8$^{+1.4}_{-0.8}$ & 0.68(9) & 1.32(14)  & 1.4$_{-0.11}^{+0.14}$ & 1.51(14) & 6050(110) & 26.7(8) & 36$\pm$1.2 & 2.4   & 4.90&  &  & \citet{Hebrard2013}\\
Kepler-96b & Confirmed & 16.2385 & 21.2 & 0.027(11) & 0.238(20) & 1.00(6)  & 1.02(9) & 5690(73) & 14.922(83) & 42.7$\pm$1.7 & 2.3  & 2.63 &  &  & \citet{Marcy2014}\\
Kepler-485b & Confirmed & 3.2432598(18) & 8.2 &  & 1.286$^{+0.302}_{-0.145}$ & 1.07(8)  & 1.09(26) & 5958(153) & 30.72(45) & 22.8$\pm$1.1 & 2.3  & 7.03  & & & ExoFOP\\
CoRoT-7b & Confirmed & 0.85353(2) & 4.27(20) & 0.0149(30) & 0.149(8) & 0.93(3) & 0.87(4) & 5275(75) & 22.4$\pm$3.6 & 2.5(2) & 2.3  & 2.93 &  &  & \citet{Leger2009}\\
CoRoT-29b & Confirmed & 2.850570(6) & 9.22(19) & 0.85(20) & 0.90(16)  & 0.98(14) & 0.90(12) & 5260(100) &  & 16.1$\pm$1.5 & 2.3  & 5.65 &  &  &\citet{Cabrera2015} \\
WASP-107b & Confirmed & 5.7214742(43) & 18.164(37) & 0.12(1) & 0.948(30)  & 0.69(5)  & 0.66(2) & 4430(120) & 17.1$\pm$1.0 & 10.0(5) & 2.3  & 1.75 &  & 40-140 & \citet{Anderson2017,Dai2017}\\
Kepler-7b & Confirmed & 4.885525(40) & 6.623(49) & 0.443(42) & 1.614(15) & 1.36(3) & 2.02(2) & 5933(44) & 15.02(21) & 54.1(9) & 2.2  & 11.07  &  & &\citet{Demory2011} \\
Kepler-8b$^{\rm \ref{note1}}$ & Confirmed & 3.52254(5) & 6.97(24) & 0.60(19) & 1.419(58)  & 1.213(63) & 1.486(62) & 6213(150) & 28.64(32) & 27.1(3) & 2.2  & 7.69 &    & 5$\pm$7 & \citet{Jenkins2010}; \citet{Albrecht2012}\\
CoRoT-23b & Confirmed & 3.6313(1) & 6.85(60) & 2.8(3) & 1.08(13) & 1.14(8) & 1.61(18)  & 5900(100) &  & 7.3$\pm$1.5 & 2.2   & 2.01 &  &  & \citet{Rouan2012}\\
Kepler-522b & Confirmed & 38.58422(46) & 28.4 &  & 0.616(134) & 1.54(14) & 1.98(43) & 6267(109) & 4.895(34) & 27.7(5) & 2.2  & 0.72 &  &  & ExoFOP\\
CoRoT-26b & Confirmed & 4.20474(5) & 6.28$^{+0.01}_{-0.52}$ & 0.52(5) & 1.26(13) & 1.09(6) & 1.79(18) & 5590(100) &  & 15.1(9) & 2.2  & 3.59 &  &  & \citet{Almenara2013}\\
CoRoT-27b & Confirmed & 3.57532(6) & 9.48(95) & 10.39(55) & 1.007(44)  & 1.05(11) & 1.08(18) & 5900(120) &  & 35.5(5) & 2.2  & 9.93 &  & & \citet{Parviainen2014}\\
HATS-11b & Confirmed & 3.6191613(99) & 6.88(27) & 0.85(12) & 1.510(78) & 1.00(6) & 1.44(6)  & 6060(150) &  & 14(3) & 2.1 & 3.87 &  &  & \citet{Rabus2016}\\
Kepler-539b & Confirmed & 125.63243(71) & 103.3$\pm$1.1 & 0.97(29) & 0.747(16)  & 1.048(45) & 0.952(20) & 5820(80) & 11.769(16) & 16.0(2) & 2.0  & 0.13 & &  & \citet{Mancini2016} \\
CoRoT-4b & Confirmed & 9.20205(37) & 16.54(18) & 0.72(8) & 1.19(5) & 1.10(2) & 1.15(3) & 6190(60) & 8.7$\pm$1.1 & 31(2) & 2.0   & 3.37 &  &  & \citet{Moutou2008} \\
HAT-P-7b & Confirmed & 2.204737(17) & 4.1502(39) & 1.741(28) & 1.431(11) & 1.51(5) & 2.00(2) & 6259(32) & 28.10(36) & 0.8(2) & 2.0  & 0.36 &  &   155$\pm$37 & \citet{Albrecht2012}; \citet{Masuda2015}\\
WASP-75b & Confirmed & 2.484193(3) & 6.40(14) & 1.07(5) & 1.270(48) & 1.14(7) & 1.26(4) & 6100(100) & 13.7$\pm$1.1  & 22$\pm$3 & 1.9  & 8.87  & & & \citet{Gomez2013}\\
Kepler-548b & Confirmed & 4.45419434(38) & 14.7 &  & 1.070(195 )  & 0.93(9) & 0.90(16) & 5359(178) & 31.6(3.1) & 115(2) & 1.9   & 25.82 &  &  & ExoFOP\\
CoRoT-10b & Confirmed & 13.2406(2) & 31.33 $\pm$ 2.15 & 2.75(16) & 0.97(5) & 0.89(5)  & 0.79(5) & 5075(75) &  & 30(2) & 1.9  & 2.27 &  &  & \citet{Bonomo2010}\\
CoRoT-1b & Confirmed & 1.5089557(64) & 4.92(8) & 1.03(12) & 1.49(8)  & 0.95(15) & 1.11(5) & 5950(150) &  & 14(3) & 1.8  & 9.28 &  &  & \citet{Barge2008}\\
CoRoT-15b & Brown Dwarf & 3.06036(3) & 6.68$^{+0.49}_{-1.04}$ & 63.3 $\pm$ 4.1 & 1.12(30) & 1.32(12) & 1.46(31)& 6350(200) &  & 19.7(6) & 1.7  & 6.44 &  &  &\citet{Bouchy2011}\\
KOI-425.01 & Candidate & 5.42834472(49) & 22.1 &  & 1.23(53) & 1.075(187) & 0.965(419) & 5936(166) & 5.301(177) & 11.2(8) & 1.7  & 2.06 &  &  & ExoFOP\\
WASP-118b & Confirmed & 4.0460435(44) & 6.92(11) & 0.514(20) & 1.440(36) & 1.320(35) & 1.696(29) & 6410(125) &  & 7.8(6) & 1.6  & 1.93 &  &  & \citet{Hay2016}\\
CoRoT-5b & Confirmed & 4.0378962(19) & 8.970(47) & 0.467(47) & 1.330(47) & 1.00(2)  & 1.186(40)  & 6100(65) &  & 2.5(2) & 1.5   & 0.62 &  &  & \citet{Rauer2009} \\
CoRoT-20b & Confirmed & 9.24285(30) & 18.95(73) & 4.24(23) & 0.84(4) & 1.14(8)  & 1.02(5) & 5880(90) &  & 12(4) & 1.1  & 1.30 &  &  & \citet{Deleuil2012}\\
CoRoT-9b & Confirmed & 95.2738(14 )& 93(3) & 0.84(7) & 0.94(4) & 0.99(4) & 0.94(4) & 5625(80) &  & 26(1) & 0.7  & 0.27 &  &  & \citet{Deeg2010}\\

\hline
\end{tabular}
\label{tab:planet}
\end{sidewaystable}

\clearpage
\begin{sidewaystable}

\fontsize{6}{6}\selectfont

\LongTables
\caption{List of Eclipsing Binary Systems Searched}
\begin{tabular}{lcccccccccccccccc}
\hline
\hline
KIC & Status & $P_{\text{orb}}$ (days) & ecos($\omega$) \footnote{Eccentricity constraints from \citet{vanEylenEB}}& $a/R_{\star}$ & $R_1/R_2$& $M_\star$ ($M_{\odot}$) & $R_\star$ ($R_{\odot}$) & $T_{\text{eff}}$ (K) & $P_{\text{phot}}$ (days)& $P_{\text{tcc}}$ (days)& $N_{\sigma}$ & $P_{\text{tcc}}$/$P_{\text{orb}}$ & $\Psi_{\text{Upper}}$ ($^{\circ}$) & Ref. \\
6307537 & EB (Algol Detached) & 29.74455(33) & 0.0042	(8)& 11.54 & 0.23861(45) & 0.839(73) & 0.788(94) & 4892(192) & 78(3) & 80.4 $\pm$1.6 & 11.4  & 2.70  & <6 &   ExoFOP \\
5193386 & EB & 21.378308(138) & -0.0022(15) & 9.31 & 0.261272(288) & 0.883(115)  & 3.484(909) & 4905(66) & 26.0(8) & 25.8(3) & 9.5   & 1.21  & <14 &   ExoFOP \\
6603756 & EB (Algol Detached) & 5.204283146(797) & & 7.09 & 0.177941(42) & 0.798(134) & 0.86(34) & 5332(188) & 6.128(54) & 6.0(1) & 9.2  & 1.15  & <12 &   ExoFOP \\
5098444 & EB & 26.94923943(977)&  & 19.24 & 0.131908(196) & 0.526(30) & 0.512(26) & 4723(135) & 23.49(19) & 23.40(24) & 7.9  & 0.87  & <8 &   ExoFOP \\
7767559 & EB & 4.409409352(787) && 9.15 & 0.106974(96) & 0.811(144) & 0.975(405) & 5600(186) & 5.02(20) & 5.1(1) & 7.7   & 1.16  & <7 &    ExoFOP \\
5376836 & EB & 3.479425021(583)&  & 7.0 & 0.136233(295) & 0.929(108) & 0.878(248) & 5907(158) & 3.63(21) & 3.69(16) & 7.0    & 1.06  & <22 &   ExoFOP \\
3128793 & EB & 24.68637(57) & & 12.24 & 0.0909(5) & 0.950(201) & 5.132(1.336) & 4648(69) & 67.1 $\pm$2.2 & 67.8 $\pm$2.0 & 5.4  & 2.75  & <3 &   ExoFOP \\
5282049 & EB & 5.91037138(84)&  & 12.48 & 0.12539(85) & 1.029(134) & 0.958(291) & 6200(200) & 17.5 $\pm$1.0 & 16.9(5) & 5.0    & 2.86  & <4 &   ExoFOP \\
\hline
4737267 & EB & 9.52407775(158) &  & 8.71 & 0.416822(283) & 1.112(189) & 3.371(0.641) & 5180(62) & 9.52(7) & 9.25(10) & 12.6   & 0.97  &  &   ExoFOP \\
5270698 & EB (Algol Detached) & 3.964333792(178)&  & 7.01 & 0.15095(64) & 1.258(106) & 2.318(380) & 5758(77) & 4.004(33) & 15.38(25) & 10.8   & 3.88  &  &   ExoFOP \\
8081482 & EB (Algol Detached) & 2.819454482(498) && 5.42 & 0.254718(261) & 1.074(305) & 1.074(370) & 5767(161) & 2.762(86) & 2.8(2) & 9.0   & 0.99  &  &   ExoFOP \\
6548447 & EB & 10.76838965(176) & -0.0873(48) & 6.55 & 0.297635(114) & 0.921(101) & 1.643 (851) & 5226(173) & 5.380(39) & 9.6(1) & 7.5   & 0.89  &  &   ExoFOP \\
7940533 & EB & 3.905532661(473) &0.0070(36)& 5.04 & 0.242875(1) & 0.812(100) & 0.798(152) & 5495(169) & 3.842(96) & 7.5(2) & 5.9  & 1.92  &  &  ExoFOP \\
3955867 & EB & 33.6575(83) &  & 6.54 & 0.110724(28) & 0.871(151) & 6.447(1.781) & 4774(71) & 33.26(39) & 33.4 $\pm$1.1 & 5.2  & 0.99  & &   ExoFOP \\
8180020 & EB & 5.803127542(553)& & 8.93 & 0.231357(76) & 0.876(092) & 0.795(188) & 5533(182) & 6.54(10) & 16.2(2) & 5.1  & 2.79  &  &   ExoFOP \\
8230809 & EB (Algol Detached) & 4.078353619(475)&0.0108(113) & 5.7 & 0.182022(95) & 0.83(105) & 0.867(221) & 5872(177) & 3.989(81) & 7.8(4) & 5.1   & 1.91  &  &   ExoFOP \\
5370302 & EB & 3.904344085(366) &0.0114(118)& 7.812 & 0.127552(175) & 0.989(116) & 0.898(260) & 5810(157) & 3.63(21) & 7.8(6) & 4.3   & 2.00  &  &   ExoFOP \\
6613006 & EB (Algol Detached) & 7.388833197(504) && 10.04 & 0.117062(64) & 0.926(99) & 0.838(241) & 5845(139) &  & 34.2(2) & 4.3   & 4.63  &  &   ExoFOP \\
5306862 & EB & 2.025580507(270)& -0.0052(27)& 4.44 & 0.197412(202) & 1.431(201) & 1.990(578) & 6348(170) & ELV\footnote{unable to measure $P_{\rm phot}$ because ellipsoidal light variation dominates the flux variation} & 1.0(5) & 4.0     & 0.49  &  &   ExoFOP \\
6205460 & EB & 3.7228051(92) &0.0155(0.0018) & 1.96 & 0.324154(195) & 1.166(177) & 2.59(1.10) & 5425(296) & 2.47(1) & 0.6(2) & 3.3   & 0.16  & &  ExoFOP \\
5700330 & EB & 53.2204218(52) && 22.96 & 0.25037(46) & 1.240(96) & 2.057(292) & 5653(84) & 26.18(27) & 17.5 $\pm$1.2 & 3.0   & 0.33   &  & ExoFOP \\
6694186 & EB & 5.554226627(175) &0.0142(12)& 10.98 & 0.349048(83) & 0.756(119) & 0.843(181) & 5433(145) & 5.472(83) & 7.24(20) & 2.9  & 1.30  &  &   ExoFOP \\
5014753 & EB & 3.17062650(50) &0.0057(111)& 8.96 & 0.12876(30) & 1.042(159) & 1.034(330) & 6068(190) & 3.66(11) & 21(4) & 2.6  & 6.62  &  &   ExoFOP \\
2438502 & EB (Variable of RS CVn) & 8.36061289(25) && 7.45 & 0.16632(162) & 1.343(144) & 2.445(420) & 5463(87) & 8.31(28) & 8.4(5) & 1.9  & 1.00  &  &   ExoFOP \\ 
\hline

\hline
\label{tab:eb}
\end{tabular}

\end{sidewaystable}

\begin{table*}
\centering
\caption{Properties of Active Regions inferred from TCC}
\begin{tabular}{ccccccccc}
\hline
\hline
System &
$T_{\rm eff}$ (K) &
$P_{\rm rot}$ (days) &
Ro\footnote{Rossby Number, using the prescription for convective overturn time of \citet{Noyes}} &
Longitudinal Size\footnote{Longitudinal width of the active regions.}($^{\circ}$) &
Contrast\footnote{Average contrast between the active regions and the surrounding photosphere, in the {\it Kepler} band.} &
Lifetime (days) &
$N_{\rm AL}$\footnote{Number of the active longitudes} &
Active Latitude ($^{\circ}$)
\\
\hline
Sun\footnote{https://nssdc.gsfc.nasa.gov/planetary/factsheet/sunfact.html; \citet{BU2003}, \citet{deToma2000}, \citet{Kilcik}} & 5777 & 25.05 & 2.3 & $\lesssim$20 & $\approx$3-33\% (spots) & $\approx$200\footnote{Lifetime of active longitudes instead of sunspots \citep{deToma2000}.} & 2 (0-4) & 30-10  \\
Kepler-17 & 5800 & 11.9 & 0.81 &5-20 & 10-20\% & 100-200 & 1-4 & 16$ \pm $8 \\ 
CoRoT-2 & 5600 & 4.5 & 0.22 &10-20 & 7-17\% & >150 & 2 & 13$\pm$10 \\
Qatar-2 & 4600 & 18.5 & 0.76 &15-25 & 3-7\% & >80 & 2 & 2$\pm$9 \\
Kepler-71 & 5600 & 19.9 & 1.04 &5-15 & 10-20\% & 100-200 & 2 & 2$\pm$8 \\
KOI-883 & 4800 & 9.11 & 0.40 &10-40 & 10-20\% & $\approx$100 & 0-2 & 0$\pm$10 \\
\hline
\end{tabular}
\label{active_regions}
\end{table*} 

\end{document}